\documentclass{ws-ijqi}
\usepackage{bm,bbm}
\renewcommand{\trimmarks}{\relax}

\newcommand{\JournalReference}{%
\raisebox{10ex}[0pt][0pt]{\makebox[0pt][l]{\normalfont\small\hspace{-7.8em}%
\underline{Journal reference: International Journal of Quantum Information \textbf{8} (2010) 535--640}}}}

\renewcommand{\draftnote}{\relax}

\def\currenttime{\hour=\time \divide\hour by 60 \number\hour:%
  \multiply\hour by 60 \minute=\time \global\advance\minute by -\hour%
  \ifnum\minute<10 0\number\minute\else\number\minute\fi}

\newcommand{\ket}[1]{\vert #1 \rangle}
\newcommand{\bra}[1]{\langle #1 \vert}
\newcommand{\braket}[2]{\langle #1 | #2 \rangle}

\newcommand{\Exp}[1]{\mathrm{e}^{\mbox{\footnotesize$#1$}}}
\newcommand{\power}[1]{^{\mbox{\footnotesize$#1$}}}
\newcommand{\subscr}[1]{_{\mbox{\footnotesize$#1$}}}
\newcommand{\I}{\mathrm{i}}

\newcommand{\D}{\mathrm{d}}
\newcommand{\ds}{\displaystyle}
\newcommand{\tr}[2][]{\mathrm{tr}_{#1}\bigl\{ #2 \bigr\}}

\newcommand{\m}{\mathnormal{\textsc{m}}}

\newcommand{\repr}{\mathrel{\widehat{=}}}

\numberwithin{equation}{section}
\renewcommand{\theequation}{\thesection.\arabic{equation}}

\newcommand{\JournalTitle}[1]{\textit{#1}\ }
\newcommand{\IJQI}{\JournalTitle{Int.\ J. Quant.\ Inf.}\ }
\newcommand{\PR}{\JournalTitle{Phys.\ Rev.}}
\newcommand{\PRA}{\JournalTitle{Phys.\ Rev.\ A}}
\newcommand{\PRD}{\JournalTitle{Phys.\ Rev.\ D}}
\newcommand{\PRL}{\JournalTitle{Phys.\ Rev.\ Lett.}}
\newcommand{\PLA}{\JournalTitle{Phys.\ Lett.\ A}}
\newcommand{\ZPhys}{\JournalTitle{Z.~Phys.}}
\newcommand{\PNAS}{\JournalTitle{Proc.\ Natl.\ Acad.\ Sci.\ U.~S.~A.}}

\newcommand{\PhRep}{\JournalTitle{Phys.\ Rep.}}
\newcommand{\JPhysAmg}{\JournalTitle{J.~Phys.\ A: Math.\ Gen.}}
\newcommand{\JPhysAmt}{\JournalTitle{J.~Phys.\ A: Math.\ Theor.}}
\newcommand{\OC}{\JournalTitle{Opt.\ Commun.}}
\newcommand{\JMO}{\JournalTitle{J.~Mod.\ Opt.}}
\newcommand{\JMaPh}{\JournalTitle{J.~Math.\ Phys.}}
\newcommand{\OSID}{\JournalTitle{Open Syst.\ Inf.\ Dyn.}}
\newcommand{\QIC}{\JournalTitle{Quant.\ Inf.\ Comp.}}
\newcommand{\LAA}{\JournalTitle{Lin.\ Alg.\ Appl.}}

\newcommand{\eprint}[2][quant-ph]{\mbox{e-print} arXiv:#1/\linebreak[0]#2.}

\newcommand{\Eprint}[3][quant-ph]{%
\mbox{e-print} arXiv:#2\linebreak[0][#1] (#3).}

\usepackage{overcite}\let\refcite\citen\relax

\makeatletter
\newcommand\tableofcontents{%
    \section*{Contents}%
    \@starttoc{toc}%
    }
\newlength{\seclabwidth}
\newcommand*\l@section{\@dottedtocline{1}{0em}{\seclabwidth}}
\newcommand*\l@subsection{\@dottedtocline{2}{1em}{2.1em}}
\newcommand*\l@subsubsection{\@dottedtocline{3}{3.1em}{2.8em}}
\newcommand*\l@paragraph{\@dottedtocline{4}{5.6em}{3.5em}}
\newcommand*\l@subparagraph{\@dottedtocline{5}{9.1em}{3.5em}}
\newcommand\@pnumwidth{1.55em}
\newcommand\@tocrmarg{2.55em}
\newcommand\@dotsep{4.5}
\makeatother

\usepackage{curves}

\newcommand{\MK}[2]{\put(#1){\makebox(0,0)[c]{#2}}}

\newcommand{\MKtable}[1]{%
\ifthenelse{\equal{#1}{0}}{%
\MK{0,0}{0}\MK{0,1}{0}\MK{0,2}{0}\MK{0,3}{0}%
\MK{1,0}{1}\MK{1,1}{1}\MK{1,2}{\textbf{1}}\MK{1,3}{1}%
\MK{2,0}{2}\MK{2,1}{2}\MK{2,2}{2}\MK{2,3}{2}%
\MK{3,0}{3}\MK{3,1}{3}\MK{3,2}{3}\MK{3,3}{3}%
}{\relax}
\ifthenelse{\equal{#1}{1}}{%
\MK{0,0}{0}\MK{0,1}{1}\MK{0,2}{2}\MK{0,3}{3}%
\MK{1,0}{1}\MK{1,1}{0}\MK{1,2}{\textbf{3}}\MK{1,3}{2}%
\MK{2,0}{2}\MK{2,1}{3}\MK{2,2}{0}\MK{2,3}{1}%
\MK{3,0}{3}\MK{3,1}{2}\MK{3,2}{1}\MK{3,3}{0}%
}{\relax}
\ifthenelse{\equal{#1}{2}}{%
\MK{0,0}{0}\MK{0,1}{2}\MK{0,2}{3}\MK{0,3}{1}%
\MK{1,0}{1}\MK{1,1}{3}\MK{1,2}{\textbf{2}}\MK{1,3}{0}%
\MK{2,0}{2}\MK{2,1}{0}\MK{2,2}{1}\MK{2,3}{3}%
\MK{3,0}{3}\MK{3,1}{1}\MK{3,2}{0}\MK{3,3}{2}%
}{\relax}
\ifthenelse{\equal{#1}{3}}{%
\MK{0,0}{0}\MK{0,1}{3}\MK{0,2}{1}\MK{0,3}{2}%
\MK{1,0}{1}\MK{1,1}{2}\MK{1,2}{\textbf{0}}\MK{1,3}{3}%
\MK{2,0}{2}\MK{2,1}{1}\MK{2,2}{3}\MK{2,3}{0}%
\MK{3,0}{3}\MK{3,1}{0}\MK{3,2}{2}\MK{3,3}{1}%
}{\relax}
\ifthenelse{\equal{#1}{4}}{%
\MK{0,0}{0}\MK{0,1}{1}\MK{0,2}{2}\MK{0,3}{3}%
\MK{1,0}{0}\MK{1,1}{1}\MK{1,2}{\textbf{2}}\MK{1,3}{3}%
\MK{2,0}{0}\MK{2,1}{1}\MK{2,2}{2}\MK{2,3}{3}%
\MK{3,0}{0}\MK{3,1}{1}\MK{3,2}{2}\MK{3,3}{3}%
}{\relax}
}

\newcommand{\mnGrid}[1]{%
\setlength{\unitlength}{12pt}
\begin{picture}(0,0)(1.8,1.5)
\renewcommand{\xscale}{0}\renewcommand{\xscaley}{1}
\renewcommand{\yscale}{0}\renewcommand{\yscalex}{1}
\MKtable{#1}
\renewcommand{\xscale}{1}\renewcommand{\xscaley}{0}
\renewcommand{\yscale}{1}\renewcommand{\yscalex}{0}
\end{picture}
\setlength{\unitlength}{1pt}
}

\begin{document}

%%%% file name: MUB-Front.tex
%%%% input file for MUB.tex 
%%%%
%%%% last changes on 20 April 2010 by Berge
%%%% typo corrected on 27 April 2010
%%%%  
%%%%%%%%%%%%%%%%%%%%%%%%%%%%%%%%%%%%%%%%%%%%%%

\markboth{T. Durt, B.-G. Englert, I. Bengtsson, and K. \.Zyczkowski}
{On mutually unbiased bases}

\title{\JournalReference%
\uppercase{On mutually unbiased bases}}

\author{\uppercase{Thomas Durt}}
\address{TONA Vrije Universiteit Brussel,
Pleinlaan 2, B-1050 Brussels, Belgium\\
thomdurt@vub.ac.be}

\author{\uppercase{Berthold-Georg Englert}}
\address{Centre for Quantum Technologies, %
National University of Singapore\\
3 Science Drive 2, Singapore 117543, Singapore\\
and Department of Physics, National University of Singapore\\
2 Science Drive 3, Singapore 117542, Singapore\\
cqtebg@nus.edu.sg}

\author{\uppercase{Ingemar Bengtsson}}
\address{Stockholms Universitet, Fysikum, Alba Nova, %
106\,91 Stockholm, Sweden\\
ingemar@physto.se}

\author{\uppercase{Karol \.Zyczkowski}}
\address{Instytut Fizyki Uniwersytetu Jagiello\'nskiego, %
ul.\ Reymonta 4, 30-059 Krak\'ow, Poland\\
and  Centrum Fizyki Teoretycznej PAN,  Al. Lotnik{\'o}w 32/44, %
02-668 Warszawa, Poland \\
karol@tatry.if.uj.edu.pl}

\maketitle

\begin{history}
(Posted on the arXiv on 20 April 2010)
\end{history}

\begin{abstract} 
Mutually unbiased bases for quantum degrees of freedom are central to
all theoretical investigations and practical exploitations of complementary
properties. 
Much is known about mutually unbiased bases, but there are also a fair number
of important questions that have not been answered in full as yet.
In particular, one can find maximal sets of ${N+1}$ mutually unbiased bases in
Hilbert spaces of prime-power dimension ${N=p^\m}$, with $p$ prime and $\m$ a
positive integer, and there is a continuum of mutually unbiased bases for a
continuous degree of freedom, such as motion along a line. 
But not a single example of a maximal set is known if the dimension is another
composite number ($N=6,10,12,\dots$).

In this review, we present a unified approach in which the basis states are
labeled by numbers ${0,1,2,\dots,N-1}$ that are both elements of a Galois
field and ordinary integers.
This dual nature permits a compact systematic construction of maximal sets of
mutually unbiased bases when they are known to exist but throws no light on the
open existence problem in other cases.
We show how to use the thus constructed mutually unbiased bases in
quantum-informatics applications, including dense coding, teleportation,
entanglement swapping, covariant cloning, and state tomography, all of which
rely on an explicit set of maximally entangled states (generalizations of the
familiar two--q-bit Bell states) that are related to the mutually unbiased
bases. 

There is a link to the mathematics of finite affine planes.
We also exploit the one-to-one correspondence between unbiased bases and
the complex Hadamard matrices that turn the bases into each other.
The ultimate hope, not yet fulfilled, is that open questions about mutually
unbiased bases can be related to open questions about Hadamard matrices or
affine planes, in particular the notorious existence problem for dimensions
that are not a power of a prime. 
   
The Hadamard-matrix approach is instrumental in the very recent advance,
surveyed here, of our understanding of the ${N=6}$ situation.
All evidence indicates that a maximal set of seven mutually unbiased bases
does not exist --- one can find no more than three pairwise unbiased bases ---
although there is currently no clear-cut demonstration of the case.
\end{abstract}


\tableofcontents

\section*{Acronyms}
\addcontentsline{toc}{section}{Acronyms}
\begin{tabular}{rl}
  MU & mutually unbiased\\
 MUB & mutually unbiased bases\\
MUHM & mutually unbiased Hadamard matrices\\
POVM & positive operator valued measure\\
 SIC & symmetric informationally complete
\end{tabular}

\section*{Introduction}
\addcontentsline{toc}{section}{Introduction}

Two orthonormal bases of a Hilbert space are said to be
\emph{mutually unbiased} (MU) if the transition probabilities from each state
in one basis to all states of the other basis are the same irrespective of
which pair of states is chosen.
Put differently, if the physical system is prepared in a state of
the first basis, then all outcomes are equally probable when we conduct a
measurement that probes for the states of the second basis.
This situation is symmetrical, it does not matter from which of the two bases
we choose the prepared state and which is the other basis that is measured:
Unbiasedness of bases is a mutual property, possessed jointly by both bases.
Familiar examples are the bases of position and momentum eigenstates for a
particle moving along a line, and the spin states of a spin-$\frac{1}{2}$
particle for two perpendicular directions.  

When the Hilbert space dimension $N$ is a prime power, ${N=p^\m}$, there
exist sets of ${N+1}$ mutually unbiased bases (MUB).   
These sets are maximal in the sense that it is not possible to find more than
${N+1}$ MUB in any $N$-dimensional Hilbert
space, there is simply no room for the ${(N+2)}$th
basis. 
Such a maximal set of MUB is also complete because when we know all the
probabilities of transition of a given quantum state towards the states of the
bases of this set --- exceptional situations aside, there are
$(N+1)(N-1)=N^2-1$ independent probabilities --- we can reconstruct the
statistical operator that characterizes this quantum state; in other words we
can perform full tomography or complete quantum state 
determination.

The existence of a maximal set of MUB for ${N=p^\m}$ is demonstrated by an
explicit construction, not by an abstract existence proof. 
Various methods have been used for the construction of maximal sets of
MUB, including the Galois--Fourier approach of this review.
Other constructions are based on generalized Pauli matrices, discrete Wigner
functions, abelian subgroups, mutually orthogonal Latin squares, and
finite-geometry methods.

All known constructions rely on the fact that $N$ is the power of a prime and,
therefore, they say nothing about other dimensions, of which ${N=6}$ is the
smallest one and also the one that has been studied most intensely.
At present, there is a widely shared conviction that one cannot have a maximal
set of seven MUB for ${N=6}$ and that the largest sets of MUB have no more
than three bases. 
This conviction is strongly founded in a solid body of evidence but, strictly
speaking, it is an unproven conjecture.

This situation is reminiscent of seemingly similar existence questions about
finite affine planes, Graeco-Latin squares, and related geometrical structures
where prime-power dimensions also play a privileged role.
As suggestive as these similarities may be, there is, however, no known
connection as yet between the two kinds of existence problems.

There is a plethora of applications whenever maximal sets of MUB are
available, in particular when the physical system is composed of many q-bits
(${N=2^\m}$), the building blocks of devices for quantum information
processing. 
Not surprisingly, then, the rise of quantum information science has triggered
fresh interest in MUB and, as a consequence, our knowledge about MUB and their
applications is much richer now.
But the various facts are scattered over a large number of publications, and
the many pieces of the puzzle do not readily fit together and do not compose a
uniform picture.

We are here reviewing the state of affairs in an attempt to offer a unified
view, with emphasis on both the structural properties of MUB and their use
in quantum-information applications.  
As in all constructions of MUB in prime power dimension, a crucial element is
a finite commutative division ring --- a Galois field of $N$ elements.%
\footnote{\label{fn:field}A \emph{ring} is a set that is closed under two
  operations: addition and multiplication.
  They obey the usual rules, associativity and commutativity of both
  operations, the distributive law, existence of a unique neutral element $0$
  for the addition and a neutral element $1$ for the multiplication.
  A \emph{field}, or division ring, is a ring with multiplicative inverses for
  every nonzero element.}\ 
Finite fields with $N$ elements exist if and only if $N$ is a power of a
prime, and the mathematical properties of Galois fields are exploited in all
constructions of maximal sets of MUB.
Modifications of these constructions in the absence of a finite field 
do not yield maximal sets of MUB for other dimensions.

The paper is structured as follows.
We begin with a brief survey of elements of quantum kinematics in
Sec.~\ref{sec0}.
The legacy of Weyl and Schwinger: the notion of complementary observables and
their algebraic completeness, the MUB associated with them, and the
${N\to\infty}$ limit of continuous degrees of freedom --- all these are
central to the story told in Sec.~\ref{sec:Weyl-Schwinger}.
It is supplemented by remarks on the Heisenberg--Weyl group of unitary
operators and the related Clifford group as well as, in Sec.~\ref{section0}, a
geometrically motivated ``measure of unbiasedness'' of two bases, a distance
in a real euclidean vector space. 

Section~\ref{section2} deals with the construction of a maximal set of MUB in
prime power dimension, ${N=p^\m}$, systematically treated as a composite
system of $\m$ $p$-dimensional subsystems.
For the purpose of introducing some notational conventions, but also for the
benefit of the typical working physicist for whom Galois fields are hardly the
daily bread, we recall the most important and most
relevant properties of Galois fields in Sec.~\ref{sec2.0}.
We are making extensive use of a formalism in which the numbers
$0,1,2,\dots,N-1$ play a dual role --- they are elements of a Galois field,
but also ordinary integers. 
This somewhat unconventional approach enables us to give a compact,
transparent construction of a maximal set of MUB in
Sec.~\ref{sec2.1}--\ref{sec2.3}.
A fitting version of the discrete Heisenberg--Weyl group, also known as the
generalized Pauli group, is an important tool for the construction; its
abelian subgroups define the MUB. 
In passing, we establish the contact between these MUB and the
complementary observables of the Weyl--Schwinger methodology 
(Sec.~\ref{sec2.4}).   

The survey of applications of the maximal set of MUB in Sec.~\ref{section3}
begins with the construction of a complete set of maximally entangled states,
the analogs of the familiar Bell states of two--q-bit systems, in
Sec.~\ref{sec3.1}. 
After brief accounts of their use for quantum dense coding (Sec.~\ref{sec3.2})
and teleportation (Sec.~\ref{sec3.3}), we discuss in Sec.~\ref{sec3.4} how
the generalized Bell states facilitate quantum cryptography and eavesdropping
with the aid of covariant cloning machines and comment on the role of the
Heisenberg--Weyl operators in error correction.
Section~\ref{section3} closes with a brief discussion of entanglement swapping
(Sec.~\ref{sec3.5}).

The prime-power version of the so-called Mean King's problem
(Sec.~\ref{sec4.1}) opens the section on quantum state tomography. 
The Mean King's problem is, in fact, very closely related to the discrete
analog of Wigner's 
continuous phase space function which --- jointly with its Fourier partner,
the analog of Weyl's characteristic function --- is the subject matter of
Sec.~\ref{sec4.2}. 
We comment on the covariance of the Wigner-type operator basis and discuss the
$N\to\infty$ limit of continuous degrees of freedom.
The relation to finite affine planes in Sec.~\ref{secaffin} provides further
insights into the underlying geometry.
  
Section~\ref{section5} is devoted to the matrices that transform pairs of MUB
into each other: the complex Hadamard matrices. 
Pairs of bases may be equivalent or not, in the sense that one can map the
basis states of one pair on those of the other pair by a unitary
transformation in conjunction with permutations of the basis states  
(Sec.~\ref{sec5.1}). 
The equivalence of triplets of MUB is more difficult to check
(Sec.~\ref{sec5.2}). 
Mutually unbiased Hadamard matrices (MUHM) are encountered when there are more
than two MUB. 
Accordingly, one can investigate sets of MUB by studying the corresponding
sets of MUHM, and vice versa.
All Hadamard matrices of size ${N\leq5}$ have been classified
(Sec.~\ref{sec5.3}), and all sets of MUB are known for ${N<6}$
(Sec.~\ref{sec:allMUB-Nle5}).  
The situation is not so clear, and thus more interesting, for ${N=6}$; we
report what is known about the families of ${6\times6}$ Hadamard matrices in
Sec.~\ref{sec5.5}, after a general discussion of affine families and tensor
products in Sec.~\ref{sec5.4}, and we deal with MUB for ${N=6}$ in
Secs.~\ref{sec5.8}--\ref{sec5.10}. 
Hadamard matrices for ${N>6}$ get their share of attention in
Sec.~\ref{sec5.6}.

We close with a brief summary and concluding remarks (Sec.~\ref{section6}) and
provide some additional technical details in four appendixes. 
The standard set of MUHM for prime dimension is given in \ref{sec:app1}, 
and a prime-distinguishing function related to this standard set 
is introduced in \ref{sec:app2}.
Finally, \ref{sec:app3} deals with MUB for the two--q-bit case of ${N=4}$.

     %% Title page, Abstract, Introduction
%%%% file name: MUB-1.tex
%%%% input file for MUB.tex 
%%%%
%%%% last changes on 20 April 2010 by Berge
%%%% minor corrections on 27 April 2010
%%%%
%%%%%%%%%%%%%%%%%%%%%%%%%%%%%%%%%%%%%%%%%%%%%%

\section{Elements of quantum kinematics}\label{sec0}

\subsection{The Weyl--Schwinger legacy}
\label{sec:Weyl-Schwinger}

\subsubsection{Complementary observables and mutually unbiased bases}
As emphasized by Bohr in his 1927 Como lecture,\cite{Bohr27} quantum systems
have properties that are \emph{complementary}: equally real but mutually
exclusive. 
If one such property is known accurately, then the complementary property is
completely unknown. 
Here, ``known accurately'' means that the outcome of a measurement can be
predicted with certainty, whereas ``completely unknown'' means that all
outcomes are equally likely --- the two properties are maximally incompatible.
Familiar examples are the position and momentum of a particle moving along a
line, and the $x$ and $z$ spin components of a spin-$\frac{1}{2}$ object.
These are, in fact, the extreme cases of a continuous degree of freedom and a
binary degree of freedom --- the latter being the ``q-bit'' of recent quantum
information terminology. 

Intermediate are ``q-nits,'' $N$-dimensional quantum degrees of freedom
($N>1$), for which the measurement of a physical property can have at most $N$
exclusive outcomes. 
Following Weyl\cite{Weyl1,Weyl2} and 
Schwinger,\cite{Schwinger,KinDyn,SchwingerQMbook}  
we call a pair of observables, $A$ and $B$, complementary if their eigenvalues
are not degenerate (there is the full count of $N$ different possible
measurement results) and the sets of normalized kets $\ket{a_j}$ and
$\ket{b_k}$ that describe states with predictable measurement outcomes for $A$
and $B$, respectively, are MU,
\begin{equation}
  \label{eq1:MUstates}
  \bigl|\braket{a_j}{b_k}\bigr|^2=\frac{1}{N}
\quad\mbox{for all $j,k=0,1,\dots,N-1$}\,.
\end{equation}
The important detail is not the value on the right, which is implied by the
normalization to unit total probability, but that the transition probabilities
on the left do not depend on the quantum numbers $a_j$ and $b_k$.%
\footnote{\label{fn:contDFnorm}%
  In fact, there can be different right-hand sides for infinite degrees
  of freedom, when normalization is more subtle; 
  see Secs.~\ref{sec:WScont1}--\ref{sec:WScont4}. 
  We will mostly deal with finite degrees of freedom.}  

Technically speaking, $A$ and $B$ are normal operators%
\footnote{A normal operator $A$ commutes with its adjoint $A^\dagger$:
  $AA^\dagger=A^\dagger A$, and can be regarded either as a function of a more
  fundamental hermitian operator or as a function of a unitary operator.}\ 
and $\ket{a_j}$, $\ket{b_k}$ are their eigenkets,
which make up two bases that are orthonormal and complete,
\begin{equation}
  \label{eq1:ab-complete}
  \braket{a_j}{a_k}=\delta_{j,k}=\braket{b_j}{b_k}\,,\qquad
  \sum_{j=0}^{N-1}\ket{a_j}\bra{a_j}=\mathbf{1}= 
    \sum_{k=0}^{N-1}\ket{b_k}\bra{b_k}\,,
\end{equation}
where $\mathbf{1}$ is the identity operator.
We recognize that the complementarity of $A$ and $B$ is in fact a property of
their respective eigenket bases. 
The particular eigenvalues are irrelevant, we just
need to know that they are not degenerate.
It follows in particular that,
if $A$ and $B$ are complementary, then $\alpha A$ and $\beta B$ with
${\alpha\beta\neq0}$ are complementary as well.
And if a unitary transformation turns $A$ into $A'$ and $B$ into $B'$, then
the pair $A',B'$ is complementary if the pair $A,B$ is.
Therefore, we can shift the focus from the pair $A,B$ of complementary
observables to the pair $\{\ket{a_j}\},\{\ket{b_k}\}$ of MUB.

Whenever is it expedient to be specific about the observables associated with
a basis, we will follow the guidance of Weyl and Schwinger%
\footnote{A brief account of the history of the subject can be found in
  Ref.~\refcite{SchwingerOnWeyl}.}\  
and choose unitary operators to represent physical quantities. 
In the present context, these will be nondegenerate cyclic operators with
period $N$, 
\begin{equation}
  \label{eq1:cyclic-AB}
  A^N=\mathbf{1}\,,\quad B^N=\mathbf{1}\,,
\end{equation}
with products of fewer than $N$ factors not equaling the identity.
The eigenvalues of $A$ and $B$ are then the $N$ different $N$th roots of
unity,
\begin{equation}
  \label{eq1:AB-eigen}
  A\ket{a_j}=\ket{a_j}\gamma_N^j\,,\quad B\ket{b_k}=\ket{b_k}\gamma_N^k\quad
\mbox{with $\gamma_N^{\ }=\Exp{\I2\pi/N}$}\,.
\end{equation}
That these cyclic operators are a pair of complementary operators can be
stated as
\begin{equation}
  \label{eq1:AB-trace}
  \frac{1}{N}\tr{A^mB^n}=\delta_{m,0}\delta_{n,0}\qquad
  \mbox{for $m,n=0,1,\dots,N-1$}\,,
\end{equation}
which is the operator version of (\ref{eq1:MUstates}).
Indeed, (\ref{eq1:MUstates}) and (\ref{eq1:AB-trace}) imply each 
other.\cite{Englert}

\subsubsection{Existence of a basic pair of complementary observables}
\label{sec:WSexist}
The first question we address is whether there always is a pair of
complementary observables for each quantum degree of freedom.
The affirmative answer begins with selecting an orthonormal reference basis 
$\ket{0}$, $\ket{1}$, \dots, $\ket{N-1}$ --- we will refer to it as the
\emph{computational basis} from Sec.~\ref{sec2.1} onwards.
Then we define a second orthonormal basis 
$\ket{\widehat{0}},\ket{\widehat{1}},\dots,\ket{\widehat{N-1}}$ by means
of the discrete quantum Fourier transformation,
\begin{equation}
  \label{eq1:q-Fourier}
  \ket{\widehat{j}}=\frac{1}{\sqrt{N}}\sum_{k=0}^{N-1}\ket{k}\gamma_N^{-jk}\,,
\end{equation}
so that
\begin{equation}
  \label{eq1:j2k-amplitude}
  \braket{\widehat{j}}{k}=\frac{1}{\sqrt{N}}\gamma_N^{jk}
  \quad\mbox{for $j,k=0,1,\dots,N-1$}
\end{equation}
by construction --- the two bases are MU, indeed.

In analogy with the Pauli operators $\sigma_x$ and $\sigma_z$, we introduce
the cyclic operators $X$ and $Z$ in accordance with
\begin{equation}
  \label{eq1:-defXZ}
  X\ket{\widehat{j}}=\ket{\widehat{j}}\gamma_N^j\,,\quad X^N=\mathbf{1}
\qquad\mbox{and}\qquad
  Z\ket{k}=\ket{k}\gamma_N^k\,,\quad Z^N=\mathbf{1}\,.
\end{equation}
As an immediate consequence of (\ref{eq1:j2k-amplitude}),
we note that $X$ and $Z$ are unitary shift operators that permute the kets or
bras of the respective other basis cyclically,
\begin{eqnarray}
  \label{eq1:def-X}
  X\ket{k}=\ket{k+1}\quad\mbox{for $k=0,1,\dots,N-2$}\,,\qquad 
  X\ket{N-1}=\ket{0}\phantom{\,,}
\end{eqnarray}
as well as
\begin{eqnarray}
  \label{eq1:def-Z}
  \bra{\widehat{j}}Z=\bra{\widehat{j+1}}\quad
\mbox{for $j=0,1,\dots,N-2$}\,,\qquad
\bra{\widehat{N-1}}Z=\bra{\widehat{0}}\,,
\end{eqnarray}
and (\ref{eq1:AB-trace}) holds for $(A,B)=(X,Z)$, as it must.
The fundamental Weyl commutation rule ${ZX=\gamma_N^{\ }XZ}$ follows.
It is the analog of the familiar $N=2$ identity
${\sigma_z\sigma_x=-\sigma_x\sigma_z}$ and
is more generally, and more usefully, stated as
\begin{equation}
  \label{eq1:commWeyl}
  X^mZ^n=\gamma_N^{-mn}Z^nX^m\,,
\end{equation}
valid for all integer values of $m$ and $n$, both positive and negative.

When we change the kets of the reference basis by phase factors, 
$\ket{k}\to\ket{k}\Exp{\I\phi_k}$, the resulting second basis will change
accordingly and we get another complementary partner $X$ to the same
observable $Z$. 
This freedom to adjust phases that do not affect the projectors
$\ket{k}\bra{k}$ of the reference basis but modify the projectors
$\ket{\widehat{j}}\bra{\widehat{j}}$ of the Fourier transformed basis is 
crucial for quantifying Einstein's\cite{Einstein:05,Einstein:06} and 
de Broglie's\cite{deBroglie:24} wave-particle duality in the context of 
two-path\cite{Englert:96,Englert+1:00}
and multi-path\cite{Englert+3:08} interferometers.

\subsubsection{Algebraic completeness of the basic pair of operators}
The second question, which also has an affirmative answer, is whether the pair
$X,Z$ of complementary observables parameterizes the degree of freedom
completely. 
Put differently: Are all other operators functions of $X$ and $Z$? 

As a first step, we observe that the projectors onto the respective
eigenstates are polynomials of $X$ or $Z$,
\begin{eqnarray}
  \label{eq1:projectors}
  \delta_{X,\gamma^j_N}&=&\ket{\widehat{j}}\bra{\widehat{j}}
                     =\frac{1}{N}\sum_{n=0}^{N-1}\Bigl(\gamma_N^{-j}X\Bigr)^n\,,
\nonumber\\
  \delta_{Z,\gamma^k_N}&=&\ket{k}\bra{k}
                     =\frac{1}{N}\sum_{m=0}^{N-1}\Bigl(\gamma_N^{-k}Z\Bigr)^m\,,
\end{eqnarray}
where the Kronecker delta symbols are to be understood in the usual sense of
an operator function, as exemplified by
\begin{equation}
  \label{eq1:opfunc}
  f(Z)=\sum_{k=0}^{N-1}\ket{k}f\bigl(\gamma_N^k\bigr)\bra{k}\,.
\end{equation}
The second step in writing an arbitrary operator $F$ as a function of $X$ and
$Z$ is to exploit the completeness of the two bases,
\begin{equation}
  \label{eq1:arbF}
 F=\sum_{j,k}\ket{\widehat{j}}\bra{\widehat{j}}F\ket{k}\bra{k}
     =\sum_{j,k}\delta_{X,\gamma_N^j}f_{j,k}\delta_{Z,\gamma_N^k}
\quad\mbox{with}\quad f_{j,k}=\frac{\bra{\widehat{j}}F\ket{k}}
                             {\braket{\widehat{j}}{k}}\,,
\end{equation}
where the denominator is assuredly nonvanishing.%
\footnote{Numbers of the form of $f_{j,k}$ are known as ``weak values'' of $F$
  in the context of ``weak measurements.''\cite{weakvalue}}\ 
This answers the second question by giving an explicit expression for $F$ as a
polynomial of $X$ and $Z$, here written in a unique way as an $XZ$-ordered
function: 
In products, all $X$ operators stand to the left of all $Z$ operators.  
Of course, quite analogously, we can also write $F$ in a unique $ZX$-ordered
form --- as an example recall the equivalence of the $XZ$-ordered operator on
the left of (\ref{eq1:commWeyl}) with the $ZX$-ordered product on the right.
In summary, there is not just one function of $X$ and $Z$ that equals the
given operator $F$, there are many such functions.

The lesson of these considerations is that the pair $X,Z$ is algebraically
complete, there are no operators that are not linear combinations of products
of powers of $X$ and $Z$.
Accordingly, we can phrase Bohr's Principle of Complementarity, the fundamental
principle of quantum kinematics, in the following technical terms: 
\emph{For each degree of freedom the dynamical variables are a pair of
  complementary observables.}\cite{SEW91}
For a textbook discussion, see Ref.~\refcite{QMnotes-PE}.

\subsubsection{The Heisenberg--Weyl group; the Clifford group}
\label{sec:WSgroups}
Supplemented with powers of $\gamma_N^{\ }$, the $XZ$-ordered products that
are implicit in (\ref{eq1:arbF}),
\begin{equation}
  \label{eq1:HWgroup1}
  Y_{l,m,n}^{\ }=\gamma_N^lX^mZ^n\quad\mbox{with}\quad l,m,n=0,1,\dots,N-1\,,
\end{equation}
make up the \emph{Heisenberg--Weyl group} of unitary operators, also called the
generalized Pauli group, with operator multiplication as the composition,
\begin{equation}
  \label{eq1:HWgroup2}
   Y_{l_1,m_1,n_1}^{\ } Y_{l_2,m_2,n_2}^{\ }=Y_{l_1+l_2+n_1m_2,m_1+m_2,n_1+n_2}^{\ }\,,
\end{equation}
where we understand all subscripts as integers modulo $N$, and the same
convention applies in
\begin{equation}
  \label{eq1:HWgroup3}
   Y_{l,m,n}^{-1}=Y_{l,m,n}^\dagger=Y_{mn-l,-m,-n}^{\ }\,.
\end{equation}
We could also use the $ZX$-ordered products to enumerate the group
elements, or consider the set of all products of powers of $X$ and $Z$ without
additional powers of $\gamma_N^{\ }$ as phase factors.
Each recipe gives the same set of $N^3$ unitary operators, but double counting
of group elements is most easily avoided when the ordered products are used.
In the ${N=2}$ example of ${X=\sigma_x}$ and ${Z=\sigma_z}$, the eight group
elements are $\pm\mathbf{1}$, $\pm\sigma_x$, $\pm\sigma_z$, and
${\pm\sigma_x\sigma_z=\mp\I\sigma_y}$. 
If we use the standard real $2\times2$ Pauli matrices to represent $\sigma_x$
and $\sigma_z$, then all eight unitary operators of the q-bit Heisenberg--Weyl
group are represented by real matrices.

In addition to this notion of the Heisenberg--Weyl group as a group of unitary
operators that are composed by multiplication, there is also the notion of the
Heisenberg--Weyl group as a group of unitary transformations
\begin{equation}
  \label{eq1:HWgroup4}
  F\to YFY^\dagger
\end{equation}
that are composed by sequential execution. 
There is no difference in (\ref{eq1:HWgroup4}) between $Y=X^nZ^m$ and
$Y=Z^mX^n$, 
\begin{equation}
  \label{eq1:HWgroup5}
  X^mZ^nF(X,Z)Z^{-n}X^{-m}=F(\gamma_N^nX,\gamma_N^{-m}Z)
 =Z^nX^mF(X,Z)X^{-m}Z^{-n}\,.
\end{equation}
More generally, the powers of $\gamma_N^{\ }$ in
(\ref{eq1:HWgroup1}) are irrelevant here, and therefore the group of unitary
transformations has $N^2$ elements and is abelian.
By contrast, the group of unitary operators is nonabelian; its abelian
subgroups play a crucial role in Sec.~\ref{sec:WSprime} below.
Weyl's view of ``quantum kinematics as an abelian group of rotations'' with
its utter disregard of phase factors in the ``ray fields'' should be
understood in this context; see Ch.~IV, Sec.~14 in Ref.~\refcite{Weyl2}.

We shall pay due attention to the phase factors in (\ref{eq1:HWgroup1}) where
they are relevant, but otherwise remember that the physically more essential
factors in (\ref{eq1:HWgroup1}) are the powers of $X$ and $Z$, and thus we
will not be overly pedantic when referring to the Heisenberg--Weyl group. In
the given context, it will be clear whether we mean the group of unitary
operators with its $\gamma_N^l$ phase factors or the group of unitary
transformations. 
An example is the observation that the $N$th power of $Y_{l,m,n}^{\ }$ can
differ from the identity operator,
\begin{equation}
  \label{eq1:HWgroup6}
  \bigl(Y_{l,m,n}\bigr)^N=\left\{
    \begin{array}{c@{\quad}l}
      \mathbf{1}&\mbox{if $N$ is odd,}\\[1ex]
      (-1)^{mn} \mathbf{1}&\mbox{if $N$ is even.}\\[1ex]
    \end{array}\right.
\end{equation}
For the group of unitary operators, the appearance of $(-1)^{mn}$ is crucial,
telling us that one quarter of the $Y_{l,m,n}^{\ }$s have period $2N$ for even
$N$, whereas this is of no concern for the group of unitary transformations. 
As an example, consider once more the ${N=2}$ situation with ${X=\sigma_x}$
and ${Z=\sigma_z}$, for which
\begin{equation}
  \label{eq1:HWgroup7}
  (\sigma_x\sigma_z)^2=-\mathbf{1}\,,\qquad
  (\sigma_x\sigma_z)^2F(\sigma_x,\sigma_z)(\sigma_x\sigma_z)^2
  =F(\sigma_x,\sigma_z)\,.
\end{equation}

The unitary operators $C$ that map the Heisenberg--Weyl group onto itself
under conjugation,%
\footnote{Anti-unitary operators could be, and often are, included 
--- see Ref.~\refcite{Appleby}, for example --- but we have no use for them
here.}\   
that is: $Y_{l,m,n}\to CY_{l,m,n}C^\dagger$ equals one of the $Y_{l,m,n}$s, 
constitute the so-called 
\emph{Clifford group}.\cite{Fivel,Gottesman98,Zauner,Appleby}
It contains the Heisenberg--Weyl group as a subgroup, but is truly larger.
For ${N=2}$, the Clifford group contains $24$ unitary transformations
(and is isomorphic to the symmetry group of the cube) whereas the
Heisenberg--Weyl group contains only four unitary transformations.
An example of a transformation belonging to the former but not the latter is
the ``q-bit Hadamard gate'' $(\sigma_x+\sigma_z)/\sqrt{2}$ that is represented
by the familiar Hadamard matrix 
\begin{equation}
  \label{eq1:qbitHada}
 H = \frac{1}{\sqrt{2}}
\left(\begin{array}{rr}
   1 & 1 \\
   1 & -1    
  \end{array}\right), 
\end{equation}
if we use the standard $2\times2$ matrices for $\sigma_x$ and $\sigma_z$.

\subsubsection{Composite degrees of freedom}\label{sec:WScomp}
If $N$ is a composite number, $N=N_1N_2$ with $N_1>1$ and $N_2>1$, then some
of the Heisenberg--Weyl operators have a shorter period, as exemplified by 
$(Y_{0,N_1,0})^{N_2}=(X^{N_1})^{N_2}=X^N=\mathbf{1}$. 
As a consequence, there are Heisenberg--Weyl operators that have different
spectral properties and are not related to each other by a unitary
transformation. 

It is then methodical to regard the $N$-dimensional degree of freedom as
composed of a $N_1$-dimensional and a $N_2$-dimensional degree of freedom.
Accordingly, the labels $k$ of the kets $\ket{k}$ of the reference basis are
understood as pairs $k_1,k_2$ with ${k=k_1+k_2N_1}$ whereby
${k_1=0,1,\dots,N_1-1}$ and  ${k_2=0,1,\dots,N_2-1}$.
The action of the corresponding cyclic operators $X_1$ and $X_2$ is given by
\begin{eqnarray}
  \label{eq1:compDF1}
  X_1\ket{k}&=&X_1\ket{k_1,k_2}=\ket{k_1+1,k_2}=\ket{k+1}\hphantom{N_1}
\quad\mbox{for $k_1=0,1,\dots,N_1-2$}\,,\nonumber\\
  X_2\ket{k}&=&X_2\ket{k_1,k_2}=\ket{k_1,k_2+1}=\ket{k+N_1}\hphantom{1}
\quad\mbox{for $k_2=0,1,\dots,N_2-2$}\,,\nonumber\\&&
\end{eqnarray}
and the respective $k_1=N_1-1$ and $k_2=N_2-1$ statements are
\begin{eqnarray}
  \label{eq1:compDF2}
  X_1\ket{k=(k_2+1)N_1-1}&=&X_1\ket{N_1-1,k_2}=\ket{0,k_2}=\ket{k_2N_1}\,,
\nonumber\\
  X_2\ket{k=k_1+N_1(N_2-1)}&=&X_2\ket{k_1,N_2-1}=\ket{k_1,0}=\ket{k_1}\,.
\end{eqnarray}
By construction, $X_1$ and $X_2$ have periods $N_1$ and $N_2$, respectively,
and as a consequence of the algebraic completeness of the pair $X,Z$ of
complementary observables, we can express $X_1$ and $X_2$ quite explicitly as
functions of $X$ and $Z$, with the outcome
\begin{equation}
  \label{eq1:compDF3}
  X_1=X-\bigl(\mathbf{1}-X^{-N_1}\bigr)\delta_{Z^{N_2},1}^{\ }X\,,
  \qquad X_2=X^{N_1}\,.
\end{equation}
Clearly, $X_1$ commutes with $X_2$ because $Z^{N_2}$ commutes with $X^{N_1}$
when $N_1N_2=N$, as is the case here.

Likewise one constructs the complementary partners $Z_1$ and $Z_2$ as the
operators that cyclically advance the respective quantum numbers of the common
eigenbras $\bra{\widehat{j_1},\widehat{j_2}}$ of $X_1$ and $X_2$, which
are related to the kets $\ket{k_1,k_2}$ through the analog of
(\ref{eq1:j2k-amplitude}), 
\begin{equation}
  \label{eq1:compDF4}
  \braket{\widehat{j_1},\widehat{j_2}}{k_1,k_2}=
  \frac{1}{\sqrt{N_1}}\gamma_{N_1}^{j_1k_1}
  \frac{1}{\sqrt{N_2}}\gamma_{N_2}^{j_2k_2}\,.
\end{equation}
In summary, then, the original $N$-dimensional degree of freedom,
parameterized by the pair $X,Z$, is decomposed into the product of two degrees
of freedom, a $N_1$-dimensional and a $N_2$-dimensional one, parameterized by
the pairs $X_1,Z_1$ and $X_2,Z_2$, respectively. 

In passing, we note that the two bases of product kets $\ket{k_1,k_2}$ and
$\ket{\widehat{j_1},\widehat{j_2}}$ are MU.
This illustrates how one can construct MUB of a composite
degree of freedom from such bases of its constituents.

If $N_1$ or $N_2$ are composite numbers themselves, this reasoning can be
applied again, if necessary repeatedly, until one has one degree of freedom
for each prime factor of $N$. 
These prime degrees of freedom are fundamental and cannot be decomposed
further. 
As emphasized by Schwinger in his teaching,\cite{SchwingerQMbook} they are the
elementary quantum degrees of freedom.

\subsubsection{Prime degrees of freedom}\label{sec:WSprime}
The simplest prime degree of freedom is the q-bit case $N=2$, for which we
have $X=\sigma_x$, $Z=\sigma_z$, and $XZ=-\I\sigma_y$. 
With $\ket{0}$ and $\ket{1}$ denoting the eigenkets of $\sigma_z$ to eigenvalues
$+1$ and $-1$, respectively, the eigenkets of $\sigma_x$ are 
${2^{-\frac{1}{2}}(\ket{0}\pm\ket{1})}$, and the eigenkets of $\sigma_y$ are
${2^{-\frac{1}{2}}(\ket{0}\I\pm\ket{1})}$.
These three bases are \emph{pairwise} MU, and the three operators
$X$, $Z$, and $XZ$ are pairwise complementary.

More generally, we can consider any two components
${A=\vec{a}\cdot\vec{\sigma}}$ and ${B=\vec{b}\cdot\vec{\sigma}}$ 
of Pauli's vector operator $\vec{\sigma}$ whose cartesian components are
$\sigma_x$, $\sigma_y$, and $\sigma_z$. 
Operators $A$ and $B$ are complementary if the nonvanishing three-dimensional
numerical vectors $\vec{a}$ and $\vec{b}$ are orthogonal to each other,
${\vec{a}\cdot\vec{b}=0}$. 
Since there are at most three pairwise orthogonal vectors, there are at most
three pairwise complementary operators and at most three MUB.
The choice $\sigma_x$, $\sigma_y$, $\sigma_z$ for the three operators is,
therefore, not particular, but typical.

If $N$ is an odd prime, $N=3,5,7,11,13,\dots$, then all unitary
Heisenberg--Weyl operators $Y_{l,m,n}^{\ }$ of (\ref{eq1:HWgroup1}) are cyclic
with period $N$, except for the identity $\mathbf{1}=Y_{0,0,0}^{\ }$. 
Further, we observe that the $N+1$ operators
\begin{equation}
  \label{eq1:prime1}
  X\,,\ XZ\,,\ XZ^2\,,\ \dots\,,\ XZ^{N-1}\,,\ Z
\end{equation}
are pairwise complementary,\cite{Englert} as one verifies most directly
with the aid of (\ref{eq1:AB-trace}) and (\ref{eq1:HWgroup2}) in conjunction
with 
\begin{equation}
  \label{eq1:prime2}
  \tr{Y_{l,m,n}^{\ }}=N\gamma_N^l\delta_{m,0}^{\ }\delta_{n,0}^{\ }\,.
\end{equation}
It follows that the ${N+1}$ bases of eigenkets, one for each of the operators in
(\ref{eq1:prime1}), are MU.
In addition to the eigenbases of $X$ and $Z$ that we met in
Sec.~\ref{sec:WSexist}, there are thus ${N-1}$ more such bases.

And there cannot be a ${(N+2)}$th basis because a counting argument shows that
one can at most have ${N+1}$ bases that are~MU.\cite{Ivanovic} 
One way of seeing this is presented in Sec.~\ref{section0} below.  

In this context, we note here that the powers of the operators in
(\ref{eq1:prime1}) make up ${N+1}$ abelian cyclic subgroups of the 
Heisenberg--Weyl group with $N$ unitary operators in each subgroup.
Remembering that the identity is contained in each subgroup, this gives a
total count of ${(N+1)(N-1)+1=N^2}$ operators, one representative for each set
of $Y_{l,m,n}^{\ }$s with common $m,n$ values, that is: one count for each
$X^mZ^n$ product.

Explicitly, ket $\ket{i,k}$, the $k$th eigenket of the $i$th basis, 
$XZ^i\ket{i,k}=\ket{i,k}\gamma_N^k$, is given by
\begin{equation}
  \label{eq1:prime3}
 \mbox{$N$ odd:}\quad
  \ket{i,k}=\frac{1}{\sqrt{N}}\sum_{l=0}^{N-1}\ket{l}
            \gamma_N^{-kl}\gamma_N^{il(l-1)/2}\quad
  \mbox{for $i=0,1,2,\dots,N-1$}
\end{equation}
in terms of the reference basis of eigenkets of $Z$.
For $i=0$ we have the eigenstates of $X$, $\ket{\widehat{k}}=\ket{0,k}$.
While (\ref{eq1:prime3}) correctly states the eigenkets of $XZ^i$ for all odd
$N$, these bases are pairwise MU only if $N$ is prime.
With due attention to the extra phase factors required by (\ref{eq1:HWgroup6})
one can give a similar expression for $\ket{i,k}$ when $N$ is even. 

In summary, we can systematically construct ${N+1}$ bases that are MU if $N$
is prime. 
As noted, the construction based on the cyclic operators in (\ref{eq1:prime1})
does not work if $N$ is composite; 
try $N=4$ to see what goes wrong.
We return to the case of $N=6$ in Sec.~\ref{sec5.10},
and a general discussion for arbitrary $N\geq2$ is given in \ref{sec:app2}. 

Yet, this is not the end of the story. 
If $N=p^\m$ is the power of a prime, for which ${N=8=2^3}$ and ${N=9=3^2}$ are
examples, it is possible to modify the construction such that it does work in
a closely analogous way.  
The clue is to replace the modulo-$N$ shifts of (\ref{eq1:def-X}) and
(\ref{eq1:def-Z}) by shifts of a Galois field arithmetic that treats the
$N$-dimensional degree of freedom systematically as composed of $\m$
$p$-dimensional constituents.  
This is the theme of Sec.~\ref{section2}, followed by applications in
Secs.~\ref{section3} and \ref{section4}. 

This Galois cure is, however, not available for ${N=6}$ and ${N=10}$ or other
composite $N$ values that are not powers of a prime, simply because the number
of elements in a finite field is always a prime power.
Section~\ref{section5} contains a report on what is known about these cases,
in particular about ${N=6}$. 
The question whether there are seven MUB for ${N=6}$ is
currently unanswered, but there is a lot of evidence, and a growing conviction
in the community, that there are no more than three such bases. 
And three such bases are immediately available by pairing each of the three
q-bit bases ($N_1=2$) with one of the four q-trit bases ($N_2=3$) to product
bases as in (\ref{eq1:compDF4}).

\subsubsection{The continuous limit of $N\to\infty$}\label{sec:WSlim}
Since composite values of $N$ refer to composite quantum degrees of freedom,
we take the limit $N\to\infty$ through prime values of $N$, thereby dealing
with a single degree of freedom of increasing complexity.
The prime nature of $N$ will not be so crucial, however, but we make use of
the fact that large primes are odd numbers and relabel the kets of the
reference basis $\ket{k}$ and the bras $\bra{\widehat{j}}$ of the
Fourier-transformed basis such that now 
${j,k=0,\pm1,\pm2,\dots,\pm\frac{1}{2}(N-1)}$.

Next, we introduce a small, eventually infinitesimal, parameter $\epsilon$ by
\begin{equation}
  \label{eq1:lim1}
  N=\frac{2\pi}{\epsilon^2}
\end{equation}
to account for the fact that the basic unit of complex phase $2\pi/N$ gets
arbitrarily small when $N\to\infty$.
Aiming at a continuous degree of freedom in this limit, we also relabel the
states in accordance with
\begin{eqnarray}
  \label{eq1:lim2}
  j&\longrightarrow& j\epsilon=a=0,\pm\epsilon,\pm2\epsilon,\dots,
          \pm\Bigl(\frac{\pi}{\epsilon}-\frac{\epsilon}{2}\Bigr)\,,
\nonumber\\
  k&\longrightarrow& k\epsilon=b=0,\pm\epsilon,\pm2\epsilon,\dots,
          \pm\Bigl(\frac{\pi}{\epsilon}-\frac{\epsilon}{2}\Bigr)\,.
\end{eqnarray}
The numbers $a$ and $b$ will cover the real axis, ${-\infty<a,b<\infty}$, when
${N\to\infty}$, ${\epsilon\to0}$. 

The unitary operator $X$ acting on $\ket{k}$ increases $k$ by unity, so that it
effects ${b\to b+\epsilon}$.
Likewise $Z$ applied to $\bra{\widehat{j}}$ results in ${a\to a+\epsilon}$.
This suggests the identification of hermitian operators $A$ and $B$ such that
\begin{eqnarray}
  \label{eq1:lim3}
  X&=&\Exp{\I\epsilon A}\quad\mbox{with $A=A^\dagger$}\,,  \nonumber\\
  Z&=&\Exp{\I\epsilon B}\quad\mbox{with $B=B^\dagger$}\,.
\end{eqnarray}
The Weyl commutation relation (\ref{eq1:commWeyl}) then appears as
\begin{equation}
  \label{eq1:lim4}
  X^kZ^j=\Exp{-\I\frac{2\pi}{N}jk}Z^jX^k\longrightarrow
  \Exp{\I k\epsilon A}\Exp{\I j\epsilon B}
 =\Exp{-\I j\epsilon k\epsilon}\Exp{\I j\epsilon B}\Exp{\I k\epsilon A}
\end{equation}
or
\begin{equation}
  \label{eq1:lim5}
  \Exp{\I bA}\Exp{\I aB}=\Exp{-\I ab}\Exp{\I aB}\Exp{\I bA}\,.
\end{equation}
The two equivalent versions
\begin{eqnarray}
  \label{eq1:lim6}
  \Exp{\I b(A-a\mathbf{1})}&=&\Exp{-\I aB}\Exp{\I bA}\Exp{\I aB}
                            = \Exp{\I b \,\Exp{-\I aB}A\,\Exp{\I aB}}\,,
\nonumber\\
  \Exp{\I a(B-b\mathbf{1})}&=&\Exp{\I bA}\Exp{\I aB}\Exp{-\I bA}
                            = \Exp{\I a \,\Exp{\I bA}B\,\Exp{-\I bA}}
\end{eqnarray}
seem to imply that
\begin{eqnarray}
  \label{eq1:lim7}
  \Exp{-\I aB}A\,\Exp{\I aB}&=&A-a\mathbf{1}\,,\nonumber\\
  \Exp{\I bA}B\,\Exp{-\I bA}&=&B-b\mathbf{1}\,,
\end{eqnarray}
but this does not follow without imposing a restricting condition, just as
${\Exp{\I\alpha}=\Exp{\I\beta}}$ does not imply ${\alpha=\beta}$, but only that
${\alpha-\beta}$ is an integer multiple of $2\pi$.

The said restriction is that, for large $N$, only $a,b$ values from a finite
vicinity of~$0$ matter, which is to say that we break the cyclic nature of the 
labels $a,b$,
\begin{equation}
  \label{eq1:lim8}
  \bra{a}\Exp{\I a'B}=\bra{a+a'\,(\mbox{mod}\,2\pi/\epsilon)}\,,\qquad
  \Exp{\I b'A}\ket{b}=\ket{b+b'\,(\mbox{mod}\,2\pi/\epsilon)}\,,
\end{equation}
and take for granted that all relevant values of $a,a'$ and $b,b'$ are such
that we stay inside the range
$-(\pi/\epsilon-\epsilon/2)\cdots(\pi/\epsilon-\epsilon/2)$. 
Put differently, we give up the periodicity that would force us to identify
${a=+\infty}$ with ${a=-\infty}$ in the ${\epsilon\to0}$ limit.

After performing the ${N\to\infty}$, ${\epsilon\to0}$ limit with this
restriction, the statements of (\ref{eq1:lim7}) hold with continuous values
for $a$ and $b$. 
We can, therefore, exhibit the terms that are linear in $a$ or $b$ and arrive
at
\begin{equation}
  \label{eq1:lim9}
  AB-BA=[A,B]=\I\mathbf{1}\,.
\end{equation}
We recognize, of course, Heisenberg's commutation relation for a pair of
complementary hermitian observables of a continuous degree of freedom, such as
position $A$ and momentum $B$ (in natural units) for the motion along a line.

These $N\to\infty$ considerations for $X$ and $Z$ have to be supplemented by
counterparts for their respective kets and bras. 
We need to identify
\begin{equation}
  \label{eq1:lim10}
  \bra{a}=\frac{1}{\sqrt{\epsilon}}\bra{\widehat{j}}\Biggr|_{\epsilon\to0}
\quad\mbox{with $j\epsilon=a$, and}\quad
\ket{b}=\ket{k}\frac{1}{\sqrt{\epsilon}}\Biggr|_{\epsilon\to0}
\quad\mbox{with $k\epsilon=b$,}
\end{equation}
and then get
\begin{equation}
  \label{eq1:lim11}
  \braket{a}{b}=\frac{1}{\sqrt{2\pi}}\Exp{\I ab}
\end{equation}
as the analog of (\ref{eq1:j2k-amplitude}) as well as
\begin{equation}
  \label{eq1:lim12}
  \braket{a}{a'}=\delta(a-a')\,,\quad\braket{b}{b'}=\delta(b-b')
\end{equation}
and
\begin{equation}
  \label{eq1:lim13}
  \int\limits_{-\infty}^{\infty}\!\D a\,\ket{a}\bra{a}=\mathbf{1}=
  \int\limits_{-\infty}^{\infty}\!\D b\,\ket{b}\bra{b}
\end{equation}
as the continuum versions of the orthogonality and completeness relations in
(\ref{eq1:ab-complete}). 

This discussion of the $N\to\infty$ limit is a variant of Schwinger's
treatment in Sec.~1.16 of Ref.~\refcite{SchwingerQMbook}; 
see also Sec.~1.2.5 in Ref.~\refcite{QMnotes-PE}. 
It should be appreciated that ${N\to\infty}$ is not a limit in the precise
sense that one has in calculus.
Rather, it is a systematic method for inferring the properties of the basic
operators for continuous degrees of freedom, but these operators then stand on
their own and the consistency of the inferred algebraic properties must be
verified. 
 
We note that, in addition to the standard symmetric limit that
treats $X$ and $Z$ on equal footing and results in the Heisenberg
pair of $A$ and $B$ (position and momentum for motion along a line), 
there are also asymmetric limits. 
For instance, if the position variable --- the analog of the hermitian $A$ of
(\ref{eq1:lim3}) --- is kept periodic over a finite range in
the limit, one obtains the pair of azimuth-angle operator and angular-momentum
operator for motion on a circle,\cite{asymLimit} with which we deal in
Sec.~\ref{sec:WScont2}. 
In a third way of taking the $N\to\infty$ limit, the hermitian
position variable is kept positive throughout and one arrives at a continuous
quantum degree of freedom of the kind that parameterizes radial motion; 
see Sec.~\ref{sec:WScont3}.
Finally, there is a fourth procedure, in which the position values cover a
finite range without, however, retaining the cyclic nature by identifying the
boundaries with each other; this results in a degree of freedom of the kind
associated with the polar angle in spherical coordinates
(Sec.~\ref{sec:WScont4}).

\subsubsection{Continuous degree of freedom 1: Motion along a line}
\label{sec:WScont1}
Knowing that there are ${N+1}$ pairwise complementary observables for prime
degrees of freedom, we expect to find an infinite number of them for a
continuous degree of freedom. 
Indeed, there is a continuum of pairwise complementary observables and,
therefore, a continuum of MUB, although an interesting complication can be
observed too.\cite{WeigertWilkinson} 

Harking back to Sec.~\ref{sec:WSprime}, we recall that each basis in the set of
MUB consists of the joint eigenstates of the unitary operators that one gets
by taking products of \emph{one} of the unitary operators in the list
(\ref{eq1:prime1}) with itself. 
Translated into the continuum case of Sec.~\ref{sec:WSlim}, the corresponding
unitary operators are those of (\ref{eq1:lim5}), and since
\begin{equation}
  \label{eq1:conti1}
  \Exp{\I b_1A}\Exp{\I a_1B}\,\Exp{\I b_2A}\Exp{\I a_2B}=
  \Exp{\I(a_1b_2-b_1a_2)}\,
  \Exp{\I b_2A}\Exp{\I a_2B}\,\Exp{\I b_1A}\Exp{\I a_1B}
\end{equation}
tells us that two of these unitary operators commute if ${a_1b_2=b_1a_2}$,
the operators $\Exp{\I bA}\Exp{\I aB}$ for which $(a,b)=(\alpha t,\beta t)$
with common values of $\alpha$ and $\beta$ make up an abelian subgroup of
Heisenberg--Weyl operators.
The elements of the subgroup specified by the pair $(\alpha,\beta)$ are
labeled by parameter $t$, which takes on all real values.
For ${\alpha=\beta=0}$, we have the one-element subgroup of the identity; 
this case is of no further interest and excluded from the following
considerations. 

It is expedient to choose the single-exponent form
\begin{equation}
  \label{eq1:conti2}
  Y(\alpha,\beta;t)=\Exp{\I t(\beta A+\alpha B)}
\end{equation}
for the subgroup elements, so that the subgroup composition rule 
\begin{equation}
  \label{eq1:conti2'}
   Y(\alpha,\beta;t_1)Y(\alpha,\beta;t_2)= Y(\alpha,\beta;t_1+t_2)
\end{equation}
involves no additional phase factors, and we denote the common eigenkets and
eigenbras of all unitary operators in the $(\alpha,\beta)$ subgroup by
$\ket{\alpha,\beta;y}$ and  $\bra{\alpha,\beta;y}$, 
\begin{equation}
  \label{eq1:conti3}
  Y(\alpha,\beta;t)\ket{\alpha,\beta;y}
  =\ket{\alpha,\beta;y}\Exp{\I ty}\,,\qquad
\bra{\alpha,\beta;y}Y(\alpha,\beta;t)=\Exp{\I ty}\bra{\alpha,\beta;y}\,.
\end{equation}
If one wishes, one can regard  $\ket{\alpha,\beta;y}$ and
$\bra{\alpha,\beta;y}$ as eigenstates of the hermitian operator 
${\beta A+\alpha B}$ with eigenvalue $y$, but we prefer to work with the sets
of bounded unitary operators rather than the unbounded hermitian operators.

As usual, the eigenstates are normalized to the Dirac delta function,
\begin{equation}
  \label{eq1:conti4}
  \braket{\alpha,\beta;y}{\alpha,\beta;y'}=\delta(y-y')\,,
\end{equation}
which implies that, up to a phase factor of no consequence, 
\begin{equation}
  \label{eq1:conti5}
  \ket{\lambda\alpha,\lambda\beta;\lambda y}\sqrt{|\lambda|}=
  \ket{\alpha,\beta;y}
\end{equation}
for $\lambda\neq0$, consistent with 
$Y(\lambda\alpha,\lambda\beta;t/\lambda)=Y(\alpha,\beta;t)$.
The subgroup for $(\lambda\alpha,\lambda\beta)$ is identical with the subgroup
for $(\alpha,\beta)$, with the elements parameterized differently.
The respective eigenstates are in one-to-one correspondence, but differ from
each other by a normalization factor (except when $\lambda=-1$).

These statements have no analogs for finite $N$, when the normalization of
states is unambiguous and the parameterization of the abelian subgroups is
essentially unique.
In the continuous case, by contrast, there is more than one way of
parameterizing the continuous abelian subgroups, and one would have to impose
constraints on $\alpha$ and $\beta$ to avoid this innocuous ambiguity, such
as insisting on $\alpha=\cos\theta$ and $\beta=\sin\theta$ with
${0\leq\theta<\pi}$ or, equivalently, permitting only $(\alpha,\beta)=(0,1)$
and $\alpha=1$ with arbitrary $\beta$.
Clearly, constraints of this sort are a bit awkward, 
and they are not necessary. 

The projector $\ket{\alpha,\beta;y}\bra{\alpha,\beta;y}$ is given by
\begin{equation}
  \label{eq:conti6}
 \ket{\alpha,\beta;y}\bra{\alpha,\beta;y}
 =\int\limits_{-\infty}^{\infty}\!\frac{\D t}{2\pi}\,Y(\alpha,\beta;t)\Exp{-\I ty}
\end{equation}
as one verifies by, for instance, checking that 
\begin{equation}
  \label{eq1:conti7}
  \bigl(\ket{\alpha,\beta;y}\bra{\alpha,\beta;y}\bigr)\ket{\alpha,\beta;y'}
  =\ket{\alpha,\beta;y}\,\delta(y-y')\,.
\end{equation}
The completeness relation
\begin{equation}
  \label{eq1:conti8}
  \int\limits_{-\infty}^\infty\!\D y\,\ket{\alpha,\beta;y}\bra{\alpha,\beta;y}
  =Y(\alpha,\beta;0)=\mathbf{1}
\end{equation}
follows and confirms that we have a basis for each of the abelian subgroups.
  
Next, we consider two different abelian subgroups, specified by
$(\alpha,\beta)$ and  $(\alpha',\beta')$, respectively,
with  ${\alpha\beta'\neq\beta\alpha'}$, and evaluate the
transition probability density%
\footnote{It is a \emph{density} because we need to multiply with 
${\D y\,\D y'}$ to get the probabilities referring to infinitesimal 
intervals of $y$ and $y'$.}
between their respective eigenstates by means of
\begin{eqnarray}
  \label{eq1:conti9}
  \bigl|\braket{\alpha,\beta;y}{\alpha',\beta';y'}\bigr|^2
 &=&\tr{\bigl(\ket{\alpha,\beta;y}\bra{\alpha,\beta;y}\bigr)
        \bigl(\ket{\alpha',\beta';y'}\bra{\alpha',\beta';y'}\bigr)}
\nonumber\\
&=&\int\frac{\D t\,\D t'}{(2\pi)^2}\,
    \tr{Y(\alpha,\beta;t)Y(\alpha',\beta';t')}\,
    \Exp{-\I (ty+t'y')}
\nonumber\\
&=&\int\frac{\D t\,\D t'}{(2\pi)^2}\,
   \frac{2\pi\delta(t)\delta(t')}
        {\bigl|\alpha\beta'-\beta\alpha'\bigr|}\,
    \Exp{-\I (ty+t'y')}
\nonumber\\
&=&\frac{1}{2\pi\bigl|\alpha\beta'-\beta\alpha'\bigr|}\,,
\end{eqnarray} 
which is Eq.~(11) in Ref.~\refcite{WeigertWilkinson}.
Since the value of $\bigl|\braket{\alpha,\beta;y}{\alpha',\beta';y'}\bigr|^2$ 
does not depend on the quantum numbers $y$ and $y'$ that label the states of
the two bases, the two bases are MU.
This is true for the bases to any two different abelian subgroups.
Indeed, we have a continuum of MUB for a continuous degree of freedom.

As a consequence, the hermitian operators ${\beta A+\alpha B}$ and
${\beta'A+\alpha'B}$ are complementary observables if their commutator 
${\I[\beta A+\alpha B,\beta' A+\alpha'B]}%
={(\alpha\beta'-\beta\alpha')\mathbf{1}}$ does not vanish.
The absolute value of this commutator appears in the denominator of
(\ref{eq1:conti9}). 
Not unexpectedly, for a continuous degree of freedom, there is a continuum of
pairwise complementary observables.

We could have arrived at the same conclusion by the following more direct
argument that exploits the observations made after (\ref{eq1:ab-complete}).
There are unitary transformations that turn ${\beta A+\alpha B}$ into
$\kappa A$ and ${\beta' A+\alpha'B}$ into $\kappa' B$ with 
${\kappa\kappa'=\alpha\beta'-\beta\alpha'}\neq0$.
Now, since the pair $A,B$ is complementary, so is the pair $\kappa A,\kappa' B$,
which implies that the pair ${\beta A+\alpha B},{\beta' A+\alpha'B}$ is
complementary as well, and their bases of eigenstates are MU.

Whereas the right-hand side of (\ref{eq1:MUstates}) has the same value of
$N^{-1}$ for any pair of MUB for a $N$-dimensional degree of freedom, this is
not the case for the right-hand side of (\ref{eq1:conti9});
recall footnote `\ref{fn:contDFnorm}'.
For a given pair of bases specified by the coefficients
$(\alpha,\beta)$ and $(\alpha',\beta')$, we can either choose
$(\alpha'',\beta'')=(\alpha+\alpha',\beta+\beta')$ or  
$(\alpha'',\beta'')=(\alpha-\alpha',\beta-\beta')$ to supplement them with
a third basis such that these three MUB have the same numerical value for the
constant transition probability densities between each pair of bases.
The basis for any fourth choice $(\alpha''',\beta''')$ will have a
different value for one or more of its transition probability densities with
the earlier three bases.
This observation by Weigert and Wilkinson\cite{WeigertWilkinson} means that
the continuous set of MUB, composed of the bases of (\ref{eq1:conti3}),  
contains three-element subsets that 
are polytopes of MUB in the sense of Sec.~\ref{section0}.

\subsubsection{Continuous degree of freedom 2: Motion along a circle}
\label{sec:WScont2}
We parameterize the position around the circle by the $2\pi$-periodic azimuth
$\varphi$ --- with $\ket{\varphi}=\ket{\varphi+2\pi}$, for instance --- 
and normalize the corresponding bras and kets in accordance with
the orthogonality and completeness relations
\begin{equation}
  \label{eq1:contb1}
  \braket{\varphi}{\varphi'}=2\pi\delta^{(2\pi)}(\varphi-\varphi')\,,
\qquad
\int\limits_{(2\pi)}\!\frac{\D\varphi}{2\pi}\,\ket{\varphi}\bra{\varphi}
=\mathbf{1}\,,
\end{equation}
where the integration range is any $2\pi$ interval and $\delta^{(2\pi)}(\ )$
denotes the $2\pi$-periodic version of Dirac's delta function,
\begin{equation}
  \label{eq1:contb2}
  \delta^{(2\pi)}(\varphi-\varphi')=\frac{1}{2\pi}\sum_{l=-\infty}^{\infty}
\Exp{\I l (\varphi-\varphi')}\,.  
\end{equation}
We regard the azimuthal states $\ket{\varphi}$ as eigenstates of a unitary
operator $E$,
\begin{equation}
  \label{eq1:contb2'}
  E\ket{\varphi}=\ket{\varphi}\Exp{\I\varphi}\,,\qquad
  E=\int\limits_{(2\pi)}\!\frac{\D\varphi}{2\pi}\,
    \ket{\varphi}\Exp{\I\varphi}\bra{\varphi}\,.
\end{equation}
This $E$ is the proper $N\to\infty$ limit of $X$ in the present context.

All azimuthal wave functions 
$\psi(\varphi)=\braket{\varphi}{\ }=\psi(\varphi+2\pi)$ 
are periodic, 
and the Fourier series of $\bra{\varphi}$ identifies the
eigenstates of the associated angular momentum operator $L$,
\begin{equation}
  \label{eq1:contb3}
  \bra{\varphi}=\sum_{l=-\infty}^{\infty}\Exp{\I l\varphi}\bra{l}\,,\quad
L\ket{l}=\ket{l}l\,.
\end{equation}
Their orthonormality and completeness relations are
\begin{equation}
  \label{eq1:contb4}
  \braket{l}{l'}=\delta_{l,l'}\,,\qquad
  \sum_{l=-\infty}^{\infty}\ket{l}\bra{l}=\mathbf{1}\,,
\end{equation}
consistent with (\ref{eq1:contb1}).

In view of
\begin{equation}
  \label{eq1:contb5a}
  \braket{\varphi}{l}=\Exp{\I l\varphi}\,,\qquad
  \bigl|\braket{\varphi}{l}\bigr|^2=1\,,
\end{equation}
the $\varphi$-basis and the $l$-basis are MU.
The respective unitary shift
operators are powers of $E$ and exponential functions of $L$,
\begin{equation}
  \label{eq1:contb5b} 
  E^m\ket{l}=\ket{l+m}\,,\qquad
  \bra{\varphi}\Exp{\I\phi L}=\bra{\varphi+\phi}\,.
\end{equation}
Their products $E^m\Exp{\I\alpha L}$ make up the Heisenberg--Weyl group with
the basic Weyl commutation relation given by
\begin{equation}
  \label{eq1:contb6}
  E^m\Exp{\I\phi L}=\Exp{-\I m\phi}\Exp{\I\phi L} E^m\,,
\end{equation}
which is the analog of (\ref{eq1:lim5}).
For each modulo-$2\pi$ value  of $\phi$, there is an abelian subgroup composed
of the unitary operators $(E\Exp{\I\phi L})^m$ with $m=0,\pm1,\pm2,\dots\,$.

Despite these analogies and the great structural similarities with the
situation of Sec.~\ref{sec:WScont1}, there is a striking difference: There is
no third basis that is MU with respect to both the $\varphi$-basis and the
$l$-basis.   

To make this point, let us assume that ket $\ket{\ }$ belongs to such a third
basis. 
Then it must be true that
\begin{equation}
  \label{eq1:contb8}
  \bigl|\braket{\varphi}{\ }\bigr|^2=\lambda>0\quad\mbox{for all $\varphi$}
\qquad\mbox{and}\qquad
  \bigl|\braket{l}{\ }\bigr|^2=\mu>0\quad\mbox{for all $l$.}
\end{equation}
The completeness relations in (\ref{eq1:contb1}) and (\ref{eq1:contb4}) then
imply
\begin{eqnarray}
  \label{eq1:contb9}
  \braket{\ }{\ }&=&\int\limits_{(2\pi)}\!\frac{\D\varphi}{2\pi} \,
                    \bigl|\braket{\varphi}{\ }\bigr|^2
                  =\int\limits_{(2\pi)}\!\frac{\D\varphi}{2\pi}\,\lambda 
                  =\lambda\nonumber\\
\mbox{and}\qquad  
  \braket{\ }{\ }&=&\sum_{l=-\infty}^{\infty} \bigl|\braket{l}{\ }\bigr|^2
                  =\sum_{l=-\infty}^{\infty} \mu  =\infty\,,
\end{eqnarray}
which contradict each other. 
It follows that there is not even a single ket with the properties
(\ref{eq1:contb8}); indeed, there is no third basis.

This situation of a missing third basis is a unique feature of the $E,L$-type
continuous degree of freedom.
There is always a third basis for finite $N$ --- the three eigenbases to $X$,
$Z$, and $XZ$ of (\ref{eq1:prime1}) are pairwise MU for all $N>1$ --- and
there is a continuum of MUB for the continuous degrees of freedom of the three
other types.
It appears that the combination of the continuous position variable $E$ with the
discrete momentum variable $L$ is at the heart of the matter.
For the other continuous degrees of freedom, the respective position and
momentum variables are both continuous, as will be discussed below.

The nonexistence of a third basis that supplements the $\varphi$-basis and the
$l$-basis does not exclude the possibility that there are other bases that are
MU, perhaps with sets of MUB that have more than two elements.
Currently, we are not aware of any such set, however, but its bases would have
to be rather unusual.
For, two different discrete bases (such as the
$l$-basis) cannot be MU, nor can two different continuous bases 
(such as the $\varphi$-basis) be MU.
And if one basis is discrete and the other continuous, the dilemma of
(\ref{eq1:contb9}) cannot be avoided.

\subsubsection{Continuous degree of freedom 3: Radial motion}
\label{sec:WScont3}
In spherical coordinates, 
radial motion is characterized by a positive position operator $R>0$,
\begin{equation}
  \label{eq1:contc1}
  R\ket{r}=\ket{r}r\quad\mbox{with $r>0$}\,,\qquad
  \braket{r}{r'}=r\delta(r-r')\,,\quad
  \int\limits_0^\infty\frac{\D r}{r}\,\ket{r}\bra{r}=\mathbf{1}\,,
\end{equation}
whereas the eigenvalues of its complementary partner $S$ are all real numbers,
\begin{equation}
  \label{eq1:contc2}
  S\ket{s}=\ket{s}s\quad\mbox{with $-\infty<s<\infty$}\,,\qquad
  \braket{s}{s'}=\delta(s-s')\,,\quad
  \int\limits_{-\infty}^\infty\D s\,\ket{s}\bra{s}=\mathbf{1}\,.
\end{equation}
The transition amplitudes
\begin{equation}
  \label{eq1:contc3}
  \braket{r}{s}=\frac{r\power{\I s}}{\sqrt{2\pi}}
\end{equation}
confirm that the $r$-basis and the $s$-basis are MU and that $R$ and $S$ are a
pair of complementary observables.

The unitary shift operators $R\power{\I t}$ and $\Exp{\I\lambda S}$ have the
expected effect when applied to the states of the other basis,
\begin{equation}
  \label{eq1:contc4}
  \bra{r}\Exp{\I\lambda S}=\bra{\Exp{\lambda}r}\,,\qquad
  R\power{\I t}\ket{s}=\ket{s+t}\,,
\end{equation}
as follows from (\ref{eq1:contc3}). 
The resulting Weyl commutation relation
\begin{equation}
  \label{eq1:contc5}
  R\power{\I t}\Exp{\I\lambda S}
  =\Exp{-\I\lambda t}\Exp{\I\lambda S}R\power{\I t}
\end{equation}
and the Heisenberg commutator
\begin{equation}
  \label{eq1:contc6}
  \bigl[R,S\bigr]=\I R
\end{equation}
tell us that $S$ is the hermitian generator of scaling transformations,
\begin{equation}
  \label{eq1:contc7}
  \Exp{-\I\lambda S}R\,\Exp{\I\lambda S}=\Exp{-\lambda}R\,,
\end{equation}
fitting to the positive nature of $R$.

The unitary operator products in (\ref{eq1:contc5}) make up the
Heisenberg--Weyl group here, and the abelian subgroups can be characterized by
common values of $\tau$ and $\mu$ in $(t,\lambda)=\kappa(\tau,\mu)$. 
In full analogy with (\ref{eq1:conti3})--(\ref{eq1:conti9}), then, the bases
defined by 
\begin{equation}
  \label{eq1:contc8}
  \Exp{\I\kappa^2\tau\mu/2}R\power{\I\kappa\tau}\Exp{\I\kappa\mu S}
  \ket{\tau,\mu;\alpha}=\ket{\tau,\mu;\alpha}\Exp{\I\kappa\alpha}
\end{equation}
for $(\tau,\mu)\neq(0,0)$ are pairwise MU,
\begin{equation}
  \label{eq1:contc9}
  \bigl|\braket{\tau,\mu;\alpha}{\tau',\mu';\alpha'}\bigr|^2=
  \frac{1}{2\pi\bigl|\tau\mu'-\mu\tau'\bigr|}\,.
\end{equation}
Just as in Sec.~\ref{sec:WScont1}, here too we have a continuum of
pairwise complementary observables and a continuum of MUB, and the set of MUB
has three-element polytopes in the sense of Ref.~\refcite{WeigertWilkinson}. 
The $R,S$-type degree of freedom  is really quite similar to the $A,B$-type
degree of freedom of the Heisenberg kind, because $\log R$ and $S$ are a
Heisenberg pair of operators: $[\log R,S]=\I\mathbf{1}$.
This commutator is a particular case of 
\begin{equation}
  \label{eq1:contc10}
  \bigl[f(R),S\bigr]=\I R\frac{\partial f(R)}{\partial R}\,,
\end{equation}
which follows from (\ref{eq1:contc6}) or from (\ref{eq1:contc7}).

\subsubsection{Continuous degree of freedom 4: Motion within a segment}
\label{sec:WScont4}
After dealing with the azimuthal and radial degrees of freedom in
Secs.~\ref{sec:WScont2} and \ref{sec:WScont3}, we now turn to the degree of
freedom associated with the polar angle $\vartheta$ of spherical coordinates,
$(x,y,z)=%
(r\sin\vartheta\,\cos\varphi,r\sin\vartheta\,\sin\varphi,r\cos\vartheta)$.
Since the values of $\vartheta$ are restricted to a finite interval
$0\leq\vartheta\leq\pi$, where the endpoints are not identified with each
other as is the case for the periodic azimuth $\varphi$, we speak of ``motion
within a segment,'' very much like the popular textbook example of the
``particle in a box,'' about which some non-textbook material is reported in
Ref.~\refcite{Englert:98}.   
The relations between the position and momentum operators for cartesian and
spherical coordinates are discussed in \ref{sec:app0}. 

The eigenstates of the position variable $\Theta$ and its complementary
partner $\Omega$ are related to each other by
\begin{equation}
  \label{eq1:contd1}
  \braket{\vartheta}{\omega}
  =\frac{1}{\sqrt{2\pi}}\Bigl(\tan\frac{\vartheta}{2}\Bigr)\power{\I\omega}
\qquad\mbox{with $0<\vartheta<\pi$ and $-\infty<\omega<\infty$}\,,
\end{equation}
and the respective orthonormality and completeness relations are
\begin{equation}
  \label{eq1:contd2}
  \braket{\vartheta}{\vartheta'}=\sin\vartheta\,\delta(\vartheta-\vartheta')\,,
\qquad \int\limits_0^\pi\frac{\D\vartheta}{\sin\vartheta}\,
       \ket{\vartheta}\bra{\vartheta}=\mathbf{1}
\end{equation}
for the $\vartheta$-basis as well as
\begin{equation}
  \label{eq1:contd3}
  \braket{\omega}{\omega'}=\delta(\omega-\omega')\,,\qquad
  \int\limits_{-\infty}^{\infty}\D\omega\,\ket{\omega}\bra{\omega}=\mathbf{1}
\end{equation}
for the $\omega$-basis.
Accordingly, the unitary shift operators are specified by
\begin{equation}
  \label{eq1:contd4}
\bra{\vartheta}\Exp{\I\lambda\Omega}=\bra{\vartheta'}\Bigr|
\subscr{\vartheta'=2\arctan(\mathrm{e}^{\lambda}\tan\frac{\vartheta}{2})}\,,
\qquad
\Bigl(\tan\frac{\Theta}{2}\Bigr)\power{\I\omega'}\ket{\omega}
=\ket{\omega+\omega'}\,,
\end{equation}
telling us that the unitary transformation effected by $\Exp{\I\lambda\Omega}$
has no simple geometrical meaning. 

The Weyl commutation relation reads
\begin{equation}
  \label{eq1:contd5}
 \Bigl(\tan\frac{\Theta}{2}\Bigr)\power{\I\omega}\Exp{\I\lambda\Omega}
=\Exp{-\I\omega\lambda} 
\Exp{\I\lambda\Omega} \Bigl(\tan\frac{\Theta}{2}\Bigr)\power{\I\omega}\,,
\end{equation}
from which we get the Heisenberg commutator
\begin{equation}
  \label{eq1:contd6}
  \bigl[\Theta,\Omega\bigr]=\I\sin\Theta\,.
\end{equation}
More generally, we have the analog of (\ref{eq1:contc10}),
\begin{equation}
  \label{eq1:contd7}
    \bigl[f(\Theta),\Omega\bigr]
    =\I\sin\Theta\frac{\partial f(\Theta)}{\partial \Theta}\,,
\end{equation}
and the particular case
\begin{equation}
  \label{eq1:contd8}
      \Bigl[\log\tan\frac{\Theta}{2},\Omega\Bigr]=\I\mathbf{1}
\end{equation}
identifies $\log\tan\frac{\Theta}{2}$ and $\Omega$ as a Heisenberg pair of
complementary observables. 
Remembering the lessons of Secs.~\ref{sec:WScont1} and \ref{sec:WScont3}, we
conclude that the abelian subgroups of the Heisenberg--Weyl group composed of
the unitary operators of (\ref{eq1:contd5}) define a continuum of MUB, with
the set of MUB having three-element subsets that are MUB polytopes in the
sense of Ref.~\refcite{WeigertWilkinson}. 

We close this excursion into the realm of continuous degrees of freedom
with a comment on the completeness and orthonormality relations
(\ref{eq1:contc1}) and (\ref{eq1:contd2}).
Why did we not absorb the factors $r$ and $\sin\vartheta$ into the
normalization of the respective bras and kets? 
There are two good reasons: 
(i) Such a change of normalization would spoil the
relations (\ref{eq1:contc3}) and (\ref{eq1:contd1}); 
(ii) these factors would re-appear in a disturbing way when the orthonormality
and completeness relations are rewritten in terms of the eigenstates 
for the Heisenberg partners $\log R$ and $\log\tan\frac{\Theta}{2}$ of $S$ and
$\Omega$, respectively. 
In other words, it is very natural to have the factors  $r$ and
$\sin\vartheta$ in (\ref{eq1:contc1}) and (\ref{eq1:contd2}).

In view of the various subtle issues regarding the normalization of eigenkets
and eigenbras for continuous degrees of freedom, the definition of what
constitutes a pair of complementary observables --- given above in the context
of (\ref{eq1:MUstates}) --- should perhaps be modified to state more carefully
that two nondegenerate observables $A$ and $B$ are complementary if one can
normalize their respective eigenstates consistently such that
$\bigl|\braket{a}{b}\bigr|^2$ has the same value for all eigenbras $\bra{a}$
of $A$ and all eigenkets $\ket{b}$ of $B$.

\subsection{A geometrically motivated measure of mutual unbiasedness}
\label{section0}
The kets $\ket{\ }$ in $N$-dimensional Hilbert space, and their adjoint bras
$\bra{\ }=\ket{\ }^\dagger$, are rather abstract geometrical objects, 
and so are the linear operators that map kets on kets and bras on bras, among
them the statistical operator $\rho$ that summarizes our knowledge about the
state of the physical $N$-dimensional degree of freedom under consideration.  
With reference to a specified basis, the kets are represented by numerical
column vectors $\psi$ ($N\times1$ matrices), the bras by row vectors
$\psi^\dagger$ ($1\times N$ matrices), and the linear operators by 
$N\times N$ matrices, among them the density matrix $\varrho$ for the
statistical operator $\rho$.
We denote these relationships by $\psi\repr\ket{\ }$, 
$\psi^\dagger\repr\bra{\ }$, and $\varrho\repr\rho$, respectively.

There are many density matrices, one for each reference basis, to one and the
same statistical operator, much like there are many trios of components for
the velocity vector of the moon, one trio for each coordinate system.
One should not confuse the velocity vector with its components, or the
statistical operator with the density matrix used to represent it numerically.

When they exist, maximal sets of MUB form a very distinct geometrical
pattern in the set of hermitian matrices of unit trace --- the real euclidean 
space that contains the set of density matrices. 
This is where we begin our story about maximal sets of MUB, although in most
of what follows we will prefer to work directly in Hilbert space.  
The two pictures ought to be considered as complementary, 
each of them possessing advantages and drawbacks.

The set $\{\varrho\}$ of density matrices is a convex body in the set of
hermitian matrices of unit trace. 
Its pure states are the one-dimensional projectors. 
The set of its pure states has real dimension $2(N-1)$, and can be identified
with the complex projective Hilbert space.  
The dimension of $\{\varrho\}$ is $N^2-1$, and the space in which it sits can
be regarded as a vector space, with its origin at the maximally mixed state 
\begin{equation} 
\varrho_\star = \frac{1}{N}\mathbbm{1}\repr\frac{1}{N}\mathbf{1}=\rho_\star \,, 
\end{equation}
where $\mathbf{1}$ is the identity operator of (\ref{eq1:ab-complete}) 
and $\mathbbm{1}$ is the unit matrix that represents it.

With any hermitian matrix $M$ of unit trace we associate a traceless matrix 
\begin{equation} \label{M-to-m}
\mathbf{m} = M - \varrho_\star \,. 
\end{equation}
The set of these traceless matrices forms a vector space, and we will think 
of them as vectors. 
The matrix representation is used to define the inner product 
\begin{equation} 
\mathbf{m}_1 \cdot \mathbf{m}_2 
= \frac{1}{2}\tr{(M_1 - \varrho_\star )(M_2-\varrho_\star )} \,. 
\end{equation}
Thus the squared distance between the tips of the two vectors 
$\mathbf{m}_1$ and $\mathbf{m}_2$ is 
\begin{equation} 
D(\mathbf{m}_1,\mathbf{m}_2)^2 = 
\frac{1}{2}\tr{(M_1 - M_2)^2} \,. 
\end{equation} 
With any unit ket $\ket{e}$ in Hilbert space we associate a 
vector $\mathbf{e}$ in $\mathbf{R}^{N^2-1}$, the space of $(N^2-1)$-component
real vectors, through 
\begin{equation} \label{eq1:e} 
\mathbf{e}=\psi_e^{\ }\psi_e^\dagger-\varrho_\star\repr  
\ket{e}\bra{e} - \rho_\star 
\end{equation}
so that the squared length of $\mathbf{e}$ is
\begin{equation} 
|\mathbf{e}|^2 = \frac{N-1}{2N} \,. 
\end{equation}
All vectors in $\mathbf{R}^{N^2-1}$ with this specific length sit on the
surface of the outsphere of the body $\{\varrho\}$, the smallest
sphere containing the body. 
But it is important to realize that it is only a small $2(N-1)$-dimensional
subset of this outsphere that corresponds to vectors in Hilbert space --- 
most of the outsphere lies outside the body. 
The case ${N = 2}$ is an exception: 
In this case the outsphere is the familiar Bloch sphere, which is identical to
the boundary of the body of density matrices.

Note furthermore that the relations 
\begin{equation} 
\braket{e_i}{e_j}= \delta_{i,j}\,,\qquad 
\mathbf{e}_i\cdot\mathbf{e}_j =\frac{1}{2}\delta_{i,j} - \frac{1}{2N} 
\end{equation}
imply each other.
If $\ket{e_i}$ is an orthonormal basis of kets, 
the corresponding vectors $\mathbf{e}_i$ form a regular simplex that spans 
an $(N-1)$-plane, and clearly  
\begin{equation} \label{su} 
\sum_{i = 0}^{N-1} \mathbf{e}_i = 0 \,. 
\end{equation}
Hence the simplex is centered at the origin. 
We have normalized its edge lengths to unity.
  
Next consider two MUB with kets $\ket{e_i}$ and $\ket{f_j}$, respectively,
represented by the vectors $\mathbf{e}_i$ and $\mathbf{f}_j$.
The two equations 
\begin{equation} 
\bigl|\braket{e_i}{f_j}\bigr|^2 = \frac{1}{N}\,, 
\qquad \mathbf{e}_i\cdot \mathbf{f}_j = 0 
\label{krav} 
\end{equation}
are equivalent ways of stating that the bases are MU and, therefore, the two
planes spanned by a  pair of MUB are totally orthogonal: 
Each vector in one plane is orthogonal to all vectors in the other plane. 
Since the dimension of our space is $N^2-1 = (N+1)(N-1)$, we can fit at most
$N+1$ totally orthogonal $(N-1)$-planes into it. 
This is one way of seeing that the maximal number of MUB is $N+1$. 

Let us now momentarily forget that our vectors $\mathbf{e}_i$, $\mathbf{f}_i$,
and so on, are supposed to come from unit vectors in Hilbert space. 
Whatever the value of $N$, we can always find $N+1$ totally orthogonal 
$(N-1)$-planes in $\mathbf{R}^{N^2-1}$, and if we place a regular simplex in
each we will obtain a quite interesting convex polytope with $N(N+1)$ 
vertices.\cite{BE05} 
When $N = 2$, it is in fact a regular octahedron, but for other values of $N$
it needs a name of its own. 
We will call it the MUB polytope, without implying that there
exists a maximal set of MUB in the $N$-dimensional Hilbert space. 
The MUB polytope and the body of density matrices share the same outsphere 
and, in this manner, the existence problem for MUB 
can be turned into the problem of rotating the MUB polytope in such a way 
that all its vertices fit into the small subset of pure quantum states that 
are present in that outsphere. 
This is a hard problem (unless $N = 2$). 
Indeed, from this perspective it is not obvious that we can find even one pair
of MUB but, as we have seen in Sec.~\ref{sec:WSexist}, we can always do this. 
It is the existence of a \emph{maximal} set, with $N+1$ bases that are
pairwise MU, which is in doubt for general $N$. 

Viewing bases as $(N-1)$-planes in $\mathbf{R}^{N^2-1}$ gives us the means to 
quantify how close a given pair of bases is to being MU. 
The trick is to regard $n$-planes in $\mathbf{R}^m$ as rank-$n$ projectors in
a vector space of real $m\times m$ matrices, in analogy to the way we go from
vectors in Hilbert space to density matrices. 
This gives us an embedding of the Grassmannian of 
$n$-planes into a flat vector space equipped with a natural euclidean distance, 
and hence a natural notion of distance between vectors in Hilbert space. 
To derive it, consider the $N$ vectors $\mathbf{e}_i$. 
Then form the $(N^2-1)\times N$ matrix
\begin{equation} \label{eq1:B}
B = \bigl[ \mathbf{e}_1 \ \mathbf{e}_2 \ \dots \ \mathbf{e}_N\bigr]\,.
\end{equation}
It has rank $N-1$ because of (\ref{su}).
Next form the projector onto the $(N-1)$-plane 
spanned by the linearly dependent vectors $\mathbf{e}_i$. 
It is 
\begin{equation} \label{eq1:Pi}
\Pi = 2 BB^{\rm T} \,.
\end{equation}
Finally, the square of the chordal Grassmannian distance between a pair of
planes is\cite{BBELTZ07} 
\begin{eqnarray}\label{distgras1}  
D_c(\Pi_e,\Pi_f)^2 \equiv \frac{1}{2}\tr{(\Pi_e - \Pi_f)^2}
&=&N-1-\sum_{a,b}\left(\bigl|\braket{e_a}{f_b}\bigr|^2-\frac{1}{N}\right)^2
\nonumber\\ 
&=&\sum_{a,b} 
\bigl|\braket{e_a}{f_b}\bigr|^2\Bigl(1-\bigl|\braket{e_a}{f_b}\bigr|^2\Bigr)\,,
\end{eqnarray}
where the kets $\ket{e_a}$ are related to $\Pi_e$ through (\ref{eq1:e}),
(\ref{eq1:B}), and (\ref{eq1:Pi}), and the kets $\ket{f_b}$ are analogously
related to $\Pi_f$.
The last expression of (\ref{distgras1}) shows that ${D_c=0}$ if the
projectors $\ket{f_b}\bra{f_b}$ are a permutation of the projectors
$\ket{e_a}\bra{e_a}$, in which case we have the same basis twice, possibly
with different labeling.
 
One can check that 
\begin{equation} 
0 \leq D^2_c \leq N-1 \,, 
\end{equation}
and that the distance is maximal if and only if the two bases are MU.  
This notion of distance has been used to study packing problems for 
$n$-planes,\cite{conway96} and as a measure of ``MUness''.\cite{BBELTZ07}
If we pick our bases at random, using the 
unitarily invariant Fubini--Study measure to define ``random,'' 
we find that the average squared distance is given by
\begin{equation} 
\langle D^2_c\rangle_{\mathrm{FS}}^{\ } = \frac{N}{N+1}(N-1) \,. 
\end{equation}
If the dimension is large, $N\gg1$, two bases picked at random are likely to 
be almost MU.  
Let us finally mention that entropic uncertainty relations in effect provide
an interesting alternative measure of 
``MUness''.\cite{Deutsch:83,Bialynicki-B:84,Wehner+1:09}

         %% Section 1
%%%% file name: MUB-2.tex
%%%% input file for MUB.tex 
%%%%
%%%% last changes on 20 April 2010 by Berge
%%%% 
%%%%%%%%%%%%%%%%%%%%%%%%%%%%%%%%%%%%%%%%%%%%%%%%%

\section{Construction of mutually unbiased bases 
in prime power dimensions}\label{section2}
\subsection{Galois fields}\label{sec2.0}

In what follows, we work in a Hilbert space of
prime power dimension $N=p^\m$ with $p$ a prime number and $\m$ a positive
integer. 
These are the dimensions for which maximal sets of MUB are known to exist. 
Moreover, and not coincidentally, there is a finite Galois field 
with $N= p^\m$ elements. 
We shall label these elements by integer numbers $i$, $0\leq i\leq N-1$,
or, equivalently, by $\m$-tuples $(i_{0},i_{1},\ldots,i_{\m-1})$ of integers,
each integer running from 0 to $p-1$, 
that we get from the $p$-ary expansion of~$i$: 
\begin{equation}\label{def-coeff}
 i=(i_{0},i_{1},\ldots,i_{\m-1})\qquad\mbox{if}\qquad 
i=\sum_{n=0}^{\m-1}i_{n}p^n\,.  
\end{equation}

Each field is characterized by two operations, a multiplication and an
addition, that we shall denote by $\odot$  
and $\oplus$ respectively. 
As in footnote~`\ref{fn:field}', we shall use the symbols $0$ and $1$ for
the neutral elements of addition 
and multiplication, respectively, throughout the paper, consistent with their
meaning as integers.
 
Further, we adopt the particular convention that the elements of the field are
labeled in such a way that the addition is equivalent to the 
component-wise addition modulo $p$, that is
\begin{equation}\label{eq2:fieldadd}
\mbox{$i=j\oplus k$ is tantamount to $i_n=j_n+k_n\,(\bmod\ p)$}  
\end{equation}
for $n=0,1,\dots,\m-1$, where $i_n,j_n,k_n$ are the respective coefficients of
(\ref{def-coeff}). 
As a consequence, the summation in (\ref{def-coeff}) is also a field
summation,
\begin{equation}
  \label{def-coeff'}
  i=(i_0p^0)\oplus(i_1p^1)\oplus\cdots\oplus(i_{\m-1}p^{\m-1})
   =\bigoplus_{n=0}^{\m-1}i_{n}p^n\,.  
\end{equation}

All fields with the same number of elements are equivalent up to a relabeling,
and there is no strict obligation for the convention (\ref{eq2:fieldadd}), 
but it is natural and convenient in the present context, because it allows us
to regard the elements of the field both as labels of basis states and as
integer numbers that we can use for getting powers of complex numbers in
accordance with the usual computation rules. 

Actually, that there exists a relabeling such that the addition is
equivalent to the addition modulo $p$ component-wise is a direct consequence
of the fact that for all finite fields the characteristics of the field ---
the smallest number of times that we must add the element~$1$
(neutral for the multiplication) to itself before we obtain the element~$0$
(neutral for the addition) --- is always equal to a prime number ($p$ when
$N=p^\m$). 

Unfortunately, there is no similarly simple convention for the
field multiplication~$\odot$, and --- the exceptions $N=p$ and $N=4$ aside
--- one has a choice between several equally good ways of defining the field
multiplication $\odot$ such that it is consistent with the component-wise
definition of the field addition $\oplus$.
In view of the associative and distributive nature of $\odot$, that is:
$(a\odot b)\odot c=a\odot (b\odot c)$ and $(a\oplus b)\odot c=(a\odot
c)\oplus(b\odot c)$, respectively, we only need to state the values of
$p^j\odot p^k$, the products of powers of $p$, with $j,k=0,1,\ldots,\m-1$.

For $\m=1$, $N=p$, the field multiplication is just multiplication modulo $p$.
For $\m>1$, we have the Galois construction
\begin{equation}
  \label{eq1:odot-def1}
  p^j\odot p^k=\left\{
    \begin{array}{l}
      p^{j+k}\quad\mbox{if $j+k<\m\,,$}\\[1ex]
      \ds\sum_{l=0}^{\m-1}\mu^{\ }_lp^l
              =(\mu^{\ }_0,\mu^{\ }_1,\dots,\mu^{\ }_{\m-1})
      \quad\mbox{if $j+k=\m\,,$}\\[3ex]
     p\odot (p^{j-1}\odot p^k) \quad\mbox{recursively, if $j+k>\m\,.$}
    \end{array}\right.
\end{equation}
Hereby, the coefficients that define the $j+k=\m$ products are restricted by
the requirement that 
\begin{equation}
  \label{eq1:odot-def2}
  x\mapsto x^\m-\sum_{l=0}^{\m-1}\mu^{\ }_lx^l
\end{equation}
is an \emph{irreducible polynomial} over the Galois field with $p$ elements,
which is to say that it cannot be factored into two nonconstant polynomials
whose coefficients are modulo-$p$ integers.

In a standard textbook parameterization of the Galois field with ${N=p^{\m}}$
elements,\cite{Karpilovski} 
one identifies the field elements with polynomials that are defined
by the coefficients of the $p$-ary expansion of (\ref{def-coeff}),
\begin{equation}
  \label{eq2:textbookGF}
  i=(i_{0},i_{1},\ldots,i_{\m-1})\longleftrightarrow\sum_{m=0}^{\m-1}i_mx^m\,.
\end{equation}
Addition and multiplication of the field elements are then carried out as
addition and multiplication of the corresponding polynomials modulo the
polynomial of (\ref{eq1:odot-def2}), with the resulting sums and products
stated as polynomials of degree $\m-1$ with modulo-$p$ integer coefficients.
Clearly, this gives the component-wise addition of (\ref{eq2:fieldadd}) 
and multiplication in accordance with (\ref{eq1:odot-def1}).
Since the field multiplication is not familiar to readers with a typical
theoretical-physics background, we now discuss it in some detail.

For instance, the choice $2\odot2=3$ is unique for $N=4$, and for $p$ odd and
$N=p^2$, one can always choose $p\odot p=\mu^{\ }_0$ with $\mu^{\ }_0$ not a
square, such as $3\odot3=2$, $5\odot5=2$ or $5\odot5=3$, $7\odot7=3$ or
$7\odot7=5$ or $7\odot7=6$, and so forth. For higher powers of $p=2$, there
are several choices too; they include $2\odot4=5$ for $N=8$, $2\odot8=3$ for
$N=16$, and $2\odot16=5$ for $N=32$.

As a final example, we mention $3\odot9=(1,2,2)=25$ for $N=3^3$.%
\footnote{The choice $3\odot9=25$ is the largest one of the eight permissible
  values.
  The other seven values for $(\mu_0,\mu_1,\mu_2)$ are $(1,1,0)=4$,
  $(2,1,0)=5$, $(2,0,1)=11$, $(1,1,1)=13$, $(2,2,1)=17$, $(1,0,2)=19$, and
  $(2,1,2)=23$. 
Each of them yields a consistent implementation of the field multiplication.} 
This implies first $9\odot9=(2,2,0)=8$ and then
\begin{eqnarray}
  \label{eq1:odot-N=27}
  N=27\,:&\quad&(a_0,a_1,a_2)\odot(b_0,b_1,b_2)=a \odot b=c=(c_0,c_1,c_2)
\nonumber\\
&&\quad\mbox{with}\quad
\begin{array}[t]{rcl}
  c_0&=&a_0b_0+a_1b_2+a_2b_1-a_2b_2\ (\mbox{mod}\ 3)\,,\\
  c_1&=&a_0b_1+a_1b_0-a_1b_2-a_2b_1-a_2b_2\ (\mbox{mod}\ 3)\,,\\
  c_2&=&a_0b_2+a_1b_1+a_2b_0-a_1b_2-a_2b_1\ (\mbox{mod}\ 3)\,,
\end{array}
\end{eqnarray}
for the multiplication of two arbitrary field elements.
The special cases ${3\odot13=1}$ and ${9\odot17=1}$ may serve as illustrations.

More generally, when writing
\begin{equation}
  \label{eq1:odot-M}
   p^j\odot p^k=\bigl(M^{(j+k)}_0,M^{(j+k)}_1,\ldots, M^{(j+k)}_{\m-1}\bigr)\,,
\end{equation}
we have
\begin{equation}
  M^{(j+k)}_m=\delta_{j+k,m}\quad\mbox{for $j+k=0,1,\dots,\m-1\,,$ and}\quad
  M^{(\m)}_m=\mu_m\,,
\end{equation}
and the coefficients for $j+k=\m+1,\m+2,\ldots,2\m-2$ are successively
calculated with the aid of the recurrence relation
\begin{equation}
  \label{eq1:odot-recur}
  M^{(j+k)}_m=(1-\delta_{m,0})M^{(j+k-1)}_{m-1}+\mu_m M^{(j+k-1)}_{\m-1}\ 
  (\mbox{mod}\ p)\,,
\end{equation}
which is valid for $j+k=1,2,\dots,2\m-2$.
The field product of two arbitrary elements is then given by
\begin{equation}
  \label{eq1:odot-arb}
  a\odot b=\left(a\mathcal{M}_0^{\ }b^T,a\mathcal{M}_1^{\ }b^T,\ldots,
                 a\mathcal{M}_{\m-1}^{\ }b^T\right)\,,
\end{equation}
where $\mathcal{M}_m^{\ }=\mathcal{M}_m^T$ is the symmetric $\m\times\m$ matrix
\begin{equation}
  \label{eq1:odot-matr}
  \mathcal{M}_m^{\ }=\left(\begin{array}{ccccccc}
    M^{(0)}_m &   M^{(1)}_m &   M^{(2)}_m & \cdots &\cdots &&  \\ 
    M^{(1)}_m &   M^{(2)}_m &            & &&& \\ 
    M^{(2)}_m &            &    & \hphantom{M^{(2\m-4)}_m} &&& \vdots\\ 
    \vdots&\hphantom{M^{(2\m-4)}_m}&\hphantom{M^{(2\m-4)}_m}& \ddots &&& \vdots\\
    \vdots&&&  &         &            & M^{(2\m-4)}_m  \\
    &&&  &              & M^{(2\m-4)}_m &   M^{(2\m-3)}_m  \\
    &&\hdots&\hdots & M^{(2\m-4)}_m &   M^{(2\m-3)}_m &   M^{(2\m-2)}_m \\
    \end{array}\right)\,,
\end{equation}
and in the products $a\mathcal{M}_mb^T$ we regard $a=(a_0,a_1,\dots)$ 
as a row of $p$-ary coefficients and $b^T$ as a column.
These row$\,\times\,$matrix$\,\times\,$column products are ordinary matrix
products with the outcome evaluated modulo~$p$.
The matrices $\mathcal{M}_0^{\ }$, $\mathcal{M}_1^{\ }$, \dots, 
$\mathcal{M}_{\m-1}^{\ }$ are invertible, in the sense of modulo-$p$
arithmetic, because there is a unique multiplicative inverse for each non-zero
field element.
For instance, we have
\begin{equation}
  \label{eq1:odot-M27}
  \mathcal{M}_0^{\ }=
  \left(\begin{array}{ccc}
  1&0&0 \\ 0&0&1 \\ 0&1&2  
  \end{array}\right)\,,\qquad
  \mathcal{M}_1^{\ }=
  \left(\begin{array}{ccc}
  0&1&0 \\ 1&0&2 \\ 0&2&2  
  \end{array}\right)\,,\qquad
  \mathcal{M}_2^{\ }=
  \left(\begin{array}{ccc}
  0&0&1 \\ 0&1&2 \\ 1&2&0  
  \end{array}\right)\,,
\end{equation}
and
\begin{equation}
  \label{eq1:odot-M27inv}
  \mathcal{M}_0^{-1}=
  \left(\begin{array}{ccc}
  1&0&0 \\ 0&1&1 \\ 0&1&0  
  \end{array}\right)\,,\quad\,\,
  \mathcal{M}_1^{-1}=
  \left(\begin{array}{ccc}
  2&1&2 \\ 1&0&0 \\ 2&0&2  
  \end{array}\right)\,,\quad\,\,
  \mathcal{M}_2^{-1}=
  \left(\begin{array}{ccc}
  1&1&1 \\ 1&1&0 \\ 1&0&0  
  \end{array}\right)\,,
\end{equation}
for the $N=27$ example in (\ref{eq1:odot-N=27}).

Having thus established how the field addition $a\oplus b$ and the field
multiplication $a\odot b$ are implemented for any two field elements
${a,b=0,1,\ldots,N-1}$ with $N=p^{\m}$, we can put the Galois field to use.
For notational simplicity, let us denote by $\gamma$ the basic $p$th root of
unity,  
\begin{equation}
\gamma =\Exp{\I2\pi/p}\,,
\end{equation}
rather than writing $\gamma_p$ as in (\ref{eq1:AB-eigen}).
Exponentiating  $\gamma$ with elements $g$ of 
the field --- regarding now, as noted above, the field elements as integers
--- we obtain complex phase factors of the type 
$\gamma^g$ with $0\leq g\leq N-1\,$.
As $g$ is an integer, such phase factors can take on only $p$ different
values, which are completely determined by the first component $g_{0}$ of the
$p$-ary expansion of $g$,
\begin{equation}
  \label{eq2:gamma-g0}
  \gamma^g=\gamma^{g_0}\quad\mbox{for}\ g=\sum_{m=0}^{\m-1}g_mp^m\,,
\end{equation}
because $g_{0}$ is just the remainder of $g$ 
when dividing by $p$ in the usual sense. 
The phase factor $\gamma^g$ can be considered as a $p$-tuple 
generalization of the (binary) parity operation $\Exp{\I(2\pi/2)g}=(-1)^g$ 
of the q-bit case (that is $p=2$).
   
The following identity plays a fundamental role:   
\begin{equation}\label{identi1}
\sum_{j=0}^{N-1} \gamma ^{j\odot i}=N\delta_{i,0}\,.
\end{equation} 
Indeed, if $i=0$, then 
$\ds\sum_{j=0}^{N-1} \gamma ^{j\odot i}=\sum_{j=0}^{N-1}1=N$. 
Otherwise, 
\begin{equation}\label{identi1a}
i\neq0\,:\qquad
 \sum_{j=0}^{N-1} \gamma ^{j\odot i}
    =\sum_{j'=0}^{N-1} \gamma ^{ j'} 
\end{equation}
because the field multiplication is invertible, and then
\begin{equation}
\sum_{j'=0}^{N-1} \gamma ^{ j'}=p^{\m-1}\sum_{j'_0=0}^{p-1} \gamma ^{ j'_0}
=p^{\m-1}\frac{(1-\gamma ^p)}{(1-\gamma )}=0\,,  
\end{equation}
where the first step exploits (\ref{eq2:gamma-g0}) and recognizes that there
are $p^{\m-1}$ field elements $j'$ with the same value of $j'_0$.
  
The fact that the field addition is the component-wise addition modulo $p$, 
combined with the rule (\ref{eq2:gamma-g0}), implies the following useful
identity:  
\begin{equation}\label{identi2}
\gamma^{i}\gamma^{j}=\gamma^{i+j}=\gamma^{i_{0}+ j_{0}}=
 \gamma^{(i\oplus j)_{0}}=\gamma^{i\oplus j}\,.
\end{equation}
In the final expression on the right, the sum $i\oplus j$ is the Galois sum
of $i$ and $j$, which is then regarded as an integer, just as we regard the
result of the Galois multiplication $j\odot i$ in (\ref{identi1}) and
(\ref{identi1a}) as an integer, and so get integer powers of $\gamma$. 
Relation (\ref{identi2}) expresses, in the language of mathematicians, 
that the $p$th roots of unity are additive characters of the Galois 
field.\cite{Karpilovski}
  
It is important to note, in order to avoid confusions, that different types of
operations are present at this level: 
The internal field operations ($\oplus$ and $\odot$) must not be confused with
the modulo-$N$ operations.  
As an illustration of the differences between these operations, we consider
the case $p=2$, $\m=2$, $N=p^\m=4$  
and give the tables for field addition ($\oplus$) and field multiplication
($\odot$) in Table~\ref{tbl:4-field}(a) as well as the tables for 
modulo-$N$ addition and multiplication 
($\oplus_4$ and $\odot_4$, respectively) in Table~\ref{tbl:4-field}(b). 

One can check that the field and modulo-$4$ multiplications are distributive
with respect to the associated addition, but that there are no 
non-zero dividers of $0$ only in the case of the field 
multiplication, whereas we have $0=2\odot_42$ for the modulo-$4$ multiplication.
As a consequence, the field multiplication table exhibits an invertible
group structure when the first line and first column are removed.
All operations are commutative as can be seen from the invariance of all four
tables under transposition.  

Let us express q-quarts as products of two
q-bits, in accordance with the binary encoding of $i=(i_0,i_1)$ for $i=0,1,2,3$
as stated by
\begin{eqnarray}\label{eq1:4=2x2a}\hspace*{-4em}
\ket{i}_4=\ket{i_0}_2\otimes\ket{i_1}_2\,:\quad
\ket{0}_{4}&=&\ket{0}_{2}\otimes\ket{0}_{2}\,,\nonumber\\
\ket{1}_{4}&=&\ket{1}_{2}\otimes\ket{0}_{2}\,,\nonumber\\
\ket{2}_{4}&=&\ket{0}_{2}\otimes\ket{1}_{2}\,,\nonumber\\
\ket{3}_{4}&=&\ket{1}_{2}\otimes\ket{1}_{2}\,.  
\end{eqnarray}
With the aid of the $\oplus$ subtable in Table~\ref{tbl:4-field}, 
it is easy to verify that 
\begin{eqnarray}\label{eq1:4=2x2b}
&\ket{i\oplus j}_{4}=\ket{i_{0}\oplus_2j_{0}}_{2}
\otimes\ket{i_{1}\oplus_2j_{1}}_{2}&
\nonumber\\&\mbox{for $i=(i_0,i_1)$ and $j=(j_0,j_1)$}\,.&
\end{eqnarray}
This illustrates that the field addition is equivalent to the
component-wise modulo-$p$ addition.

\begin{table}[tb]
\tbl{%
(a)~Addition and multiplication tables for the field with $N=4$ elements.
(b)~Addition and multiplication modulo $N=4$.
\label{tbl:4-field}}
{\begin{tabular}{@{\hspace*{41pt}}lll@{\hspace*{41pt}}}
\textbf{(a)}&
\begin{tabular}{@{}c@{}|cccc}
\makebox[14.5pt]{$\oplus$}  & $0$ & $1$ & $2$  & $3$\\
\hline 0 & 0 & $1$ & $2$  & $3$ \\
1 & $1$ & $0$ & $3$  & $2$\\
2 & $2$ & $3$ & $0$  & $1$\\
3 & $3$ & $2$ & $1$  & $0$ 
\end{tabular}
 & 
\begin{tabular}{@{}c@{}|cccc}
\makebox[14.5pt]{$\odot$}   & $0$ & $1$ & $2$  & $3$\\
\hline  0 & 0 & $0$ & $0$  & $0$ \\
1 & $0$ & $1$ & $2$  & $3$\\
2 & $0$ & $2$ & $3$  & $1$\\
3 & $0$ & $3$ & $1$  & $2$ 
\end{tabular}
\\ & & \\
\textbf{(b)}& 
\begin{tabular}{@{}c@{}|cccc}
\makebox[14.5pt]{$\oplus_4$} & $0$ & $1$ & $2$  & $3$\\
\hline  0 & 0 & $1$ & $2$  & $3$ \\
1 & $1$ & $2$ & $3$  & $0$\\
2 & $2$ & $3$ & $0$  & $1$\\
3 & $3$ & $0$ & $1$  & $2$ 
\end{tabular}
&
\begin{tabular}{@{}c@{}|cccc}
\makebox[14.5pt]{$\odot_4$} & $0$ & $1$ & $2$  & $3$\\
 \hline  0 & 0 & $0$ & $0$  & $0$ \\
1 & $0$ & $1$ & $2$  & $3$\\
2 & $0$ & $2$ & $0$  & $2$\\
3 & $0$ & $3$ & $2$  & $1$ 
\end{tabular}
\end{tabular}}
\end{table}

It is also worth reminding that the properties
\begin{equation}\label{eq2:gammaN}
\gamma_N^i\gamma_N^j=\gamma_N^{i\oplus_N j}\quad\mbox{and}\quad
 \sum_{p=0}^{N-1} \gamma_N^{p\odot_N q}=N\delta_{q,0} 
\end{equation}
with $\ds\gamma_N=\Exp{\I2\pi/N}$ as in (\ref{eq1:AB-eigen})
are true for the modulo-$N$ addition and multiplication as well, but note that
$\gamma_N$ is the basic $N$th root of unity in these analogs of (\ref{identi2})
and (\ref{identi1}). 
In prime dimensions ($\m=1$, $N=p^1=p$) we have $\gamma=\gamma_N$ so that the
characteristics of the modulo-$p$ ring and the Galois field coincide. 
Indeed, both structures are rigorously identical in prime dimensions.
In prime-power but non-prime dimensions, for instance when $N=4$, 
this is not true.

\subsection{The computational basis}\label{sec2.1}
Consider now a quantum degree of freedom of prime-power dimension $N=p^\m$
--- a \emph{q-nit} composed of $\m$ \emph{q-pits}.
The corresponding Hilbert space of kets has a conveniently chosen orthonormal
reference basis consisting of $\ket{0}$, $\ket{1}$, \dots, ${\ket{N-1}}$, 
which we  regard as the \emph{computational basis} of kets.
The adjoint basis of bras comprises all $\bra{n}=\ket{n}^\dagger$ with
$n=0,1,\dots,N-1$.
As usual, the inner products $(\,\cdot\,,\,\cdot\,)$ of two kets or two bras
are given by Dirac brackets ($\equiv$ bra-kets), for which the orthonormality
relations 
\begin{equation}
  \bigl(\ket{i},\ket{j}\bigr)=\bigl(\bra{i},\bra{j}\bigr)
                             =\braket{i}{j}=\delta_{i,j}
\end{equation}
are an elementary illustration.

\subsection{The dual basis}\label{sec:dual}
Let us now consider the unitary transformations $V^0_{l}$ that shift each
label of the states of the computational basis
$\{\ket{0},\ket{1},\dots,\ket{i},\dots,\ket{N-1}\}$ by $l$,
\begin{equation}\label{def-V0l}
\ket{i} \to V^0_l\ket{i}=\ket{i\oplus l}\,,  
\end{equation}
so that each $ V^0_l$ implements a permutation among the kets of the
computational basis, but does not change the basis as a whole.
The shift in (\ref{def-V0l}) is a shift modulo $N$ in prime dimensions only
($N$=$p$) and then $V^0_l$ is identical with $X^l$ of Sec.~\ref{sec:WSexist}; 
in prime power dimensions ($N=p^\m$, $\m>1$) the shift consists of $\m$ shifts
modulo $p$, component-wise.  
The transformations effected by $V^0_{l}$ with $l=0,1,\dots,N-1$ make up a
commutative group of permutations with $N$ elements that is isomorphic to the
Galois addition. 

Generalizing the procedure outlined in Ref.~\refcite{DurtNagler}, 
we employ a suitable discrete Fourier-type transformation --- the inverse
\emph{Galois--Fourier} transformation --- to define the dual basis as follows: 
\begin{eqnarray} 
\ket{\tilde j}=\frac{1}{\sqrt{N}}
\sum_{k=0}^{N-1} \ket{k} \gamma ^{\ominus k\odot j}
\label{dual}
\end{eqnarray} 
where the symbol $\ominus$ represents the inverse of the Galois 
addition $\oplus $, that is: $x=\ominus y$ if $x\oplus y=0$.
It is easy to check that these dual kets are joint eigenkets of the unitary
permutation operators $V^0_{l}$. 
Indeed, we have
\begin{eqnarray}\label{V0l-eigen}
V^0_{l}\ket{\tilde  j}&=&\frac{1}{\sqrt{N}}\sum_{k=0}^{N-1}
 \ket{k \oplus l} \gamma^{\ominus k\odot j }\nonumber\\
&=&\frac{1}{\sqrt{N}}\sum_{k'=0}^{N-1}
\ket{k' }\gamma^{\ominus(k'\ominus l)\odot j}
=\ket{\tilde{j}}\gamma ^{l\odot j }\,,
\end{eqnarray}  
which identifies the eigenvalues $\ds\gamma ^{l\odot j }$.
These are $p$ different eigenvalues, each occurring $p^{\m-1}$ times.

Obviously, the dual basis and the computational basis are MU by construction,
\begin{equation}
  \bigl|\braket{\tilde{j}}{k}\bigr|^2
=\left|\frac{1}{\sqrt{N}}\gamma ^{j\odot k}\right|^2=\frac{1}{N}
\qquad\mbox{for all $j,k=0,1,\dots,N-1\,.$}
\end{equation}
When the dimension is prime ($N=p$), the dual basis is the standard discrete
Fourier transform of the computational basis, as in (\ref{eq1:q-Fourier}); 
when $N$ is a power of~$2$, the Galois--Fourier transform is a real Hadamard
transform.\cite{DurtNagler}  
  
Let us denote by $V_0^{l}$ the unitary transformations that shift each label of
the states of the dual basis 
$\{\ket{ \tilde 0},\ket{\tilde 1},\dots,\ket{\tilde i},\dots,%
\ket{\widetilde{N-1}}\}$ by $\ominus l$,
\begin{equation}
\ket{\tilde i} \to V_0^l\ket{\tilde i}= 
\ket{\widetilde{i\ominus l}}\,,   \qquad
\bra{\tilde{i}}\to\bra{\tilde{i}}V_0^l=\bra{\widetilde{i\oplus l}}\,,
\end{equation}
so that each $ V_0^l$ implements a permutation among the kets of the
dual basis, but does not change the basis as a whole.
In perfect analogy with the permutation operators $V^0_l$ of (\ref{def-V0l}),
the transformations effected by $V_0^{l}$ with $l=0,1,\dots,N-1$ upon the bras
$\bra{\tilde{i}}$ also compose
a commutative group of permutations with $N$ elements that is isomorphic to
the Galois addition.

These permutation operators are diagonal in the computational basis,
\begin{equation} \label{transla1}
V^l_{0}=  \sum_{k=0}^{N-1} \ket{\widetilde{k}}\bra{\widetilde{k\oplus l}}
 =  \sum_{k=0}^{N-1} \ket{ k}\gamma^{k\odot l}\bra{ k}\,.
\end{equation} 
This is the dual counterpart of the analogous expression for the shifts in the
computational basis,  
\begin{equation} \label{transla2} 
V^0_{l}=  \sum_{k=0}^{N-1} \ket{ k\oplus l}\bra{  k}
 =  \sum_{k=0}^{N-1} \ket{\tilde k}\gamma^{k\odot l}\bra{\tilde k}\,,
\end{equation} 
which is equivalent to (\ref{V0l-eigen}) and follows from that
eigenket statement. 

The unitary operators $V^0_l$ and $V^l_0$ are obviously analogs of the
operators $X^l$ and $Z^l$ of Sec.~\ref{sec:WSexist}, but for $\m>1$ these
operators are markedly different.
In particular, the period of $V^0_1$ and $V^1_0$ is $p$, not $N=p^\m$.
We indicate the difference by writing $\bra{\tilde{i}}$ for the dual basis
here, whereas the notation $\bra{\,\widehat{i}\,}$ is employed in
Sec.~\ref{sec:WSexist}. 

As mentioned above, it is immediately clear that 
$\ket{k}\to V^0_l\ket{k}=\ket{k\oplus l}$ is a component-wise addition, where
the components of the q-nit ket $\ket{k}$ are the $\m$ \mbox{q-pits} 
that compose it,
as is illustrated by (\ref{eq1:4=2x2a}) and (\ref{eq1:4=2x2b}) 
for $p=2$ and $\m=2$.
More generally,
\begin{eqnarray}
  \label{eq2:comp-shift1}
  V^0_l\ket{k}&=&V^0_l
\Bigl(\ket{k_0}\otimes\ket{k_1}\otimes\cdots\otimes\ket{k_{\m-1}}\Bigr)
 \nonumber\\
  &=&\ket{k_0+l_0}\otimes\ket{k_1+l_1}\otimes
               \cdots\otimes\ket{k_{\m-1}+l_{\m-1}}\,,
\end{eqnarray}
where each factor $\ket{k_m}$ in the tensor product is a q-pit ket, and the
sums $k_m+l_m$ are modulo-$p$ sums.
It follows that $V_l^0$ is a product of factors, each of which referring to
one of the q-pits,
\begin{equation}
  \label{eq2:comp-shift2}
  V^0_l=\left(V^0_1\right)^{l_0}\left(V^0_p\right)^{l_1}
        \left(V^0_{p^2}\right)^{l_2}\cdots
       =\prod_{m=0}^{\m-1}\left(V^0_{p^m}\right)^{l_m}\,,
\end{equation}
where the $m$th factor affects the $m$th q-pit only, with $V^0_{p^m}$ giving a
unit shift of the $m$th modulo-$p$ label.

In order to see that
$\bra{\tilde{k}}\to\bra{\tilde{k}}V^l_0=\bra{\widetilde{k\oplus l}}$ is a
q-pit--wise shift as well, we first observe that the Galois--Fourier
transformation (\ref{dual}) factorizes,
\begin{eqnarray}
  \label{eq2:dual-factors}
  \bra{\tilde{k}}&=&\frac{1}{\sqrt{N}}\sum_{j=0}^{N-1}\gamma^{k\odot j}\bra{j}
      =\frac{1}{\sqrt{N}}\sum_{j=0}^{N-1} \gamma^{k\mathcal{M}_0j^T}
        \bra{j_0}\otimes\bra{j_1}\otimes\bra{j_2}\otimes
        \cdots\otimes\bra{j_{\m-1}}
\nonumber\\ &=&
\frac{1}{\sqrt{p}}\sum_{j_0=0}^{p-1}\gamma^{(k\mathcal{M}_0)_0j_0}\bra{j_0}\otimes
\frac{1}{\sqrt{p}}\sum_{j_1=0}^{p-1}\gamma^{(k\mathcal{M}_0)_1j_1}\bra{j_1}\otimes
\cdots
\nonumber\\ &=&
\bra{\widetilde{\underline{k}_0}}\otimes
\bra{\widetilde{\underline{k}_1}}\otimes
\cdots\otimes\bra{\widetilde{\underline{k}_{\m-1}}}\,,
\end{eqnarray}
where $\mathcal{M}_0$ is the $0$th multiplication matrix in
(\ref{eq1:odot-arb}) and $\underline{k}_m$ is the $m$th component of 
$k\mathcal{M}_0=(\underline{k}_0,\underline{k}_1,\ldots)$.
Since $\mathcal{M}_0$ is invertible, we can parameterize the field element $k$
in terms of the coefficients $\underline{k}_m$,
\begin{equation}
  \label{eq2:dual-param}
  k=(\underline{k}_0,\underline{k}_1,\ldots)\mathcal{M}_0^{-1}=
    (\underline{k}_0,\underline{k}_1,\ldots)\left(
      \begin{array}{c}g_0\\g_1\\ \vdots \\g_{\m-1}\end{array}\right)=
    \sum_{m=0}^{\m-1}\underline{k}_mg_m\,,
\end{equation}
with the field elements $g_m$ defined such that their $p$-ary coefficients
make up the rows of the $\m\times\m$ matrix $\mathcal{M}_0^{-1}$.
Alternatively, we could define the $g_m$s by their basic property
\begin{equation}
  \label{eq2:dual-basis}
  \gamma^{p^m\odot g_n}=\gamma^{\delta_{m,n}}=\left\{
    \begin{array}{c@{\ \mbox{if}\ }l}
      \gamma & m=n\,,\\ 1 & m\neq n\,.
    \end{array}\right.
\end{equation}
Therefore, a unit increase of $\underline{k}_m$ means the addition of $g_m$
to $k$, and the shift operator $V^l_0$ factorizes accordingly into a product
of powers of single--q-pit Fourier operators, each of which (the $m$th, say)
acting on the single--q-pit bras $\bra{\widetilde{j_m}}$ only and leaving the
other ${\m-1}$ q-pit bras in the products of (\ref{eq2:dual-factors})
unaffected, 
\begin{equation}
  \label{eq2:comp-shift3}
  V^l_0=\left(V_0^{g_0}\right)^{\underline{l}_0}
        \left(V_0^{g_1}\right)^{\underline{l}_1}
        \left(V_0^{g_2}\right)^{\underline{l}_2}\cdots
       =\prod_{m=0}^{\m-1}\left(V_0^{g_m}\right)^{\underline{l}_m}\,,
\end{equation}
with the $m$th factor affecting the $m$th q-pit only,
\begin{equation}
  \label{eq2:comp-shift3a}
  \left(V_0^{g_m}\right)^{\underline{l}_m}\ket{k_m}
  =\ket{k_m}\gamma^{k_m\underline{l}_m}\,.
\end{equation}
For instance, we have $g_0=1$, $g_1=12$, $g_2=3$, and $\underline{k}_0=k_0$,
$\underline{k}_1=k_2$, $\underline{k}_2=k_1-k_2$ for the $N=27$ example of
(\ref{eq1:odot-N=27}), (\ref{eq1:odot-M27}), and (\ref{eq1:odot-M27inv}).

The respective unitary operator factors for unit shifts in
(\ref{eq2:comp-shift2}) and (\ref{eq2:comp-shift3}) commute if they refer to
different q-pits,
\begin{equation}
  \label{eq2:comp-shift4}
  V^0_{p^m}V_0^{g_n}=V_0^{g_n}V^0_{p^m}\quad\mbox{if $m\neq n\,,$}
\end{equation}
which essentially states that the Galois shifts with their component-wise
addition are consistent with the factorization of the ${N=p^\m}$-dimensional
degree of freedom into $\m$ $p$-dimensional degrees of freedom, as discussed
in Sec.~\ref{sec:WScomp}. 
And for the pair of operators to the same q-pit, one easily verifies the Weyl
commutation rule
\begin{equation}
  \label{eq2:comp-shift5}
   V^0_{p^m}V_0^{g_m}=\gamma^{-1}V_0^{g_m}V^0_{p^m}\,.
\end{equation}
Equations (\ref{eq2:comp-shift4}) and  (\ref{eq2:comp-shift5}) are particular
cases of (\ref{Weyl}) below.

\subsection{Construction of the remaining $N$-1 mutually 
unbiased bases}\label{sec2.3}
In the previous section we established a pair of MUB,
the computational basis, which can be chosen arbitrarily, and its dual basis,
defined by (\ref{dual}).
In this section, we shall generalize this construction in order to obtain 
the other $N-1$ bases that complement the computational basis and its dual
basis such that the bases of each of the $N(N+1)/2$ pairs are MU.
  
\subsubsection{Heisenberg--Weyl group}
Let us denote by $V^j_i$ the compositions of the shifts in the computational
and the dual bases, obtained by ordinary operator multiplication of
$V^j_{0}$ and $V^0_{i}$\,,
\begin{equation}\label{defV0}
 V^j_i= V^j_{0}V^0_{i}=\sum_{k=0}^{N-1}
  \ket{ k\oplus i}\gamma ^{( k\oplus i)\odot j}\bra{  k}
\quad\mbox{for $i,j=0,1,\dots,N-1\,,$} 
\end{equation} 
the building blocks of the Heisenberg--Weyl group.
This is consistent with the previous expressions for $i=0$ or $j=0$ because
$V_0^0$ is the identity.
In particular, for $i=0$ and $j=l$ we get the second sum of (\ref{transla1}),
and for $i=l$ and $j=0$ we have the first sum of (\ref{transla2}).

We note that the order of multiplication of $V^j_{0}$ and $ V^0_{j}$ 
matters in the definition (\ref{defV0}) because these unitary shift operators
do not commute,
\begin{equation} \label{Weyl}
V^0_{i}V^j_{0}=\gamma^{\ominus i\odot j}V^j_{0}V^0_{i}\,.
\end{equation}
We recognize here the Weyl commutation rule for the two unitary
operators $V^j_0$ and $ V^0_i$, which is their basic algebraic 
relation.\cite{Weyl1,Weyl2}

In dimension $N=p=2$, the commutation relation (\ref{Weyl}) is that of the
Pauli group (identify $V_0^1$ with $\sigma_x$ and $V_1^0$ with $\sigma_z$ once
more). 
When the dimension is a prime number, the field operations are the addition 
and multiplication modulo $p$, and the properties of MUB
are well-known;\cite{Ivanovic}
recall the discussion in Sec.~\ref{sec:WSprime} with its emphasis on the
Heisenberg--Weyl group.
 
Currently, we consider the Heisenberg--Weyl group associated with the Galois
addition and multiplication rather than the Heisenberg--Weyl group associated
with the usual modulo-$N$ operations. 
These groups coincide in prime dimensions but differ for non-prime but
prime-power dimensions. 
Notably, the Galois field is isomorphic to the modulo-$N$ ring in prime
dimensions only ($N=p$). 
Nevertheless, the Heisenberg--Weyl group factorizes in dimension $p^\m$ into
products of operators that belong to the local q-pit Heisenberg--Weyl group. 
In the case of translations of the computational basis, the factorization is
straightforward and given above in (\ref{eq2:comp-shift2}). 
And in the case of translations of the dual basis, where the mapping from
global operator labels to local operator labels is more intricate, see
(\ref{eq2:dual-factors})--(\ref{eq2:dual-basis}), the factorization is stated in
(\ref{eq2:comp-shift3}).

The composition law of the $N^2$ unitary operators introduced in (\ref{defV0})
is  
\begin{eqnarray}\label{discBH}
 V^j_iV_k^l&=&V^j_0V^0_iV^l_0V^0_k\nonumber\\
 &=&\gamma^{\ominus i\odot l}V^j_0 V^l_0 V^0_i V^0_k
 = \gamma^{\ominus i\odot l} V^{j\oplus l}_{i\oplus k}\,,
\end{eqnarray} 
which implies
\begin{equation}
  \label{inverse}
  {V_k^l}^{-1}={V_k^l}^\dagger=\gamma^{\ominus k\odot l}V_{\ominus k}^{\ominus l}
\end{equation}
and
\begin{equation}
  V_k^l V^j_i {V_k^l}^\dagger=\gamma^{l\odot i\ominus j\odot k} V^j_i
\end{equation}
for example.
Another implication is
\begin{eqnarray}
  \label{eq2:HWperiod}
  \bigl(V^j_i\bigr)^p
       &=&\Bigl(\gamma^{\ominus i\odot j}\Bigr)^{1+2+\cdots+(p-1)}V^0_0
        =\Bigl(\gamma^{\ominus i\odot j}\Bigr)^{\frac{1}{2}p(p-1)}\mathbf{1}
\nonumber\\ 
&=&\left\{
    \begin{array}{c@{\quad\textrm{for}\ }l}
      (-1)^{i\odot j}\mathbf{1} & p=2\,,\\ \mathbf{1} & p=3,5,7,11,\dots\,,
    \end{array}\right.
\end{eqnarray}
which is reminiscent of (\ref{eq1:HWgroup6}) and once again shows a difference
between the single even prime ${p=2}$ and the odd primes ${p>2}$.

Yet another implication is the orthonormality relation for the $V^j_i$s,
with respect to the Hilbert--Schmidt inner product,
\begin{equation}\label{ortho-VV}
\bigl(V_i^j,V_k^l\bigr)=\tr{{V^j_i}^\dagger V_k^l}=N\delta_{i,k}\delta_{j,l}\,,
\end{equation}
because all $V^j_i$s are traceless, except $V^0_0=\mathbf{1}$.
The other side of this coin is the relation
\begin{equation}
  \label{ergodicity}
  \frac{1}{N^2}\sum_{m,n=0}^{N-1} V_m^n\, A\, {V_m^n}^\dagger
  =\frac{1}{N}\tr{A}\mathbf{1}\,,
\end{equation}
which one may regard as a manifestation of Schur's lemma, inasmuch as the
right-hand side follows after observing that the sum on the left commutes with
all $V_j^i$ and must therefore be a multiple of the identity.
Schwinger\cite{Schwinger} calls such statements about equal-weight averages
over the whole phase space \emph{ergodic relations}. 

Equation (\ref{discBH}) is the discrete analog of the familiar
Baker--Campbell--Hausdorff relation for exponentiated position and momentum
operators that we encountered in (\ref{eq1:lim5}).
An immediate consequence of (\ref{discBH}) is
\begin{equation}
  \label{eq:commuting}
   V^j_iV^k_l=V^k_lV^j_i\quad
  \mbox{if $(i\odot k)_0=(j\odot l)_0$ and only then,}
\end{equation}
where $(\ )_0$ has the same meaning as in (\ref{identi2}).
In particular, (\ref{eq:commuting}) is fulfilled if $i\odot k=j\odot l$,
which we note for later reference.

\subsubsection{Abelian subgroups}  
Up to a global phase, (\ref{discBH}) looks like a group composition law. 
Indeed, one can show\cite{Durtsept} that there is a true analog of what we
observed in Sec.~\ref{sec:WSprime} for prime $N$: 
The $N^2$ unitary operators $V^j_{i}$ with $i,j=0,1,\dots,N-1$  make up
$N+1$ commuting sets (abelian subgroups of the Heisenberg--Weyl group)
of $N$ elements each that have only the identity $V_0^0$ in common. 
For each of these commuting sets, there is a basis of joint eigenkets of all
$V_i^j$s in the set. 
The $N+1$ bases thus identified are pairwise MU.
In passing, 
we note that this property can be shown, following an alternative approach
developed in Ref.~\refcite{india}, to be a consequence of the fact that the 
$V_i^j$ operators form what is called ``a maximally commuting 
basis of orthogonal unitary matrices.''  

It is expedient to introduce a fitting notation and terminology before we
proceed.  
We shall denote by $U^i_{l}$ the elements of these abelian subgroups, 
where $i$ labels the subgroup and runs from $0$ to $N$ to account for $N+1$
subgroups, while $l$ labels the $N$ elements in the subgroup and runs from $0$
to $N-1$.  
For the basis kets associated with the subgroups we use the 
convention that the $k$th basis ket for the $i$th subgroup is denoted by 
$\ket{e^{i}_{k}}$.

The abelian subgroups for $i=N$ and $i=0$ are composed of the two sets of
commuting operators of Sec.~\ref{sec:dual}, respectively,
\begin{eqnarray}\label{eq:defU01l}
U^N_l&=&V^l_0
=\sum_{k=0}^{N-1}\ket{k}\gamma^{k\odot l}\bra{k}
=\sum_{k=0}^{N-1}\ket{e^N_k}\gamma^{k\odot l}\bra{e^N_k}
\,,
\nonumber\\
U^0_l&=&V^0_l
=\sum_{k=0}^{N-1}\ket{\tilde{k}}\gamma^{k\odot l}\bra{\tilde{k}}
=\sum_{k=0}^{N-1}\ket{e^0_k}\gamma^{k\odot l}\bra{e^0_k}\,,
\end{eqnarray}
with $l=0,1,\dots,N-1$.
As indicated, we identify $\ket{k}$ with $\ket{e^N_k}$, and  
$\ket{\tilde{k}}$ with $\ket{e^0_k}$.
In other words, we choose the convention that the computational basis is the
$N$th basis, and the dual basis is the $0$th basis.

This suggests strongly that the other $N-1$ sets can be chosen such that
\begin{equation}\label{postul}
U^i_l=\sum_{k=0}^{N-1}\ket{e^i_k}\gamma^{k\odot l}\bra{e^i_k}
\end{equation}
with $i=1,2,\dots,N-1$ and $l=0,1,\dots,N-1$.
To complete the picture, we need to find the kets $\ket{e^i_l}$, such that
those with common label $i$ make up orthonormal sets, and the sets with
different $i$ labels are MU.
These requirements are compactly summarized by
\begin{equation}\label{postul2}
\bigl|\braket{e^i_k}{e^j_l}\bigr|^2
 =\delta_{i,j}\delta_{k,l}+\frac{1-\delta_{i,j}}{N}=\left\{
\begin{array}{cl}
\delta_{k,l}&\mbox{for $i=j$ (orthonormal),}\\[1ex]
1/N &\mbox{for $i\neq j$ (mutually unbiased),}
\end{array}\right.
\end{equation}
which have to hold for $i,j=0,1,\dots,N$ and $k,l=0,1,\dots,N-1$.

Irrespective of the choice for the $i$th orthonormal set of kets and bras in
(\ref{postul}), the $U^i_l$ are unitary and commute with each other for fixed
$i$, 
\begin{equation}\label{abelian}
  U^i_lU^i_{l'}= U^i_{l'}U^i_{l}=U^{i}_{l\oplus l'}\,,
\end{equation}
which is an immediate consequence of distributivity and the identity
(\ref{identi2}). 
In view of (\ref{eq:commuting}), we can guess that the $U^i_l$ of the $i$th set
are operators $V^k_{l}$ such that the Galois ratio $k\oslash l$ 
has the same $i$-dependent value for all of them.%
\footnote{For $l\neq0$, one naturally defines $k\oslash l$ by 
${(k\oslash l)\odot l=k}$.}\ 
For, if  $k\oslash l= k'\oslash l'$, then $k'\odot l=k\odot l'$,
and (\ref{eq:commuting}) implies that $V^k_{l}$ and $V^{k'}_{l'}$ commute.
    
We are thus invited to try the ansatz%
\footnote{For ${N=p}$ odd, we make contact with Sec.~\ref{sec:WSprime} for
  ${U^i_l=(XZ^i)^l}$, that is ${\alpha^i_l=\gamma^{-il(l+1)/2}}$.}
\begin{equation}\label{eq:defUil}
  U^i_l=\alpha_l^iV_l^{i\odot l}\quad\mbox{for $i=0,1,\dots,N-1\,,$}
\end{equation}
where the phase factors $\alpha_l^i$ have to be chosen consistently. 
In particular we have
\begin{eqnarray}\label{eq:phase1}
          &&\alpha^0_l=1\,:\quad U^0_l=V^0_l
\quad\mbox{for $l=0,1,\ldots,N-1$}
\nonumber\\
\mbox{and}&&\alpha^i_0=1\,:\quad U^i_0=V^0_0
\quad\mbox{for $i=0,1,\dots,N-1\,,$}
\end{eqnarray}
and the said consistency with (\ref{abelian}) requires 
\begin{equation}\label{eq2:phaseproduct}
  \alpha^i_k \alpha^i_l=
\alpha^i_{k\oplus l} \gamma^{i\odot k\odot l}\,,
\end{equation}
where (\ref{identi2}) and (\ref{discBH}) have been used repeatedly.
We note that all $U^i_l$s of (\ref{eq:defU01l}) and (\ref{postul}) have period
$p$, which tells us that the inclusion of $\alpha^i_l$ in (\ref{eq:defUil})
removes the even-odd distinction of (\ref{eq2:HWperiod}).

The orthonormality relation (\ref{ortho-VV}) carries over to the $U_l^i$s in
the form
\begin{equation}
  \label{ortho-UU}
  \tr{{U_k^i}^\dagger U_l^j}=N\delta_{k,l}\delta_{i\odot k,j\odot l}
    =\left\{\begin{array}{cl}
           N & \mbox{for $k=l=0\,$},\\
           N\delta_{i,j} & \mbox{for $k=l\neq0\,$},\\
           0 & \mbox{for $k\neq l\,$}.      
           \end{array}\right.
\end{equation}
This is, of course, (\ref{eq1:AB-trace}) in the present context.

Any choice for the phase factors $\alpha^i_l$ that obeys (\ref{eq:phase1})
and (\ref{eq2:phaseproduct}) is permissible in (\ref{eq:defUil}), but these
conditions do not determine the phase factors uniquely (except for $i=0$).
Just as (\ref{abelian}) remains valid when we replace $U^i_l$ by
$\gamma^{b_i\odot l}U^i_l$ with an arbitrary field element $b_i$,%
\footnote{Analogously, we could introduce a phase factor 
$\Exp{\I b(\alpha,\beta)t}$ in (\ref{eq1:conti2}) without 
affecting (\ref{eq1:conti2'}).}\   
the replacement $\alpha^i_l\to\alpha^i_l\gamma^{b_i\odot l}$ has no effect in 
(\ref{eq:phase1}) and (\ref{eq2:phaseproduct}), and in (\ref{eq:defUil}) 
it amounts to a permutation of the states in the $i$th basis: 
$\ket{e^i_k}\to\ket{e^i_{k\ominus b_i}}$, but leaves the basis as a whole
unchanged.\cite{Durtsept} 
Indeed, irrespective of the particular choice made for the phase factors in
(\ref{eq:defUil}), the set of common eigenkets $\ket{e_k^i}$ of the $N$
unitary operators in the $i$th abelian subgroup must always be the same --- a
different phase convention can only result in a different labeling of the
eigenkets.   

It remains to be shown, though, that there \emph{are} consistent choices for
all phase factors. 
This task has been completed in Ref.~\refcite{Durtsept}, from where we take
the following explicit solutions. 

In odd prime-power dimensions ($p=3,5,7,\dots$), where $1\oplus1=2$, 
the self-suggesting choice%
\footnote{Note that ${l\odot l\oslash2=l(l+p)/2\ (\mbox{mod}\ p)}$ for ${N=p}$
  odd.} 
\begin{equation} \label{conven}
\mbox{$p$ odd:}\quad\alpha_l^i=\gamma ^{\ominus(i\odot l\odot l)\oslash2} 
\end{equation} 
is simplest and indeed possible.
But in even prime-power dimensions ($p=2$), where $1\oplus1=0\neq2$, 
(\ref{conven}) does not work.

That the situation is more complicated for $p=2$ could perhaps be anticipated
because finite fields with even and odd cardinality are known to possess very
different structures. 
In the present context, the structural difference between $p=2$ and
$p=3,5,7,\dots$ manifests itself in the observation that
\begin{equation}
   \left(\alpha_l^j\right)^p=\left\{
     \begin{array}{l}
     (-1)^{j\odot l\odot l}=1\enskip\mbox{or}\enskip-1\quad\mbox{for $p=2\,,$}
     \\[1ex]  1\quad\mbox{for $p>2\,,$}
     \end{array}\right.\label{constrainteven}
\end{equation}
which combines with (\ref{eq2:HWperiod}) to ensure the $p$-periodicity of all
$U^i_l$s.  
As a consequence, we can systematically write $\alpha_l^j$ as a power of
$\gamma$ for odd $p$, as we do in (\ref{conven}).
For $p=2$ this is not possible but, instead, we can systematically write
$\alpha_l^j$ as a power of $\I=\sqrt{-1}=\Exp{\I{\frac{\pi}{2}}}$ because,
in virtue of (\ref{constrainteven}), $\alpha_l^j$ is the square root of a
power of  $\gamma=-1$ for $p=2$. 

Now, such a square root is only determined up to a global sign. 
Some extra work is thus necessary in order to fix these signs, which will
enable us to derive a $p=2$ counterpart of (\ref{conven}).
As a consequence of the group property (\ref{eq2:phaseproduct}), 
for each $j$ it is sufficient to fix $\m$ well chosen phases such that then
the values of all the $N=2^\m$ phases are determined.  

The $\m$ values of the signs of the phases $\alpha^j_l$ that we choose by
convention are $\alpha^j_1, \alpha^j_2, \dots, \alpha^j_{2^{\m-1}}$ and we
require, in agreement with (\ref{constrainteven}), that they obey 
\begin{equation}\label{eq2:alpha-even1}
     p=2:\quad       \alpha^j_{2^n}=\I^{j\odot 2^n\odot 2^n}
\quad\mbox{or}\quad  \alpha^j_{l_n2^n}=\I^{j\odot(l_n 2^n)\odot(l_n 2^n)}\,,
\end{equation}
where the latter version, with $l_n=0$ or $l_n=1$, incorporates $\alpha^j_0=1$
as well.
For
\begin{equation}
  \label{eq2:evenl-coeff}
  l=\sum_{n=0}^{\m-1}l_n2^n=\bigoplus_{n=0}^{\m-1}l_n2^n\,,
\end{equation}
we then have two ways of evaluating the product of all $\alpha^j_{l_n2^n}$s,
namely
\begin{equation}\label{eq2:alpha-even2}
  \prod_{n=0}^{\m-1}\alpha^j_{l_n2^n}
 =\prod_{n=0}^{\m-1}\I^{j\odot(l_n 2^n)\odot(l_n 2^n)}
\end{equation}
as an immediate consequence of (\ref{eq2:alpha-even1}), and
\begin{eqnarray}
    \prod_{n=0}^{\m-1}\alpha^j_{l_n2^n}&=&(-1)^{j\odot l_0
  \odot(l_12)}\alpha^j_{l_0\oplus l_12}\prod_{n=2}^{\m-1}\alpha^j_{l_n2^n}
\nonumber\\
&=&(-1)^{j\odot l_0 \odot(l_12)}(-1)^{j\odot(l_0\oplus l_12)\odot(l_22^2)} 
\alpha^j_{l_0\oplus l_12\oplus l_22^2}\prod_{n=3}^{\m-1}\alpha^j_{l_n2^n}
\nonumber\\
&=&\cdots
\nonumber\\
&=&\alpha^j_l\prod_{m=0}^{\m-2}\prod_{n=m+1}^{\m-1}(-1)^{j\odot(l_m2^m)\odot(l_n2^n)}
\end{eqnarray}
or
\begin{equation}\label{eq2:alpha-even3}
   \prod_{n=0}^{\m-1}\alpha^j_{l_n2^n}=
\alpha^j_l\mathop{\prod_{m,n=0}^{\m-1}}_{m\neq n}(-\I)^{j\odot(l_m2^m)\odot(l_n2^n)}
\end{equation}
by repeated application of (\ref{eq2:phaseproduct}).
The $n=m$ terms missing in (\ref{eq2:alpha-even3}) make up the product in
(\ref{eq2:alpha-even2}), so that we arrive at%
\footnote{\label{fn:eusebi}Owing to an oversight that was pointed out by
  Eusebi and Mancini,\cite{eusebi} the expression given in 
  Ref.~\refcite{Durtsept} is incorrect, but this inadvertence is of no
  consequence because the general properties (\ref{eq:phase1}) and
  (\ref{eq2:phaseproduct}) matter, not the explicit convention chosen for the
  values of the $\alpha^j_l$s. 
  The derivation (\ref{eq2:alpha-even1})--(\ref{eq2:even-alpha}) is
  essentially identical with the reasoning in Ref.~\refcite{eusebi}.}  
\begin{equation}\label{eq2:even-alpha}
  p=2:\quad \alpha^j_l=\prod_{m,n=0}^{\m-1}\I^{j\odot(l_m2^m)\odot(l_n2^n)}
\end{equation}
as the suitable square root of $(-1)^{j\odot l\odot l}$.
The additional option of replacing $\alpha^j_l$ by 
${\gamma^{b_j\odot l}\alpha^j_l}$, see the paragraph after
(\ref{ortho-UU}), amounts to extra factors of $(-1)^{l_n}$ in
(\ref{eq2:alpha-even1}) for some $n$ values.  
Examples of evaluating the product in (\ref{eq2:even-alpha}) can be found in
\ref{sec:app3}. 

Irrespective of the conventions adopted for the phase factors $\alpha^i_l$,
we note that the symmetry property
\begin{equation} \label{eq2:symmetry}
  \alpha_l^i=\alpha_{\ominus l}^i
\end{equation}
holds when $N$ is even, because $l=\ominus l$ for $p=2$.
It is also true for odd $N$ if the phases of (\ref{conven}) are chosen, but
not for all permissible choices.
If one imposes (\ref{eq2:symmetry}) as an additional condition, then
\begin{equation}\label{eq2:square}
  \left(\alpha_l^i\right)^2= \alpha_l^i \alpha_{\ominus l}^i 
=\gamma ^{\ominus i\odot l\odot l} 
\end{equation}
for all $N$ and all $i=0,1,\dots,N-1$, and (\ref{conven}) and
(\ref{eq2:even-alpha}) 
show how the proper square root of the right-hand side can be defined.
Unless explicitly stated, the symmetry (\ref{eq2:symmetry}) is not assumed for
$p>2$ in what follows, and neither are the explicit expressions (\ref{conven})
and (\ref{eq2:even-alpha}) for the phase factors.

\subsubsection{The remaining ${N-1}$ bases}
Having thus at our disposal the unitary operators $U^i_l$ of (\ref{postul})
and (\ref{eq:defUil}), we can also state quite explicitly the $N-1$ bases
associated with the abelian subgroups for ${i=1,2,\dots,N-1}$. 
For this purpose we exploit the analog of (\ref{eq1:projectors}),
\begin{equation} \label{MUBproj}
\ket{e_{k}^i}\bra{  e_{k}^i}
=\frac{1}{N}\sum_{l=0}^{N-1} \gamma^{\ominus k\odot l}U^{i}_{l}\,,
\end{equation}
which is an immediate consequence of (\ref{postul}) and (\ref{identi1}),
and from its implication
\begin{equation}\label{eq2:i2zero-braket}
  \braket{e^N_l}{e^i_k}\braket{e^i_k}{e^N_m}
  =\frac{1}{N}\bigl(\gamma^{k\odot l}\alpha^i_{\ominus l}\bigr)^*
              \bigl(\gamma^{k\odot m}{\alpha^i_{\ominus m}}\bigr)
\end{equation}
we find
\begin{equation} 
  \ket{e_{k}^i}=\frac{1}{\sqrt N}\sum_{l=0}^{N-1}\ket{e_{l}^N}
\gamma^{ \ominus k\odot l} {\alpha_{\ominus l}^i\!}^*\,.
\label{xxx}
\end{equation}
As a consequence, the unitary shift operators $V^n_m$ of the Heisenberg--Weyl
group, turn states of one basis into each other, but do not relate the bases
to one another,
\begin{equation}
  \label{eq2:shiftinbasis}
  V^n_m\ket{e^i_0}=\ket{e^i_{i\odot m\ominus n}}{\alpha_m^i\!}^*
  \quad\mbox{for $i=0,1,\dots,N-1$}\,,\qquad
  V^n_m\ket{e^N_0}=\ket{e^N_m}\,.
\end{equation}
   
Statements (\ref{eq2:i2zero-braket}), (\ref{xxx}), and
(\ref{eq2:shiftinbasis}), as well as
(\ref{eq2:eigen-verify1})--(\ref{eq2:tr-eV}) below, are valid both for
odd prime powers and even prime powers, whether the respective phase factors of
(\ref{conven}) and (\ref{eq2:even-alpha}) are used or any other permissible
choice, and apply also for $i=0$ when $\ket{e_k^0}=\ket{\tilde k}$ as required
by the conventions chosen in (\ref{eq:defU01l}) and (\ref{eq:phase1}).

Indeed, it is easy to establish the validity of the requirement (\ref{postul2})
for the projectors in (\ref{MUBproj}) by just exploiting (\ref{MUBproj})
itself and without relying on (\ref{xxx}):
\begin{eqnarray}\label{mubness}
\bigl|\braket{e^i_k}{e^j_l}\bigr|^2&=& 
\tr{\bigl(\ket{  e_{k}^i}\bra{  e_{k}^i}\bigr)\,
\bigl(\ket{  e_{l}^j}\bra{  e_{l}^j}\bigr)}
=\frac{1}{N^2}\sum_{m,n=0}^{N-1} 
\gamma^{k\odot m} \gamma^{\ominus l\odot n}
\tr{{U^i_m\!}^\dagger U^{j}_{n}}\nonumber\\
 &=&\frac{1}{N}\sum_{m,n=0}^{N-1} \gamma^{k\odot m\ominus l\odot n} 
\delta_{m,n}\delta_{i\odot m,j\odot n}
=\frac{1}{N}\sum_{m=0}^{N-1} \gamma^{(k\ominus l)\odot m} \delta_{i\odot m,j\odot m}
\nonumber\\
& =&\frac{1}{N}+\delta_{i,j}\frac{1}{N}\sum_{m=1}^{N-1}
\gamma^{( k \ominus l)\odot m}
=\frac{1}{N}+\delta_{i,j}\left(\delta_{k,l}-\frac{1}{N}\right)
\,,
\end{eqnarray} 
where the orthonormality relation (\ref{ortho-UU}) and the identity
(\ref{identi1}) are the main ingredients. 
The eigenvalue equations
\begin{equation}\label{eq2:eigen-value}
  U_l^i\ket{e^i_k}\bra{e^i_k}=\ket{e^i_k}\bra{e^i_k}U^i_l
  =\ket{e^i_k}\gamma^{k\odot l}\bra{e^i_k}
\end{equation}
also follow for (\ref{MUBproj}) directly from (\ref{abelian}). 

But it cannot be a mistake to check, for consistency, that $\ket{e_k^i}$ as
given in (\ref{xxx}) is the eigenket of $U^i_l$ of (\ref{eq:defUil}) to
eigenvalue $\gamma^{k\odot l}$. 
Starting from
\begin{eqnarray}\label{eq2:eigen-verify1}
  U^i_l\ket{e_m^N}=V_l^{i\odot l}\ket{m}\alpha^i_l
                  &=&\ket{m\oplus l}\gamma^{i\odot (m\oplus l)\odot l}
                                    \alpha^i_l
\nonumber\\
                  &=&\ket{m\oplus l}{\alpha^i_{\ominus(m\oplus l)}\!}^*\,
                                  \alpha^i_{\ominus m}
\end{eqnarray}
we have
\begin{eqnarray}\label{eq2:eigen-verify2}
   U^i_l\ket{e^i_k}\gamma^{\ominus k\odot l}
&=&\frac{1}{\sqrt{N}}\sum_{m=0}^{N-1} U^i_l\ket{e_m^N}
                     \gamma^{\ominus (m\oplus l)\odot k}
                     {\alpha^i_{\ominus m}\!}^*
\nonumber\\
                   &=&\frac{1}{\sqrt{N}}\sum_{m=0}^{N-1}
                      \ket{m\oplus l}{\alpha^i_{\ominus(m\oplus l)}\!}^*\,
                      \gamma^{\ominus (m\oplus l)\odot k}
                   =\ket{e^i_k}\,,\label{consistency}
\end{eqnarray}
indeed.

For later reference, we further observe that
\begin{equation}
  \label{eq2:tr-UV}
  \tr{{U^i_l}^\dagger V_m^n}
      =N\delta_{i\odot l,n}\,\delta_{l,m}\,{\alpha^i_l}^*\,,
\end{equation}
which follows from (\ref{eq:defUil}) and (\ref{ortho-VV}) and in turn implies
\begin{equation}
  \label{eq2:tr-eV}
  \bra{e^i_k}V_m^n\ket{e^i_k}=\delta_{i\odot m,n}\,\gamma^{k\odot m}
   {\alpha^i_m\!}^*\,,
\end{equation}
upon invoking the adjoint version of (\ref{MUBproj}).
And finally we note that the unitary mapping of the computational basis ($i=N$)
onto the $i$th basis is accomplished by the Clifford operator $C_i$ whose
defining property, that is:  
$C_i\ket{e^N_k}=\ket{e^i_k}$ for all $k$, implies
\begin{equation}
  \label{eq:Clifford}
  C_i=\sum_{k=0}^{N-1}\ket{e^i_k}\bra{e^N_k}\,.
\end{equation}
This includes $C_N=\mathbf{1}$.
The terminology ``Clifford operators'' refers to the Clifford 
group,\cite{planat08} which consists of all unitary operators
that map the Heisenberg--Weyl group onto itself under conjugation, that is:
$V^i_l\to C^\dagger V^i_lC$ equals one of the $V^i_l$s for each $C$ in the
Clifford group, in full analogy to the discussion in Sec.~\ref{sec:WSgroups}.

\subsection{Complementary period-$N$ observables}
\label{sec2.4}
In a sense, the $N+1$ abelian subgroups replace the $N+1$ complementary
observables of Sec.~\ref{sec:WSprime} whose powers constitute the $N+1$
abelian subgroups for prime $N$. 
But there are much closer analogs in the form of $N+1$ pairwise complementary
period-$N$ observables for which (\ref{eq1:AB-trace}) applies immediately,
rather than the analog we have in (\ref{ortho-UU}).

For each abelian subgroup, $i=0,1,2,\dots,N$, we introduce a period-$N$
observable by means of 
\begin{equation}
  \label{eq2:PCO1}
  Z_i=\sum_{k=0}^{N-1}\ket{e^i_k}\gamma_N^k\bra{e^i_k}
     =\frac{1}{N}\sum_{k,l=0}^{N-1}\gamma_N^k\gamma^{\ominus k\odot l}U^i_l\,.
\end{equation}
By construction, these observables constitute a maximal set of pairwise
complementary observables for the $N$-dimensional degree of freedom. 
See Table~\ref{tbl:qbitpair} in Sec.~\ref{sec:allMUB-Nle5} for an example of
five such observables for ${N=4}$.

         %% Section 2
%%%% file name: MUB-3.tex
%%%% input file for MUB.tex 
%%%%
%%%% last changes on 20 April 2010 by Berge
%%%% typo corrected on 27 April 2010
%%%% 
%%%%%%%%%%%%%%%%%%%%%%%%%%%%%%%%%%%%%%%%%%%%%

\section{Generalized Bell states and their applications}\label{section3}
There is a one-to-one correspondence between the elements of an orthonormal
basis of generalized Bell states and the Heisenberg--Weyl group of unitary
transformations.\cite{Fivel,DurtNagler,Durtmutu}    
This correspondence is a key concept for a uniform view of several important
applications in quantum information science, such as 
quantum dense coding (Sec.~\ref{sec3.2}), 
quantum teleportation (Sec.~\ref{sec3.3}),  
quantum cloning (Sec.~\ref{sec3.4}),
and entanglement swapping (Sec.~\ref{sec3.5}).  

The construction that we use here employs the Heisenberg--Weyl group of
Sec.~\ref{section2} whose shift operators (\ref{defV0}) change state labels
via field addition.\cite{DurtNagler,Durtmutu}     
In the context of generalized Bell states, the analogous construction 
based on the modulo-$N$ Heisenberg--Weyl operators of Sec.~\ref{sec:WSgroups}
works equally well.\cite{Fivel} 
With the necessary changes, all applications in
Secs.~\ref{sec3.2}--\ref{sec3.5} can be implemented by these other 
Bell states.\cite{cosmos}

\subsection{Generalized Bell states}\label{sec3.1}  
Following Refs.~\refcite{DurtNagler}, \refcite{Durtmutu},
and \refcite{DurtKwek}, we can define the generalized Bell states by the
following procedure.  
First, for all kets $\ket{\psi}$ and bras $\bra{\phi}$ we introduce
\emph{conjugate} kets $\ket{\psi^*}$ and bras $\bra{\phi^*}$ whose defining 
property is 
\begin{equation} 
  \label{eq:conjbas}
  \braket{\psi^*}{\phi^*}=\braket{\psi}{\phi}^*=\braket{\phi}{\psi}\,.
\end{equation}
Although this does not identify the conjugate kets and bras uniquely, any two
implementations of the map $\ket{\psi}\to\ket{\psi^*}$ are related to each 
other by a unitary transformation and, therefore, it does not matter which
convention we employ for the implementation of our choosing. 

Since the conjugate kets transform like the original bras, we have a very 
useful one-to-one correspondence of one--q-nit operators 
$\ket{\psi}\bra{\phi}$ and two--q-nit states,%
\footnote{In an experimental realization, the two different $N$-ary quantum
  degrees of freedom, the two q-nits, could just as well be carried by one
  physical object or by several.}  
\begin{equation}
  \label{eq:correspond}
  \ket{\psi}\bra{\phi} \longleftrightarrow \ket{\phi^*,\psi}\,, 
\end{equation}
which is linear in both the ket part and the bra part of the one--q-nit
operator. 
As a consequence, we have relations such as
\begin{equation}  
  \label{eq3:InnerProd}
  \mbox{if $A\longleftrightarrow\ket{a}$ and $B\longleftrightarrow\ket{b}$,
        then $\tr{A^\dagger B}=\braket{a}{b}$}
\end{equation}
as well as
\begin{equation} 
  \label{map2nd} 
    \mbox{if $A\longleftrightarrow\ket{a}$,
        then $BA\longleftrightarrow(\mathbf{1}\otimes B)\ket{a}$}
\end{equation}
and
\begin{equation} 
  \label{map2nd'} 
    \mbox{if $A\longleftrightarrow\ket{a}$,
        then $AB^\dagger\longleftrightarrow (B^*\otimes\mathbf{1})\ket{a}\,,$}
\end{equation}
where $B^*\ket{\phi^*}=\ket{\psi^*}$ if $B\ket{\phi}=\ket{\psi}$.
Take, for instance, 
\begin{equation}  
A=\ket{\psi_1}\bra{\phi_1}\longrightarrow\ket{a}=\ket{\phi_1^*,\psi_1}
\quad\mbox{and}\quad
B=\ket{\psi_2}\bra{\phi_2}\longrightarrow\ket{b}=\ket{\phi_2^*,\psi_2}\,,  
\end{equation}
for which
\begin{equation} 
  \tr{A^\dagger B}=\braket{\phi_2}{\phi_1}\braket{\psi_1}{\psi_2}
                 =\braket{\phi_1^*}{\phi_2^*}\braket{\psi_1}{\psi_2}
                 =\braket{a}{b}   
\end{equation}
as well as
\begin{equation}  
  BA=\ket{\psi_2}\braket{\phi_2}{\psi_1}\bra{\phi_1}
    \longrightarrow \ket{\phi_1^*,\psi_2}\braket{\phi_2}{\psi_1}
    =\Bigl(\mathbf{1}\otimes\ket{\psi_2}\bra{\phi_2}\Bigr)
       \ket{\phi_1^*,\psi_1}
\end{equation}
and
\begin{equation}  
  AB^\dagger=\ket{\psi_1}\braket{\phi_1}{\phi_2}\bra{\psi_2}
    \longrightarrow \ket{\psi_2^*,\psi_1}\braket{\phi_2^*}{\phi_1^*}
    =\Bigl(\ket{\psi_2^*}\bra{\phi_2^*}\otimes\mathbf{1}\Bigr)
       \ket{\phi_1^*,\psi_1}\,.
\end{equation}
Quite generally, the mapping (\ref{eq:correspond}) turns statements about
one--q-nit operators into statements about two--q-nit kets.  
 
An important example is the observation that irrespective of the basis used in 
the completeness relation, the identity operator is mapped onto one and the
same ket $\ket{B^{\ }_{0,0}}$,  
\begin{equation}
  \label{mapId}
  \mathbf{1}=\sum_k\ket{k}\bra{k}=\sum_k\ket{e_k^i}\bra{e_k^i}
\longleftrightarrow\sum_k\ket{k^*,k}=\sum_k\ket{e_k^{i*},e_k^i}
=\ket{B^{\ }_{0,0}}\sqrt{N}\,,
\end{equation}
here illustrated for the computational basis and either one of the bases of
(\ref{xxx}).   
The factor $\sqrt{N}$ normalizes $\ket{B^{\ }_{0,0}}$ to unit length, consistent
with (\ref{eq3:InnerProd}) and $\tr{\mathbf{1}}=N$.  
Owing to its basis independence, the ket $\ket{B^{\ }_{0,0}}$ plays a 
central role in tomographic protocols for quantum key distribution;
see, e.g., Refs.~\refcite{Bruss:98,Bruss+1:02,tomocrypt}.

While $\ket{B^{\ }_{0,0}}$ is basis independent in the sense of (\ref{mapId})
for a given implementation of the conjugation $\ket{\psi}\to\ket{\psi^*}$, one
should realize that different definitions of this map do result in different
forms of $\ket{B^{\ }_{0,0}}$ as expressed in the original bases.
As an example, consider the case $N=2$ of a single q-bit, and the
following four alternative ways, four of many, of defining the map 
$\ket{\psi}\to\ket{\psi^*}$:
\begin{equation}
  \label{eq3:qbit*}
 \ket{\psi}=\ket{0}\alpha+\ket{1}\beta\to\ket{\psi^*}=\left\{
    \begin{array}{l}
      \ket{0}\alpha^*+\ket{1}\beta^*\,, \\[0.5ex]
      \ket{0}\alpha^*-\ket{1}\beta^*\,, \\[0.5ex]
      \ket{0}\beta^*+\ket{1}\alpha^*\,, \\[0.5ex]
      \ket{0}\beta^*-\ket{1}\alpha^*\,.
    \end{array}\right.
\end{equation}
The respective two--q-bit kets $\ket{B^{\ }_{0,0}}$,
\begin{equation}
  \label{eq3:qbit-B00}
 \ket{B^{\ }_{0,0}}=\frac{1}{\sqrt{2}}\bigl(\ket{0^*,0}+\ket{1^*,1}\bigr)
  =\left\{
    \begin{array}{l}
   \ds\frac{1}{\sqrt{2}}\bigl(\ket{0,0}+\ket{1,1}\bigr)\,,\\[2ex]
   \ds\frac{1}{\sqrt{2}}\bigl(\ket{0,0}-\ket{1,1}\bigr)\,,\\[2ex]
   \ds\frac{1}{\sqrt{2}}\bigl(\ket{0,1}+\ket{1,0}\bigr)\,,\\[2ex]
   \ds\frac{1}{\sqrt{2}}\bigl(\ket{0,1}-\ket{1,0}\bigr)\,,
    \end{array}\right.
\end{equation}
are the familiar standard Bell states \cite{bellstate}.
The four maps in (\ref{eq3:qbit*}) differ by simple unitary transformations,
and the same unitary transformations (of the first qubit) relate the four Bell
states to each other.
For instance, $\sigma_z=\ket{0}\bra{0}-\ket{1}\bra{1}$ turns the first and
second versions of $\ket{\psi^*}$ into each other, and also the third and
fourth versions. 
Likewise, $\sigma_z\otimes\mathbf{1}$ interchanges the first and second Bell 
states, and the third and fourth.
These observations for q-bits invite us to call $\ket{B^{\ }_{0,0}}$ 
a \emph{generalized Bell state}.
   
In view of $V_0^0=\mathbf{1}$, we recognize that  
$N^{-1/2}V_0^0\longleftrightarrow\ket{B^{\ }_{0,0}}$, which identifies
$\ket{B^{\ }_{0,0}}$ as one of the $N^2$ members of the set composed of the kets
$\ket{B^{\ }_{m,n}}$ that correspond to the unitary shift operators $V_m^n$,  
\begin{equation}\label{defBmn}
V^n_m=
\sum_{k=0}^{N-1} \ket{k\oplus m}\gamma^{(k\oplus m)\odot n}\bra{ k}
\longleftrightarrow
\sum_{k=0}^{N-1} \ket{k^*,k\oplus m}\gamma^{(k\oplus m)\odot n}
=\ket{B^{\ }_{m,n}}\sqrt{N}\,.
\end{equation}
These make up the set of \emph{generalized Bell states}.
Their orthonormality follows from (\ref{eq3:InnerProd}) and (\ref{ortho-VV}),
\begin{equation}
  \braket{B^{\ }_{m,n}}{B^{\ }_{m',n'}}=\frac{1}{N}\tr{{V_m^n}^\dagger V_{m'}^{n'}}=
\delta_{m,m'}\delta_{n,n'}\,, 
\end{equation}
and (\ref{map2nd}) implies that the shift operators $V_m^n$ permute the Bell
states, 
\begin{eqnarray}\label{Bell-2}
  (\mathbf{1}\otimes V_r^s)\ket{B^{\ }_{m,n}}
      &=&\ket{B^{\ }_{m\oplus r,n\oplus s}}\gamma^{\ominus(r\odot n)}\,, 
\nonumber\\
   (V_r^{s*}\otimes\mathbf{1})\ket{B^{\ }_{m,n}}
      &=&\ket{B^{\ }_{m\ominus r,n\ominus s}}\gamma^{(m\ominus r)\odot s}\,,
\end{eqnarray}
where (\ref{discBH}) enters.
In particular, we have  
\begin{eqnarray}
  \label{B00-to-Bmn}
\ket{B^{\ }_{m,n}}&=&(\mathbf{1}\otimes V_m^n)\ket{B^{\ }_{0,0}}\,,
\nonumber\\  
\ket{B^{\ }_{\ominus m,\ominus n}}&=&(V_m^{n*}\otimes\mathbf{1})\ket{B^{\ }_{0,0}}
\gamma^{m\odot n}\,,  
\end{eqnarray}
which relate all generalized Bell states to their ``seed'' 
$\ket{B^{\ }_{0,0}}$ of (\ref{mapId}). 

We note the identity
\begin{equation}\label{eq3:MKinvar}
  (V_m^{n*}\otimes V_m^n)\ket{B^{\ }_{0,0}}=\ket{B^{\ }_{0,0}}\,,
\end{equation}
which states the invariance of the seed under simultaneous shifts of
both q-nits.  
And the analog of (\ref{ergodicity}) is
\begin{equation}
  \label{eq3:Bell-ergod}
  \frac{1}{N}\sum_{m,n=0}^{N-1} (V_m^{n*}\otimes V_m^n)\ket{\phi^*,\psi}
=\ket{B_{0,0}}\sqrt{N}\braket{\phi}{\psi}\,,
\end{equation}
which once more emphasizes the particularity of the invariant Bell seed.

Since all Bell states are related to the maximally entangled seed by a local
unitary transformation (``local'' because $\mathbf{1}\otimes V_m^n$ affects
the second q-nit only in the two--q-nit state to which $\ket{B^{\ }_{0,0}}$
refers), each of them is maximally entangled, and since they are
orthonormal and $N^2$ in number, they constitute an orthonormal, maximally
entangled basis in the Hilbert space of two--q-nit kets.
Technically speaking, this $N^2$-dimensional Hilbert
space is obtained by taking the tensor product of the $N$-dimensional Hilbert
space, in which we have the computational basis and all that, with itself.  
 
Owing to the correspondence $\ket{B_{m,n}}\longleftrightarrow N^{-1/2}V_m^n$
in (\ref{defBmn}), 
the expansion of any $N$-dimensional single--q-nit operator in the operator
basis of the $V_m^n$ shift operators 
is equivalent to the decomposition of a $N^2$-dimensional two--q-nit state
ket in the orthonormal Bell-state basis.   
This is at the heart of the quantum tomography techniques that we present in
Sec.~\ref{sec4.2} below.  

The Bell states in (\ref{B00-to-Bmn}) refer
explicitly to the computational basis because (\ref{defV0}) expresses $V_m^n$
in terms of the $\ket{k\oplus m}\bra{k}=\ket{e^N_{k\oplus m}}\bra{e^N_k}$
ket-bra products.
We get the Bell states relative to the $i$th basis by applying the Clifford
operator $C_i$ of (\ref{eq:Clifford}) to the second q-nit and its dual
analog $C_i^*$, which we define by $C_i^*\ket{e^{N*}_k}=\ket{e^{i*}_k}$, to
the first q-nit.
In view of (\ref{map2nd}) and (\ref{map2nd'}), the analog of the
correspondence (\ref{defBmn}) for the computational basis, is then
\begin{equation}
(C_i^*\otimes C_i^{\,})\ket{B_{m,n}}\longleftrightarrow
\frac{1}{\sqrt{N}}  C_i^{\,}V_m^n C_i^\dagger \,,
\end{equation}
and as a consequence of the trace rule (\ref{eq3:InnerProd}) we have
\begin{equation}
  \bra{B_{m,n}}(C_i^*\otimes C_i^{\,})\ket{B_{m',n'}}
  =\frac{1}{N}\tr{{V_m^n}^\dagger C_i^{\,} V_{m'}^{n'} C_i^\dagger}\,.
\end{equation}
Upon employing (\ref{xxx}) to express $C_i^{\,} V_{m}^{n} C_i^\dagger$ in terms
of the computational basis,
\begin{eqnarray}
  \label{eq3:ith-Vmn}
  C_i^{\,} V_{m}^{n} C_i^\dagger&=&\sum_{k=0}^{N-1} \ket{e^i_{k\oplus m}}
\gamma ^{(k\oplus m)\odot n}\bra{e^i_k}
\nonumber\\
&=&\sum_{l=0}^{N-1}\ket{n\oplus l}{\alpha^i_{\ominus(n\oplus l)}\!}^*\,
        \gamma^{\ominus l\odot m}\alpha^i_{\ominus l}\bra{l}\,,
\end{eqnarray}
the trace is readily evaluated, and we find
\begin{eqnarray}
\bra{B_{m,n}}(C_i^*\otimes C_i^{\,})\ket{B_{m',n'}}&=&
  \delta_{m,n'}\,\delta_{i\odot m\ominus n,m'}\,
   \gamma^{\ominus m\odot n}{\alpha^i_{\ominus m}\!}^*
\nonumber\\ &=&
  \delta_{m,n'}\,\delta_{n,i\odot n'\ominus m'}\,
      \gamma^{m'\odot n'}\alpha^i_{n'} \,.
\end{eqnarray}
The two Kronecker delta symbols tell us that the application of the unitary
operator $C_i^*\otimes C_i^{\,}$ to the Bell basis permutes the Bell states, 
but leaves the basis as a whole unaltered.

Quite explicitly, we have  
\begin{eqnarray}\label{eq3:differentBells}
  \ket{B_{m,n}}&=&(C_i^*\otimes C_i^{\,})\ket{B_{i\odot m\ominus n,m}}
                  \gamma^{m\odot n} \alpha^i_{\ominus m}\,,
\nonumber\\
(C_i^*\otimes C_i^{\,})\ket{B_{m,n}}&=&\ket{B_{n,i\odot n\ominus m}}
                  \gamma^{m\odot n} \alpha^i_n\,,
\end{eqnarray}
where the mappings of the indices,
\begin{equation}\label{mappings}
  \left(\begin{array}{c} m \\ n \end{array}\right)\to
  \left(\begin{array}{c@{\quad}c} i & \ominus 1 \\ 1 & 0
    \end{array}\right)
\odot\left(\begin{array}{c} m \\ n \end{array}\right)\,,\qquad
  \left(\begin{array}{c} m \\ n \end{array}\right)\to
  \left(\begin{array}{c@{\quad}c} 0 & 1 \\ \ominus 1 & i
    \end{array}\right)
\odot\left(\begin{array}{c} m \\ n \end{array}\right)\,,
\end{equation}
are each other's inverse.
The particular case of $m=n=0$,
\begin{equation}
  (C_i^*\otimes C_i^{\,})\ket{B_{0,0}}=\ket{B_{0,0}}\,,
\end{equation}
states the invariance of the Bell seed when switching from one basis to
another, which we have observed earlier in the context of (\ref{mapId}).

\subsection{Quantum dense coding}\label{sec3.2}  
The generalization of q-bit quantum dense coding\cite{dense} to an arbitrary
dimension $N$ is an immediate application of (\ref{B00-to-Bmn}). 
It goes as follows.\cite{cosmos}  
Alice and Bob initially  share the seed state $\ket{B^{\ }_{0,0}}$ of the Bell
basis, with q-nit~1 in Alice's possession and q-nit~2 in Bob's. 
Bob applies one of the $N^2$ unitary shift operators $V_m^n$ to his
q-nit~2 and then sends it to Alice who, according to (\ref{B00-to-Bmn}), 
has the q-nit pair in the Bell state $\ket{B^{\ }_{m,n}}$.
She finds out which of the states is the case by performing a von Neumann
measurement in the Bell basis.  

The measurement result tells her which one of the $N^2$ shifts was
implemented by Bob, and so she receives $2\log_2N$ bits of information, as
much as two classical $N$-valued signals could convey. 
In a manner of speaking, Bob has transmitted two c-nits by sending
one q-nit.   
This is the essence of dense coding; quite like the teleportation of the
following section, it has no classical counterpart.

Despite this ``manner of speaking,'' quantum dense coding does not violate the
\emph{Holevo bound},\cite{Holevo:73} which states that a single q-nit can only
transmit one c-nit, because of the earlier distribution of q-nit~1 to Alice
that is entangled with Bob's q-nit~2 from the beginning.
At the time when Alice carries out the measurement that discriminates the Bell
states, she has received both q-nits.

\subsection{Quantum teleportation}\label{sec3.3}  
The relation between maximal sets of orthogonal families of unitary matrices and
teleportation was already emphasized several years ago.\cite{werner}
Several generalizations of the teleportation scheme to arbitrary dimension
that were proposed in the past\cite{Zubairy:98,cosmos,Zhang+2:07} are close in
spirit to the generalization that we proceed to describe now.  

A central ingredient of the q-nit teleportation process is the
three--q-nit--states identity
\begin{eqnarray}
  \label{teleport1}
  \sum_{k=0}^{N-1}\ket{k^*,j,k}
 &=&\sum_{k,m,n}\ket{B^{\ }_{m,n},k}\braket{B^{\ }_{m,n}}{k^*,j}
\nonumber\\&=&
\sum_{k,m,n}\ket{B^{\ }_{m,n},k}\,
          \tr{N^{-1/2}{V_m^n}^\dagger\,\ket{j}\bra{k}}\nonumber\\&=&
\frac{1}{\sqrt{N}}\sum_{k,m,n}\ket{B^{\ }_{m,n},k}\bra{k}{V_m^n}^\dagger\ket{j}
\nonumber\\&=&
\frac{1}{\sqrt{N}}\sum_{m,n}
 (\mathbf{1}\otimes\mathbf{1}\otimes {V_m^n}^\dagger)\ket{B^{\ }_{m,n},j}\,,    
\end{eqnarray}
where the completeness of the Bell basis, the trace relation
(\ref{eq3:InnerProd}), and the completeness of the computational basis are
exploited.
  
Now, to teleport an unknown state $\ket{\psi}=\sum_j\ket{j}\psi_j$ from
q-nit~2 to q-nit~3, we prepare q-nits~1 and~3 in their 
Bell seed state, so that the initial three--q-nit state is
\begin{equation}
  \label{teleport2}
  \frac{1}{\sqrt{N}}\sum_{j,k}\ket{k^*,j,k}\psi_j
 =\frac{1}{N}\sum_{m,n}
   (\mathbf{1}\otimes\mathbf{1}\otimes {V_m^n}^\dagger)\ket{B^{\ }_{m,n},\psi}\,.
\end{equation}
A von Neumann measurement in the Bell basis for q-nits~1 and~2 will find
one of the generalized Bell states, $\ket{B^{\ }_{m,n}}$ say, all $N^2$ outcomes
being equally probable.
Conditioned on the said measurement result, the state ket for q-nit~3 is
then ${V_m^n}^\dagger\ket{\psi}$, which is turned into $\ket{\psi}$ by
performing the unitary transformation described by the shift operator $V_m^n$.
In effect, the unknown state $\ket{\psi}$ has been teleported successfully and
without any distortion from q-nit~2 to q-nit~3.
If, at the time of the Bell measurement on q-nits~1 and~2, they are
separated from q-nit~3 by a space-like distance, there exists no 
classical counterpart for this quantum teleportation.

\subsection{Quantum cryptography, covariant cloning machines, 
and error operators}\label{sec3.4}
In quantum cryptography, MUB play an important role
because they maximize uncertainty relations which ensures the confidentiality
of protocols for quantum  key
distribution,\cite{BB84,optimalencrypt,Bechmann,Bruss:98,Bruss+1:02} although
MUB are not really needed in arbitrary dimensions.\cite{tomocrypt}
For instance, the celebrated BB84 protocol\cite{BB84} consists of encrypting
the message in a q-bit state that is chosen at random between four states that
belong to two MUB.
The relevance of MUB for quantum cloning has also been
recognized,\cite{Cerf:98,Cerf:00a,Cerf:00b,qutrits,DurtNagler} 
which is not unexpected in view
of the close link between cloning and the security of key distribution
protocols: as a rule, the most dangerous eavesdropping attacks can be realized
with the aid of optimized one-to-two cloners --- the so-called phase-covariant
cloner,\cite{Fuchs+4:97,Griffiths+1:97,NG,bruss} for instance, when attacking
the BB84 protocol.  
 
The symmetry properties of the Bell states have important implications in the
theory of cloning machines,\cite{DurtNagler,qutrits} as we shall sketch
briefly now.   
Under very general conditions,\cite{DurtKwek} optimal cloning states obey
Cerf's  ansatz,\cite{Cerf:98,Cerf:00a,Cerf:00b}
\begin{eqnarray}  \label{CERF}
\ket{\Psi_{0-3}}&=&\sum_{m,n=0}^{N-1}\ket{B^{\ }_{m,n},B^{\ }_{\ominus m,\ominus n}}
\gamma^{\ominus m\odot n}a_{m,n}
\nonumber\\&=&\sum_{m,n=0}^{N-1}
(\mathbf{1}\otimes V_m^n\otimes\mathbf{1}\otimes {V_m^n}^\dagger)
\ket{B^{\ }_{0,0},B^{\ }_{0,0}}a_{m,n}\,,
\end{eqnarray}
which is a four--q-nit state that is constructed as a linear superposition
of states that have q-nits~0 and~1 in the $m,n$ Bell state
and q-nits~2 and~3 in the $\ominus m,\ominus n$ Bell state.
Except for the normalization constraint,
\begin{equation}
  \label{eq3:a-norm}
  \braket{\Psi_{0-3}}{\Psi_{0-3}}=\sum_{m,n=0}^{N-1}\bigl|a_{m,n}\bigr|^2=1\,,
\end{equation}
the probability amplitudes $a_{m,n}$ are arbitrary, their values specify the
particular cloning state.
In one standard scenario (see below), 
q-nit~0 will be measured and thus projected onto one of a set of chosen
states, q-nits~1 and~3 will be the clones, and q-nit~2 the anticlone
(or ``machine''). 

The expansion of the state ket (\ref{CERF}) in the biorthogonal double-Bell
basis, with only $N^2$ of the $N^4$ basis states appearing in (\ref{CERF}),
emphasizes a generic property of such cloning states, namely their
covariance when passing from one of the MUB to another. 
This covariance property, which we discussed at the end of 
Sec.~\ref{sec3.1}, is of considerable importance in various contexts, such as
cryptography protocols that treat all single--q-nit MUB on the same
footing\cite{tomocrypt,BB84,qutrits,bruss} and
phase-covariant cloning,\cite{Fuchs+4:97,Griffiths+1:97,NG} 
and also has a bearing on the Mean King's problem of Sec.~\ref{sec4.1}.

In the present context, we need yet another symmetry property, namely that the
two clones --- q-nits~1 and~3 --- play complementary roles.
To establish this point, we first recall the definition of the generalized
Bell states in (\ref{defBmn}) and note that 
\begin{equation}\label{Bell-01-23}
\ket{B^{(01)}_{m,n},B^{(23)}_{\ominus m,\ominus n}}
=\frac{1}{N}\sum_{k,l=0}^{N-1}  
\ket{k^*,k\oplus m,l^*,l\ominus m}\gamma ^{(k\oplus m)\odot n}
\gamma ^{\ominus(l\ominus m)\odot n}\,,
\end{equation}
where we now employ a notation that indicates which q-nits are paired in
the Bell states: 0 with~1, and 2 with~3, as it is the case in (\ref{CERF}).
Alternatively, we can pair 0 with~3 and 2 with~1, which gives 
\begin{equation}\label{Bell-03-21}
\ket{B^{(03)}_{m,n},B^{(21)}_{\ominus m,\ominus n}}
=\frac{1}{N}\sum_{k,l=0}^{N-1} 
\ket{{k}^*,l\ominus m,l^*,k\oplus m}\gamma ^{(k\oplus m)\odot n}
\gamma ^{\ominus(l\ominus m)\odot n}\,.
\end{equation}
In fact, the states of (\ref{Bell-01-23}) span the same $N^2$-dimensional
subspace as the states of (\ref{Bell-03-21}) in the $N^4$-dimensional
four--q-nit Hilbert space.  

To justify this remark, we evaluate the transition amplitudes, 
\begin{eqnarray}
\lefteqn{\braket{B^{(03)}_{m',n'},B^{(21)}_{\ominus m',\ominus n'}}%
{B^{(01)}_{m,n},B^{(23)}_{\ominus m,\ominus n}}}\rule{3em}{0pt}&&\nonumber\\
&=&\frac{1}{N^2}\sum_{k,k',l,l'=0}^{N-1} 
\gamma ^{(k\oplus m)\odot n\ominus(l\ominus m)\odot n
\ominus(k'\oplus m')\odot n'\oplus(l'\ominus m')\odot n'}\nonumber\\
&&\hphantom{\frac{1}{N^2}\sum_{k,k',l,l'=0}^{N-1}}\times
\braket{{k'}^*,l'\ominus m',{l'}^*,k'\oplus m'}
{k^*,k\oplus m,l^*,l\ominus m}
\nonumber\\
&=&\frac{1}{N^2}\sum_{k,k',l,l'=0}^{N-1} 
\gamma^{(k\ominus l\oplus m\oplus m')\odot n} 
\gamma^{\ominus(k'\ominus l'\oplus m'\oplus m)\odot n'}
\gamma^{(m\ominus m')\odot(n\oplus n')} \nonumber\\
&&\hphantom{\frac{1}{N^2}\sum_{k,k',l,l'=0}^{N-1}}\times
\delta_{k',k}\delta_{k'\oplus m',l\ominus m}
\delta_{l',l}\delta_{l'\ominus m',k\oplus m}\,,
\end{eqnarray}
where this product of four Kronecker delta symbols equals
$\delta_{k,k'}\delta_{l,l'}\delta_{m\oplus m',l\ominus k}$, a product of only
three, with the consequence that
\begin{equation}
  \braket{B^{(03)}_{m',n'},B^{(21)}_{\ominus m',\ominus n'}}%
{B^{(01)}_{m,n},B^{(23)}_{\ominus m,\ominus n}}=
\frac{1}{N}\gamma^{(m\ominus m')\odot(n\oplus n')}\,.
\end{equation}

For given $\ket{B^{(01)}_{m,n},B^{(23)}_{\ominus m,\ominus n}}$ these are $N^2$
transition amplitudes, each of modulus $N$, and therefore no other 
$B^{(03)}B^{(21)}$ kets can appear on the right-hand side of
\begin{equation}\label{eq3:Bell-01-23gen}
\ket{B^{(01)}_{m,n},B^{(23)}_{\ominus m,\ominus n}}
=\frac{1}{N}\sum_{m',n'=0}^{N-1}
\ket{B^{(03)}_{m',n'},B^{(21)}_{\ominus m',\ominus n'}}
\gamma^{(m\ominus m')\odot(n\oplus n')}\,.
\end{equation}
It follows that $\braket{B^{(03)}_{m',n'},B^{(21)}_{m'',n''}}%
{B^{(01)}_{m,n},B^{(23)}_{\ominus m,\ominus n}}=0$ unless both $m'\oplus m''=0$ and
${n'\oplus n''=0}$, which can be verified directly.
In particular, we have
\begin{equation} 
  \label{CERF5}
  \ket{B^{\ }_{0,0},B^{\ }_{0,0}}=\ket{B^{(01)}_{0,0},B^{(23)}_{0,0}}=
\frac{1}{N}\sum_{m,n=0}^{N-1}(\mathbf{1}\otimes V_m^n \otimes \mathbf{1}
\otimes {V_m^n}^\dagger)\ket{B^{(03)}_{0,0},B^{(21)}_{0,0}}\,,
\end{equation}
which we use in (\ref{CERF}) to arrive at the alternative expansion
\begin{equation}
  \label{CERF6}
\ket{\Psi_{0-3}}=\sum_{m,n=0}^{N-1}
(\mathbf{1}\otimes {V_m^n}^\dagger\otimes\mathbf{1}\otimes V_m^n)
\ket{B^{(03)}_{0,0},B^{(21)}_{0,0}}b_{m,n}\,,  
\end{equation}
where the probability amplitudes $b_{m,n}$ are the double Galois--Fourier
transforms of the $a_{m,n}$s,
\begin{equation}
  \label{CERF4}
  b_{m,n}=\frac{1}{N}\sum_{m',n'=0}^{N-1}\gamma^{m\odot n' \ominus n\odot m'} 
              a_{m',n'}\,.
\end{equation}

The stage is now set for a discussion of cloning.
We consider two standard scenarios.
In the first scenario, Alice and Bob believe that they share the Bell state
described by ket $\ket{B^{(01)}_{0,0}}$, but in fact eavesdropper Eve controls
the two--q-nit source and has replaced $\ket{B^{(01)}_{0,0}}$ by
$\ket{\Psi_{0-3}}$.
Alice measures her q-nit~0 and finds it in the state described by the bra
$\bra{\psi^*}$, so that the state of Bob's q-nit~1 would be described by ket
$\ket{\psi}$, but the ket for the resulting state of q-nits~1--3 is actually
given by 
\begin{eqnarray} \label{CERF1}
\ket{\Psi_{1-3}}
&=&\sum_{m,n=0}^{N-1}(V_m^n\otimes\mathbf{1}\otimes
{V_m^n}^\dagger) \ket{\psi,B^{(23)}_{0,0}}a_{m,n}\nonumber\\
&=&\sum_{m,n=0}^{N-1}({V_m^n}^\dagger\otimes\mathbf{1}\otimes V_m^n)
 \ket{B^{(21)}_{0,0},\psi}b_{m,n}\,. 
\end{eqnarray}
The resulting statistical operator for Bob's q-nit~1, the first clone, is 
\begin{equation} 
  \label{rho1} 
  \rho_1^{\ }=\tr[2\&3]{\ket{\Psi_{1-3}}\bra{\Psi_{1-3}}}=\sum_{m,n=0}^{N-1}
\ket{\psi_{m,n}}\bigl|a_{m,n}\bigr|^2\bra{\psi_{m,n}} 
\end{equation}
with $\ket{\psi_{m,n}}=V_m^n\ket{\psi}$, and for q-nit~3, the second clone,
we obtain
\begin{equation}
  \label{rho3}
  \rho_3^{\ }=\tr[1\&2]{\ket{\Psi_{1-3}}\bra{\Psi_{1-3}}}=\sum_{m,n=0}^{N-1}
\ket{\psi_{m,n}}\bigl|b_{m,n}\bigr|^2\bra{\psi_{m,n}}\,.
\end{equation}
The displacement operators $V_m^n$ appear as error operators in (\ref{rho1})
and (\ref{rho3}). 

There are two extreme complementary situations: 
If $a_{m,n}=\delta_{m,0}\delta_{n,0}$ and thus 
$\bigl|b_{m,n}\bigr|^2={1}/{N^2}$, then $\rho_1=\ket{\psi}\bra{\psi}$ is
the projector on the target state $\ket{\psi}$ and  
${\rho_3=\mathbf{1}/N}$ is the completely mixed state, as implied by
the ergodicity relation (\ref{ergodicity});
but if $b_{m,n}=\delta_{m,0}\delta_{n,0}$ and thus 
$\bigl|a_{m,n}\bigr|^2={1}/{N^2}$, we get $\rho_1=\mathbf{1}/N$
and $\rho_3=\ket{\psi}\bra{\psi}$.  
In intermediate situations, both $\rho_1$ and $\rho_3$ are imperfect copies of
$\ket{\psi}\bra{\psi}$.  

We see that, as a consequence of the Galois--Fourier relation (\ref{CERF4}),
the two clones are complementary to each other in the sense that if one of
them projects on the target state $\ket{\psi}$, then the other is completely
mixed.  
More generally, if one clone is in a pure state (not necessarily the target
state), then the other clone is in the completely mixed state. 

This complementarity is important because it helps us to understand the main
idea underlying quantum cryptography: 
If the first clone is received by Bob, to whom it appears as the target state
with an admixture of noise, and the second clone is Eve's
imperfect copy (she also has access to the anticlone), then the more Eve knows
about Alice's or Bob's signals, the less strongly their signals are
correlated.  
In other words, when the entanglement between two of the three parties becomes
stronger, the entanglement with the third party weakens, an idea that was
already central to the first entanglement-based protocol, the 1991 Ekert
protocol.\cite{Ekert} 
For obvious reasons, this property is sometimes referred to as the ``monogamy
of quantum entanglement.''  

The second scenario is that of BB84-type\cite{BB84} schemes for quantum
cryptography: Alice prepares q-nit~1 in the state described by ket
$\ket{\psi}$ and sends it to Bob.
Eve gets hold of the q-nit in transmission, combines it with her q-nits~2 and
3 that she had earlier prepared in the `00' Bell state, and realizes a unitary
transformation that effects
\begin{eqnarray}
  \label{eq3:cloning1}
  \ket{k,B^{(23)}_{0,0}}\longrightarrow&\!&\sum_{m,n=0}^{N-1}
(V_m^n\otimes\mathbf{1}\otimes{V_m^n}^\dagger) \ket{k,B^{(23)}_{0,0}}a_{m,n}
\nonumber\\=&\!&\sum_{m,n=0}^{N-1}\ket{k\oplus m,B^{(23)}_{\ominus m,\ominus n}}
\gamma^{k\odot n}a_{m,n}
\end{eqnarray}
for all kets $\ket{k}$ of q-nit~1, so that $\ket{\psi,B^{(23)}_{0,0}}$ is
turned into the ket of \eqref{CERF1},
\begin{equation}
  \label{eq3:cloning2}
  \ket{\psi,B^{(23)}_{0,0}}\longrightarrow\ket{\Psi_{1-3}}\,.
\end{equation}
Then q-nit~1, the first clone, is forwarded to Bob and Eve keeps the second
clone and the anticlone.

The unitary property of the map \eqref{eq3:cloning1} is confirmed by
\begin{eqnarray}
  \label{eq3:cloning3}
  \delta_{k,k'}&=&\braket{k,B^{(23)}_{0,0}}{k',B^{(23)}_{0,0}}\nonumber\\
&\makebox[1em][r]{$\longrightarrow$}&\sum_{m,n}^{N-1}\sum_{m',n'}^{N-1}
{a_{m,n}}^*\gamma^{\ominus k\odot n}\delta_{k\oplus m,k'\oplus m'}
\delta_{m,m'}\delta_{n,n'}\gamma^{k'\odot n'}a_{m',n'}
\nonumber\\&=&\delta_{k,k'}\,.
\end{eqnarray}
Accordingly, Eve can --- in principle, at least, if not in practice ---
implement \eqref{eq3:cloning1} by a suitable interaction between q-nits~1
and~3.

We further note that the Heisenberg--Weyl group is not only related to the
error operators that describe the imperfections of the clones, it is also
directly related to error correcting 
codes.\cite{Gottesman-1,errorcorr,nielsen,Gottesman-2} 
For instance, the Shor code for q-bits (see, e.g., Ref.~\refcite{nielsen})
exploits the fact that the Pauli $\sigma$ operators are an operator basis in
the q-bit space.  
Higher-dimensional generalizations of this code likewise exploit that the
Heisenberg--Weyl operators, essentially the shift operators of (\ref{defV0}),
constitute an operator basis, especially in the many--q-bit case ($N=2^\m$).

\subsection{Entanglement swapping}\label{sec3.5}
A system of four q-nits, prepared in the state described by one of the kets
$\ket{B^{(01)}_{m,n},B^{(23)}_{\ominus m,\ominus n}}$ of
(\ref{eq3:Bell-01-23gen}), has the q-nit pairs $(01)$ and $(23)$ in maximally
entangled states while there is no entanglement between the two pairs. 
If one then performs a Bell basis measurement on the pair $(12)$ and finds
it in the Bell state $\ket{B^{(21)}_{\ominus m',\ominus n'}}$, the state of
the pair $(03)$ is reduced to the Bell state $\ket{B^{(03)}_{m',n'}}$.
In a manner of speaking, half of the original entanglement between the pairs
$(01)$ and $(23)$ is used up in the Bell measurement on the pair $(12)$ and the
other half is transferred to the pair $(03)$ which emerges maximally
entangled.  

At the time when the pair $(12)$ is measured, q-nits~0 and 3 can be far away,
possibly at space-like separations from each other and from pair $(12)$, 
and q-nits~0 and 3 may never have been close to each other in the past.
What matters is that their partners, q-nits~1 and 2, with which they share the
maximally entangled initial Bell states, are brought into contact during the
Bell-basis measurement on the pair $(12)$.
As soon as the outcome of the measurement on pair $(12)$ is communicated
(through a classical channel) to the experimenters in possession of q-nits~0 
and 3, they can exploit the entanglement in the resulting Bell state
$\ket{B^{(03)}_{m',n'}}$. 

This \emph{entanglement swapping}\cite{Zukowski+3:93} has been demonstrated
for q-bits carried by photons in different 
experiments; see Refs.~\refcite{Pan+3:98,Halder+5:07}, for example.
In conjunction with quantum repeaters, entanglement swapping offers a
practical way of creating strong entanglement between q-nits that are far
apart.\cite{Briegel+3:98}

         %% Section 3
%%%% file name: MUB-4.tex
%%%% input file for MUB.tex 
%%%%
%%%% last changes on 20 April 2010 by Berge
%%%% typo corrected on 27 April 2010
%%%%
%%%%%%%%%%%%%%%%%%%%%%%%%%%%%%%%%%%%%%%%%%%%%%

\section{The Mean King's problem and quantum state tomography}
\label{section4}
\subsection{The Mean King's problem in prime power dimensions} 
\label{sec4.1} 
The ``Mean King's Problem'' originated in the 1987 paper by Vaidman, Aharonov,
and Albert,\cite{vaid} which deals with the $N=2$ case.
Generalizations first to $N=3$,\cite{MK3} then to $N$ prime,\cite{Englert} and
finally to prime-power values of $N$,\cite{2003,Durtmean} were completed some
15 years later. 
For further generalizations see 
Refs.~\refcite{Klappenecker+1:05} and~\refcite{Reimpell+1:07}.   
In the simplest case ($N=2$), the problem can be presented as in
Ref.~\refcite{Englert}:  
\begin{quote}
The Mean King challenges a physicist, Alice, who got stranded on the remote
island ruled by the king, to prepare a spin-$\frac{1}{2}$ atom in any state of
her choosing and to perform a control measurement of her liking. 
Between her preparation and her measurement, the king's men determine the
value of either  $\sigma_{x}$, $\sigma_{y}$, or $\sigma_{z}$. 
Only after she completed the control measurement, the physicist is told which
spin component has been measured, and she must then state the result of that
intermediate measurement correctly. 
\end{quote} 
In dimension $N$, where $N$ is a prime power, the challenge can be
summarized in this way:  
Alice prepares a q-nit system in any state of her choosing and  performs a
control measurement of her liking. 
Between her preparation and her measurement, the king's men measure the q-nit
in one of the $N+1$ MUB. 
The particular basis chosen for the intermediate measurement is communicated to
Alice only after she has completed the control measurement, and she must then
state the result of that intermediate measurement correctly.

The power of entanglement enables Alice to rise to this challenge.
Her solution consists of four stages:
\begin{enumerate}\renewcommand{\theenumi}{\roman{enumi}}
\addtolength{\labelsep}{-2.5pt}
\item 
She prepares q-nit~1, which will be handed to the king's men, jointly with
\mbox{q-nit}~0, which she will keep for herself, in the Bell state
$\ket{B_{0,0}}$ of (\ref{mapId}).
\item
The king's men measure q-nit~1 in the $i$th basis of the MUB and find it in
the $k$th state, whereafter the state ket of the q-nit pair is
$\ket{e^{i*}_k,e^i_k}$; there is a total of $N(N+1)$ states of this kind.
\item
Alice measures the q-nit pair in the entangled basis composed of the
$N^2$ pairwise orthogonal states $\ket{(m,n)}$ that are given by
\begin{eqnarray}
  \label{eq4:MKbasis}
\ket{(m,n)}&=&(V_m^{n*}\otimes V_m^n)\ket{(0,0)} 
\quad\mbox{for $m,n=0,1,\ldots, N-1$}
\nonumber\\ 
\mbox{with the ``seed''}\quad \ket{(0,0)}&=&
\frac{1}{\sqrt{N}}\sum_{i=0}^N\ket{e_0^{i*},e_0^i}-\ket{B_{0,0}}\,.
\end{eqnarray}
Alice's measurement outcome is an ordered pair of field elements $(m,n)$.
\item 
Now, being told that the $i$th basis was measured at the intermediate stage
(ii), and having her outcome $(m,n)$ of the control measurement of stage
(iii) at hand, Alice \emph{correctly} infers that the king's men found q-nit~1
in state 
$\ket{e^i_k}$ with 
\begin{equation}
  \label{eq4:MK-infer}
  k=\left\{\begin{array}{c@{\mbox{\ for\ }}l}
           i\odot m\ominus n & i=0,1,\ldots,N-1\,,\\ 
                           m & i=N\,.
           \end{array}\right.
\end{equation}
\end{enumerate}

As shown in Ref.~\refcite{Durtmean}, this solution is a special case of
Aravind's very general solution,\cite{2003} which is formulated without
a particular choice for the maximal set of MUB;
our solution exploits the specific MUB of Secs.~\ref{sec2.1}--\ref{sec2.3}.
For $N=2$, $3$, $4$, and $5$, all maximal sets of MUB are
equivalent,\cite{Kostrikin,BeBrWe} in the sense that they can be turned into
each other by unitary transformations combined with permutations of the basis
kets; more about this in Sec.~\ref{section5}.
A finer notion of equivalence, which takes entanglement properties into
account, is possible in composite dimensions.
In this finer sense there are inequivalent MUB for ${N=8}$ and 
${N=16}$.\cite{RBKS05,BRKS07,Zeil}. 

The explanation how Alice's scheme works begins with 
first noting the explicit form of the two--q-nit states $\ket{(m,n)}$ 
of Alice's measurement basis,
\begin{equation}
  \label{eq4:MKstates}
  \ket{(m,n)}=\frac{1}{\sqrt{N}}\left(\ket{e^{N*}_m,e^N_m}
     +\sum_{i=0}^{N-1}\ket{e_{i\odot m\ominus n}^{i*},e_{i\odot m\ominus n}^i}\right)
       -\ket{B_{0,0}}\,,
\end{equation}
where the sum over $i$ does not include the computational basis ($i=N$), as it
does for the seed in (\ref{eq4:MKbasis}).
With the aid of (\ref{postul2}), the invariance property (\ref{eq3:MKinvar}),  
and $\braket{e^{i*}_k,e^i_k}{B_{0,0}}=N^{-1/2}$, we then establish
\begin{equation}
  \label{eq4:MKpyra}
  \braket{B_{0,0}}{(m,n)}=\frac{1}{N}
\end{equation}
and
\begin{equation}
  \label{eq4:MKtransition}
  \braket{e^{i*}_k,e^i_k}{(m,n)}=\left\{
    \begin{array}{c@{\quad\mbox{for}\ }l}
        \delta_{k,i\odot m\ominus n}/\sqrt{N} & i=0,1,\dots,N-1\,,\\[1ex]
        \delta_{k,m}/\sqrt{N} & i=N\,,      
    \end{array}\right.
\end{equation}
from which follows the orthonormality 
\begin{equation}\label{eq4:MKortho}
\braket{(m,n)}{(m',n')}=\delta_{m,m'}\,\delta_{n,n'}\,,  
\end{equation}
thus confirming that the kets $\ket{(m,n)}$ constitute an orthonormal basis in
the $N^2$-dimensional space of two--q-nit kets.

Now, after the king's men find q-nit~1 in the $k$th state of the $i$th basis,
the q-nit pair is in the state described by the bra $\bra{e^{i*}_k,e^i_k}$.
Clearly then, the Kronecker delta symbols in (\ref{eq4:MKtransition})
enable Alice to infer the $k$ value in accordance with (\ref{eq4:MK-infer}).
For, only a single $k$ value is possible for the actual outcome $(m,n)$ of
Alice's control measurement and the $i$th basis chosen by the king's men. 

\begin{figure}[tb]
\renewcommand{\MK}[2]{\scaleput(#1){\makebox(0,0)[c]{#2}}}
\centerline{\setlength{\unitlength}{0.8pt}%
\begin{picture}(220,210)(-110,-100)
\put(0,80){\mnGrid{0}} 
\put(76.09,24.72){\mnGrid{1}} 
\put(47.02,-64.72){\mnGrid{2}}
\put(-47.02,-64.72){\mnGrid{3}}
\put(-76.09,24.72){\mnGrid{4}} 
\end{picture}}
\caption{\label{fig:MKgrids}%
The Mean King's Problem for $N=4$. 
The five $4\times4$ grids show the $k$ values for $i=0,\dots,4$ clockwise,
with $i=0$ at the top. 
In each $(m,n)$ grid, the columns are labeled by $m$ from left to right,
and the rows are labeled by $n$ from bottom to top.
For example, we have $k=2$ for $(m,n)=(2,1)$ in the grid for $i=2$.} 
\end{figure}
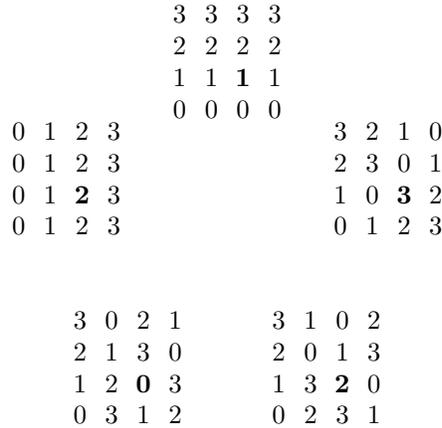

It is important that Alice can always infer the correct $k$ value with
certainty. 
This aspect can be understood, or illustrated, by a geometrical picture, in
the sense of affine geometry (more about this in Sec.~\ref{secaffin}).
When the king's men find the $k$th state of the $i$th basis (where
$i$ runs from 0 to $N$, and $k$ from $0$ to ${N-1}$) 
$N$ of the $N^2$ detectors fire with equal probability in Alice's control
measurement, namely the detectors whose $(m,n)$ values appear in
\begin{equation}
  \label{eq4:MKmarginals}
  \ket{e^{i*}_k,e^i_k}=\left\{
    \begin{array}{c@{\quad\mbox{for}\ }l}
     \ds\frac{1}{\sqrt{N}}\sum_{m=0}^{N-1}\ket{(m,i\odot m\ominus k)}
          & i=0,1,\dots,N-1\,,\\[3ex]
     \ds\frac{1}{\sqrt{N}}\sum_{n=0}^{N-1}\ket{(k,n)}& i=N\,.
    \end{array}\right.
\end{equation}
Accordingly, in the $N\times N$ discrete plane (grid) spanned by the pairs
$(m,n)$ the labels of these detectors are on the straight lines 
$m\mapsto n=i\odot m\ominus k$ with slope
$i$ when ${i=0,1,\ldots,N-1}$, and on the ``vertical'' lines $m=k$ when $i=N$. 
Figure~\ref{fig:MKgrids} shows the five grids for $N=4$ as they result from
the multiplication and addition tables in Table~\ref{tbl:4-field}(a).

In Aravind's construction,\cite{2003} the combinatorial properties offered by
an affine plane of order $N$ (properly defined in Sec.~\ref{secaffin} below) 
are a crucial ingredient.
This is also true in this geometrical picture: 
Because the addition $\oplus$ and multiplication $\odot$ form a field, 
exactly one straight line of given slope passes through each point of the
grid, which is a sine qua non condition for unambiguously inferring which 
detector fired during the king's men's measurement. 

In Alice's measurement bases (\ref{eq4:MKstates}), the
$N(N+1)$ two--q-nit states $\ket{e^{i*}_k,e^i_k}$ are grouped into $N^2$ sets
of $N+1$ states, each state appearing in $N$ sets, and each set composed of
one state from each of the $N+1$ MUB.
The states of the set associated with a measurement outcome $(m,n)$ correspond
to the respective $N+1$ grid points; such as the highlighted grid points for
$(m,n)=(2,1)$ in Fig.~\ref{fig:MKgrids}.

The normalized superposition states of the $N^2$ sets that appear in
(\ref{eq4:MKstates}), 
\begin{equation}\label{eq4:MKsuper}
    \frac{1}{\sqrt{2N+2}}\left(\ket{e^{N*}_m,e^N_m}
   +\sum_{i=0}^{N-1}\ket{e_{i\odot m\ominus n}^{i*},e_{i\odot m\ominus n}^i}\right)
    =\sqrt{\frac{N}{2N+2}}\Bigl(\ket{(m,n)}+\ket{B_{0,0}}\Bigr)\,,  
\end{equation}
are linearly independent, but they are not pairwise orthogonal.
Rather they are the edges of an acute $N^2$-dimensional pyramid, with angle
$\arccos{\frac{N+2}{2N+2}}$ between each pair of edges, and the invariant Bell
state $\ket{B_{0,0}}$ as the symmetry axis of the pyramid.
Alice's measurement is the so-called ``square-root measurement'' for this
pyramid, the natural von Neumann measurement associated with the 
pyramid.\cite{pyramids1,pyramids2}

\subsection{State tomography with discrete Weyl and Wigner 
phase-space functions}\label{sec4.2}
Owing to the correspondence (\ref{eq:correspond}), 
the expansion of any operator in a one--q-nit operator basis, 
which is at the heart of quantum tomography, is related to the
expansion of a two--q-nit state ket in the corresponding ket basis.
In the general situation, we have a positive-operator-valued measure
(POVM)%
\footnote{POVM, with its emphasis on ``measure'' and the connotations of
  measure theory, is mathematical terminology.
  The corresponding quantum-physics term POM (probability operator
  measurement) refers to the physical significance.}\ 
for the two--q-nit states,
\begin{equation}
  \label{eq4:POVM}
  \sum_k\ket{a_k}\bra{a_k}=\mathbf{1}\,,
\end{equation}
a sum of $N^2$ or more hermitian, rank-1, two--q-nit operators.
In accordance with the mapping of (\ref{eq:correspond})--(\ref{map2nd'}),
there is a single--q-nit operator $A_k$ for each ket $\ket{a_k}$,
\begin{equation}
  \label{eq4:POVMmap}
  \ket{a_k}\longleftrightarrow A_k\,,
\end{equation}
and, in view of the trace rule (\ref{eq3:InnerProd}), the expansion
\begin{equation}
  \label{eq4:POVMexpand}
  \ket{x}=\sum_k\ket{a_k}\braket{a_k}{x}
\end{equation}
of a generic ket $\ket{x}$ then implies the corresponding expansion for the
operator $X$ associated with $\ket{x}$,
\begin{equation}
  \label{eq4:complete}
  \ket{x}\longleftrightarrow X=\sum_k A_k^{\ }\,\tr{A_k^\dagger X}\,,
\end{equation}
which is valid for any single--q-nit operator $X$.
This identity is the completeness relation for the operator basis composed of
the $A_k$s.

In the more particular case of an orthonormal basis of $N^2$ kets (and its
adjoint basis of bras), $\braket{a_j}{a_k}=\delta_{j,k}$, the POVM in
(\ref{eq4:POVM}) refers to an ideal von Neumann measurement, and we have the
corresponding orthonormality statement for the operator basis:
$\tr{A^\dagger_jA_k^{\ }}=\delta_{j,k}$. 
This is the situation for the two specific two--q-nit bases that we
encountered in Secs.~\ref{sec3.1} and \ref{sec4.1}, respectively: the basis
made up by the generalized Bell states $\ket{B_{m,n}}$ of (\ref{defBmn}), 
and the basis composed of Alice's ``mean king states'' $\ket{(m,n)}$ of
(\ref{eq4:MKstates}). 
The operator basis corresponding to the ket basis of Bell states is the Galois
field version of Weyl's unitary operator basis\cite{Weyl1,Weyl2} of
Sec.~\ref{sec:Weyl-Schwinger}, and the operator
basis associated with the ket basis of mean-king states is a candidate for a
discrete analog of the familiar hermitian Wigner basis for a continuous degree
of freedom.\cite{Wigner:32,Hillery+3:84}

\subsubsection{Discrete Weyl-type unitary operator basis and phase-space
  function}\label{sec4.2.1} 

When we identify the Bell kets $\ket{B_{m,n}}$ with the basis kets $\ket{a_k}$
in (\ref{eq4:POVMmap}), the mapping (\ref{defBmn}) tells us that $N^{-1/2}V_m^n$
corresponds to $A_k$, and the completeness relation (\ref{eq4:complete})
acquires the form
\begin{equation}
  \label{eq4:Weyl-complete}
  X=\frac{1}{N}\sum_{m,n=0}^{N-1}V_m^n \,x_m^n
\qquad\mbox{with}\quad x_m^n=\tr{{V_m^n}^\dagger\,X}\,. 
\end{equation}
The unitary shift operators $V_m^n$ compose the operator basis, and the
coefficients $x_m^n$ make up the discrete phase-space function 
$(m,n)\mapsto x_m^n$ of Weyl-type.
The mapping of the operator $X$ to its Weyl-type phase-space function is
one-to-one: 
There is a unique single--q-nit operator $X$ to the given set of 
coefficients $\left\{x_m^n\right\}_{m,n=0}^{N-1}$, and all $x_m^n$s are
uniquely specified by the given operator $X$.
In particular, we have
\begin{equation}
  \label{eq4:Weyl-trace}
  x_0^0=\tr{X}\,.
\end{equation}

Since the unitary operators $U^i_l$ of the abelian subgroups of
Sec.~\ref{sec2.3} comprise all the shift operators $V_m^n$, with the identity
$\mathbf{1}=V_0^0=U^i_0$ appearing $N+1$ times, once for each subgroup
($i=0,1,\dots,N$), an alternative way of presenting (\ref{eq4:Weyl-complete}) 
is
\begin{equation}
  \label{eq4:Weyl-complete'}
  X=\frac{\mathbf{1}}{N}\tr{X}
    +\frac{1}{N}\sum_{i=0}^N\sum_{l=1}^{N-1}U^i_l\bar{x}^i_l
\qquad\mbox{with}\quad \bar{x}^i_l=\tr{{U_l^i}^\dagger\,X}\,. 
\end{equation}
The coefficients in (\ref{eq4:Weyl-complete}) and (\ref{eq4:Weyl-complete'})
are related to each other by
\begin{equation}
  \label{eq4:Weyl-complete''}
  \bar{x}^i_l=\left\{
    \begin{array}{c@{\ \mbox{for}\ }l}
      {\alpha^i_l}^*\,x_l^{i\odot l} & i=0,1,\ldots,N-1\,,\\[1ex]
      x_0^l & i=N\,,
    \end{array}\right.
\end{equation}
which is an immediate consequence of (\ref{eq:defU01l}) and (\ref{eq:defUil}).
The two expansions (\ref{eq4:Weyl-complete}) and (\ref{eq4:Weyl-complete'})
are really the same expansion twice, differing solely by the labeling of the
terms. 

Weyl tomography, on many identically prepared q-nits with statistical operator
$\rho$, amounts to measuring equal fractions of the q-nits in the
$N+1$ MUB of Secs.~\ref{sec2.1}--\ref{sec2.3}.
The measurements provide the probabilities $\bra{e^i_k}\rho\ket{e^i_k}$,%
\footnote{This is an idealization of the real physical situation.
   Any actual experiment will give the relative frequencies from which the
   probabilities can be estimated.
   The subtleties of \emph{quantum state estimation} are the subject matter of
   Ref.~\refcite{QuStateEstimation}.} 
from which the expansion coefficients
\begin{equation}
  \label{eq4:Weyl-rho}
   \bar{r}^i_l=\tr{{U_l^i}^\dagger\rho}
   =\sum_{k=0}^{N-1}\gamma^{\ominus k\odot l} \bra{e^i_k}\rho\ket{e^i_k}
\end{equation}
can then be computed, as follows from (\ref{postul}).
With $X\to\rho$, $\tr{X}\to1$, $\bar{x}^i_l\to \bar{r}^i_l$ in
(\ref{eq4:Weyl-complete''}), the statistical operator $\rho$ is
parameterized in terms of the unitary Weyl basis $U^i_l$ and the measured
coefficients $\bar{r}^i_l$. 

There are $N$ measurement outcomes for each of the $N+1$ MUB, so that one is
measuring a total of $N(N+1)$ probabilities (or relative frequencies) in order
to determine the $N^2-1$ parameters of the statistical operator.
Clearly, there is some redundancy in the data, namely that $\bar{r}^i_0=1$
for all $N+1$ values of $i$.
Nevertheless, the measurement of the $N+1$ MUB realizes state tomography that
is optimal in the sense of Ref.~\refcite{Wootters}: Other choices of $N+1$
von Neumann measurements, not composed of bases that are pairwise MU, 
give estimates for the statistical operator with larger statistical
errors when measuring finite samples, as is always the situation in practice.

Yet, when we regard the measurements of the $N+1$ bases, on equal fractions
of the q-nits, as jointly defining a POVM with $N(N+1)$ outcomes, then
these are more outcomes than are really needed to determine $N^2-1$ parameters.
More economical, and thus optimal in a different sense, are POVMs
with the minimal number of $N^2$ outcomes (the one constraint of
unit total probability is always there). 
And among those, a particularly good choice is the ``symmetric
informationally complete'' (SIC) POVM.\cite{Renes+al}
This is a different story, however, which does not need the structure of an
underlying Galois field, a ring structure suffices; see Refs.~\refcite{Grassl}
and \refcite{Appleby} for further information.  
The recent comprehensive account by Scott and Grassl\cite{Scott+1:09} is
recommended reading.

\subsubsection{The limit $N\to\infty$ of continuous degrees of freedom}
\label{sec4.2.2}
At the end of Sec.~\ref{sec:dual} --- recall (\ref{eq2:comp-shift1}) and
(\ref{eq2:comp-shift2}) --- we observed 
that the unitary shift operators $V_m^n=V_0^nV_m^0$ are products of $\m$
factors, one for each constituent q-pit,%
\footnote{As in (\ref{eq2:dual-param}), read the product $\underline{n}_jg_j$
  as the number $\underline{n}_j$ multiplying the row of $p$-ary coefficients
  for $g_j$, so that the outcome is the field element 
  $\underline{n}_j\odot g_j$. A similar remark applies to the product
  $m_jp^j$, except that in this case there is no difference between the number
  product of $m_j$ and $p^j$ and the field product.}
\begin{equation}
  \label{eq4:Weyl-product}
  V_m^n=\prod_{j=0}^{\m-1}  \Bigl(V_0^{g_j}\Bigr)^{\underline{n}_j}
                          \Bigl(V^0_{p^j}\Bigr)^{m_j}
       =\prod_{j=0}^{\m-1}V_{m_jp^j}^{\underline{n}_jg_j}\,,
\end{equation}
where the $m_j$s are the $p$-ary coefficients of $m$ as in (\ref{def-coeff}),
and the $\underline{n}_j$s are the conjugate coefficients of $n$ in the sense
of (\ref{eq2:dual-param}).   
There are $p^2$ unitary shift operators $V_{m_jp^j}^{\underline{n}_jg_j}$ for
each $j$ value, and those referring to different $j$ values commute with each
other. 
Accordingly, the factorization (\ref{eq4:Weyl-product}) is a decomposition of
$V_m^n$ into the Weyl operator bases of the individual $\m$ q-pits that make
up the q-nit. 

It is, therefore, systematic to regard the q-nit as a system of $\m$ q-pit
degrees of freedom, rather than a single q-nit degree of freedom.
The limit $N\to\infty$ is then understood as $p\to\infty$ for the given value
of $\m$, so that we obtain $\m$ continuous degrees of freedom or, put
differently, a $\m$-dimensional continuous system.

In view of the factorization observed above, the limit $p\to\infty$ is carried
out for each of the $\m$ q-pits individually. 
The details, and the subtleties, of this $p\to\infty$ limit are discussed in
Sec.~\ref{sec:WSlim}.

\subsubsection{Discrete Wigner-type hermitian operator basis and phase-space
  function}\label{sec4.2.3} 

When we identify the two--q-nit kets $\ket{(m,n)}$ of Alice's mean-king basis
in (\ref{eq4:MKstates}) with the basis kets $\ket{a_k}$ of (\ref{eq4:POVM}),
the corresponding single--q-nit operator basis is composed of the operators
$W_{m,n}$ that we get from the correspondence
(\ref{eq:correspond}),\cite{Durtmean}  
\begin{equation}
  \label{eq4:Wigner-1}
  \ket{(m,n)}\sqrt{N}\leftrightarrow W_{m,n}^{\ }=\ket{e^{N}_m}\bra{e^N_m} 
          +\sum_{i=0}^{N-1}\ket{e_{i\odot m\ominus n}^{i}}
          \bra{e_{i\odot m\ominus n}^i}-\mathbf{1}\,,
\end{equation}
with a conventional removal of the factor $1/\sqrt{N}$ from the definition of
the $W_{m,n}$s.
These operators are hermitian, normalized to unit trace, and pairwise
orthogonal, 
\begin{equation}
  \label{eq4:Wigner-2}
  W_{m,n}^\dagger=W_{m,n}^{\ }\,,\quad \tr{ W_{m,n}^{\ }}=1\,,\quad
  \tr{ W_{m,n}^{\ } W_{m',n'}^{\ }}=N\delta_{m,m'}\delta_{n,n'}\,,
\end{equation}
and their completeness relation is stated by
\begin{equation}
  \label{eq4:Wigner-3}
  \rho=\frac{1}{N}\sum_{m,n=0}^{N-1}r_{m,n}W_{m,n}   \qquad\mbox{with}\quad
  r_{m,n}=\tr{\rho W_{m,n}}
\end{equation}
for the statistical operator $\rho$, but is equally valid for any
single--q-nit operator $X$.
The coefficients $r_{m,n}$ are the discrete analog of the familiar Wigner
phase-space function for a continuous degree of freedom.

Wigner functions for finite-dimensional systems have been defined in several
different ways.\cite{Leonhardt97,Vourdas2,chatu05}
Here we choose to follow Wootters and his
collaborators,\cite{Wootters87,Wootters2,discretewigner}
who regard an operator basis as an acceptable discrete analog 
of the continuous basis underlying Wigner's phase space
function\cite{Wigner:32,Hillery+3:84} if it meets five criteria: 
\begin{equation}
  \label{eq4:Wigner-4}
  \parbox{0.78\columnwidth}{
    \begin{tabular}{@{}l@{\ }p{0.7\columnwidth}@{}}
(W1)& each basis operator is hermitian;\\
(W2)& each basis operator has unit trace;\\ 
(W3)& the basis operators are pairwise orthogonal;\\
(W4)& the basis as a whole, that is: the set of $N^2$ basis operators, is
invariant under the unitary tranformations of the $N^2$ Weyl operators;\\
(W5)& the marginals of the operator basis are rank-1 projectors, whereby
the $N$ projectors associated with parallel lines are mutually orthogonal and
thus compose a basis for the kets and bras, with MUB for
different sets of parallel lines.
    \end{tabular}}
\end{equation}
The notions of ``marginals'' and ``parallel lines'' will be explained shortly.
To the five criteria of (\ref{eq4:Wigner-4}) we add a sixth criterion:
\begin{equation}
  \label{eq4:Wigner-4'}
  \parbox{0.78\columnwidth}{
    \begin{tabular}{@{}l@{\ }p{0.7\columnwidth}@{}}
(W6)& in the limit $N\to\infty$ the sequence of discrete bases converges to
the standard continuous Wigner basis.\\
    \end{tabular}}
\end{equation}
It seems to us that (W6) is necessary to justify the term ``discrete
Wigner-type basis.''

Criteria (W1)--(W3) are the three statements in (\ref{eq4:Wigner-2}), and
criterion (W4) is an immediate consequence of (\ref{eq2:shiftinbasis}) ,
that is:
\begin{equation}
  \label{eq4:Wigner-5}
  W_{m,n}=V^n_m\,W_{0,0}\,{V^n_m}^\dagger
 =V^{n\ominus n'}_{m\ominus m'}\,W_{m',n'}\,{V^{n\ominus n'}_{m\ominus m'}}^\dagger
\end{equation}
for all $m,n$ and all $m',n'$. 
Just like $\ket{(0,0)}$ is the seed for the ket basis (\ref{eq4:MKbasis}),
$W_{0,0}$ is the seed of the operator basis (\ref{eq4:Wigner-1}).  

Regarding criterion (W5), we first note that a \emph{marginal operator}, or
simply: marginal, of the basis is the equal-weight average of all basis 
operators on an affine straight line. 
We specify a particular straight line by requiring that all $m,n$ values on 
the line obey ${a\odot m=b\odot n\oplus c}$ where ${a,b,c}$ is any given 
trio of field elements, excluding solely the choice of ${a=b=0}$. 
Clearly, the trio ${a\odot d,b\odot d,c\odot d}$ with ${d\neq0}$ specifies the
same line, and the lines for ${a_1,b_1,c_1}$ and ${a_2,b_2,c_2}$ are parallel
if ${a_1\odot b_2=a_2\odot b_1}$, whereas they intersect in one $m,n$ point if 
${a_1\odot b_2\neq a_2\odot b_1}$.

Accordingly, the marginal operators are 
\begin{equation}
  \label{eq4:Wigner-6}
  M_{a,b,c}=\frac{1}{N}\sum_{m,n=0}^{N-1}W_{m,n}\delta_{a\odot m,b\odot n\oplus c}
    =\left\{
    \begin{array}{c@{\quad\mbox{if}\quad}l}
      \ket{e^{a\oslash b}_{c\oslash b}}\bra{e^{a\oslash b}_{c\oslash b}}
      & b\neq0\,,\\[1ex]
      \ket{e^N_{c\oslash a}}\bra{e^N_{c\oslash a}}  
      & b=0\mbox{\ and\ }a\neq0\,,
    \end{array}\right.
\end{equation}
and the case ${a=b=0}$, for which 
${M_{0,0,c}=\delta_{c,0}\mathbf{1}}$, illustrates an ergodic property of the
Wigner basis,  
\begin{equation}
  \label{eq4:Wigner-7}
  \frac{1}{N}\sum_{m,n=0}^{N-1}W_{m,n}=\mathbf{1}\,.
\end{equation}
Another way of stating the explicit projector values of the marginals is
\begin{equation}
  \label{eq4:Wigner-8}
  \ket{e^i_k}\bra{e^i_k}=\left\{
    \begin{array}{l@{\quad\mbox{for}\quad}l}
  \ds M_{i,1,k}=\frac{1}{N}\sum_{m=0}^{N-1}W_{m,i\odot m\ominus k} 
        & i=0,1,2,\ldots,N-1\,,\\[3ex] 
  \ds M_{1,0,k}=\frac{1}{N}\sum_{n=0}^{N-1}W_{k,n} & i=N\,,
    \end{array}\right.
\end{equation}
which we recognize as the single--q-nit operator version of the two--q-nit
identities in (\ref{eq4:MKmarginals}). 
Indeed, the projectors for the $N$ parallel lines with slope
${a\oslash b=i}$ make up the $i$th basis for ${i=0,1,\ldots,N-1}$,  
while the computational basis (${i=N}$) is obtained for the ``vertical'' 
lines with $b=0$. 
These are, of course, the sets of parallel lines that we encountered in
Sec.~\ref{sec4.1}, as illustrated in Fig.~\ref{fig:MKgrids}.   
One could say that the relations (\ref{eq4:Wigner-1}) and
(\ref{eq4:Wigner-6}) are reciprocals of each other: 
The projectors
$\ket{e^i_k}\bra{e^i_k}$ are marginals of the basis operators $W_{m,n}$, and
the $W_{m,n}$s are marginals of the projectors (up to a subtraction of the
identity operator). 

The reciprocity of the relations (\ref{eq4:Wigner-1}) and
(\ref{eq4:Wigner-6}) is even more striking if, following the geometrical
approach emphasized in Sec.~\ref{section0}, we define the vectors of
$\mathbf{R}^{N^2-1}$ that are naturally associated with the Wigner operators
$W_{m,n}$ and the projectors $\ket{e^i_k}\bra{e^i_k}$,
\begin{eqnarray} 
\mathbf{w}_{m,n} = \mathcal{W}_{m,n}-\varrho_\star&\repr& W_{m,n}-\rho_\star \,,
\nonumber\\
\mathbf{p}^i_k = \psi^i_k{\psi^i_k}^\dagger-\varrho_\star
                   &\repr&\ket{e^i_k}\bra{e^i_k} - \rho_\star \,, 
\end{eqnarray}
where the matrix $\mathcal{W}_{m,n}$ represents $W_{m,n}$, and $\psi^i_k$ is
the column for $\ket{e^i_k}$.
It clearly results from the ergodicity condition (\ref{eq4:Wigner-7}) that
the $\mathbf{w}_{m,n}$s obey 
\begin{equation}  \label{nolla} 
\sum_{m,n=0}^{N-1}\mathbf{w}_{m,n}= 0 \,.
\end{equation}
The $\mathbf{w}_{m,n}$s are thus the vertices of a regular simplex in 
$\mathbf{R}^{N^2-1}$, and this is how we want to think about them now. 
We refer to the $\mathbf{w}_{m,n}$s as the face points. 

Equations (\ref{eq4:Wigner-1}) and (\ref{eq4:Wigner-8}) now appear as
\begin{equation}\label{embed2}
  \mathbf{w}_{m,n}=\mathbf{p}^N_m
          +\sum_{i=0}^{N-1}\mathbf{p}_{i\odot m\ominus n}^{i}\,,
\end{equation}
and
\begin{eqnarray}\label{embed1}
  \mathbf{p}^i_k&=& \mathcal{M}_{i,1,k}
                =\frac{1}{N}\sum_{m=0}^{N-1}\mathbf{w}_{m,i\odot m\ominus k} 
        \quad\mbox{for $i=0,1,2,\dots,N-1$}\,,
\nonumber\\  
  \mathbf{p}^N_k&=& \mathcal{M}_{1,0,k}
                =\frac{1}{N}\sum_{n=0}^{N-1}\mathbf{w}_{k,n} \,,
\end{eqnarray}
where matrix $\mathcal{M}_{a,b,c}$ represents $M_{a,b,c}$ of
(\ref{eq4:Wigner-6}). 

There is a natural interpretation of (\ref{embed1}) in $\mathbf{R}^{N^2-1}$:
It says that the vertices of the MUB polytope lie at the centers of certain
specially selected faces of the face point operator simplex. 
The former has been inscribed into the latter in a special way. 
Alternatively, (\ref{embed2}) says that the vertices of this simplex lie right
above the centers of certain special faces of the MUB polytope. 
These faces are orthocomplemented to the facets 
(the highest dimensional faces). 
To see this, note that $\tr{W_{m,n} M} = \mbox{constant}$
defines a hyperplane in $\mathbf{R}^{N^2-1}$, the space of vectors
$\mathbf{m}$ that (\ref{M-to-m}) associates with the unit-trace hermitian
matrices $M$. 
All the vertices of the MUB polytope lie either in the hyperplane 
$\tr{W_{m,n} M}= 0$, where they span a facet, or 
in the hyperplane $\tr{W_{m,n} M} = 1$, which is the orthocomplemented face. 
All points in the polytope obey $0 \leq\tr{W_{m,n} M} \leq 1$, for all values
of $m$ and $n$.
This underlies the construction of Wootters's analogs of Wigner's function, 
and it explains why we refer to the vectors $\textbf{w}_{m,n}$ as face points,
and to their unit trace versions $W_{m,n}$ as face point operators.

In passing we note that one can prove a remarkable result in prime 
dimensions:\cite{gross} 
All statistical operators such that 
${0\leq\tr{W_{mn}\rho}}$, which says that their Wigner coefficients are
positive, necessarily are convex combinations of projectors onto the states
$\ket{e^i_k}$ of the MUB, for which 
\begin{equation}
  \label{eq4:Wigner-8a}
  \bra{e^i_k}W_{m,n}\ket{e^i_k}=\left\{
  \begin{array}{l@{\ \textrm{for}\ }l}
   \delta_{k\oplus n,i\odot m} & i=0,1,\dots,N-1 \\[0.5ex]
   \delta_{k,m} & i=N 
  \end{array}\right\}=0\ \textrm{or}\ 1\,.
\end{equation} 
In other words, the statistical operators $\ket{e^i_k}\bra{e^i_k}$ belong to
the polytope.  
So, the equation $0 \leq \tr{W_{m,n} M} \leq 1$ is necessary 
and \emph{sufficient} for belonging to the MUB polytope. 
We conjecture that this is also true in prime power dimensions. 

With criteria (W1)--(W5) taken care of, we finally turn to (W6). 
As noted in Sec.~\ref{sec4.2.2}, the limit $N=p^\m\to\infty$ is the limit
$p\to\infty$ with a fixed value of $\m$, so that we are consistently dealing 
with a system composed of $\m$ q-pits and arrive at a $\m$-dimensional
continuous system in the limit. 
Contact with the standard Wigner basis is, therefore, established if we get%
\footnote{The integration in (\ref{eq4:Wigner-9}) is over the $\m$-dimensional 
  real space, $x=(x_0,x_1,\dots,x_{\m-1})$ with each coefficient $x_j$ taking
  on all real values.}
\begin{equation} 
  \label{eq4:Wigner-9}
  W_{0,0}\to\int\!\D^\m x\,\ket{-x}2^\m\bra{x}=P\otimes P\otimes \cdots \otimes P
\end{equation}
in the limit, that is: $\m$ copies of the one-dimensional parity operator
\begin{equation} 
  \label{eq4:Wigner-10}
  P=\int\limits_{-\infty}^\infty \!\D x \,\ket{-x}2\bra{x}\,,
\end{equation}
the seed of the Wigner basis,\cite{Glauber,wigner2}
where the factor of $2$ ensures proper normalization to unit trace,  
$\tr{P}=1$. 

Now, after expressing the projectors in  
\begin{equation}
  \label{eq4:Wigner-11}
  W_{0,0}=\sum_{i=0}^N\ket{e_0^i}\bra{e_0^i}-\mathbf{1}
\end{equation}
in terms of the unitary shift operators, we have
\begin{equation} 
  \label{eq4:Wigner-12}
  W_{0,0}=\frac{1}{N}\sum_{i=0}^{N-1}{\left(V^i_0
           +\sum_{j=1}^{N-1}\alpha^{i\oslash j}_j V_j^i\right)}\,,
\end{equation}
where (\ref{eq:defUil}) and the $k=0$ version of (\ref{MUBproj}) are the main
ingredients.  
This shows that the seed $W_{0,0}$ --- and, therefore, also all other
$W_{m,n}$s --- is an equal-weight sum of all $N^2$ operators of the unitary
Weyl basis, whereby the phase factors $\alpha^{i\oslash j}_j$ ensure that
$W_{0,0}$ is hermitian.

This is illustrated by the $N=2$ example for which
\begin{equation}
  \label{eq4:Wigner-12a}
  \begin{array}{lcr@{\qquad}lcr}
  W_{0,0}&=&\ds\frac{1}{2}(\mathbf{1}+\sigma_x+\sigma_y+\sigma_z)\,,&  
  W_{0,1}&=&\ds\frac{1}{2}(\mathbf{1}-\sigma_x-\sigma_y+\sigma_z)\,,\\[2ex]
  W_{1,0}&=&\ds\frac{1}{2}(\mathbf{1}+\sigma_x-\sigma_y-\sigma_z)\,,&  
  W_{1,1}&=&\ds\frac{1}{2}(\mathbf{1}-\sigma_x+\sigma_y-\sigma_z)\,,    
  \end{array}
\end{equation}
are well-known q-bit analogs of the Wigner basis operators.
In an ill-fated attempt, Feynman used the expectation values of these
operators to introduce probabilities of 
``$\sigma_x=1$ \emph{and} $\sigma_z=1$'' and the like.
But since the eigenvalues of the four operators in (\ref{eq4:Wigner-12a}) are
$\frac{1}{2}(1\pm\sqrt{3})$, he was forced to resort to the dubious notion of
``negative probabilities'' which, in fact, gave this paper its 
title.\cite{negprobFeyn}
A direct measurement of the said expectation values, for the polarization
q-bit of a photon, is reported in Ref.~\refcite{wignerdurtsing}.  

In the limit ${p\to\infty}$, only odd values of $p$ are relevant, and for those
${j=(j\oslash2)}{\oplus(j\oslash2)}$ is true, which allows us to write 
\begin{equation}
  \label{eq4:Wigner-13}
  V_j^i=\gamma^{i\odot j\oslash2}V^0_{j\oslash2}V_0^iV^0_{j\oslash2}
\end{equation}
with the aid of (\ref{discBH}) and, \emph{if} we choose the symmetric value of
(\ref{conven}) for $\alpha^i_l$, we have
\begin{equation} 
  \label{eq4:Wigner-14}
  \alpha^{i\oslash j}_j\gamma^{i\odot j\oslash2}=1
\end{equation}
for the product of phase factors, that is: \emph{if} we enforce the symmetry
property (\ref{eq2:symmetry}).
With this symmetry in place, then, the seed is ($j\to2\odot k$)
\begin{eqnarray}
  \label{eq4:Wigner-15}
  W_{0,0}&=&\frac{1}{N}\sum_{i,k=0}^{N-1}V^0_kV^i_0V^0_k
         =\sum_{k=0}^{N-1}V^0_k\ket{0}\bra{0}V^0_k \nonumber\\
         &=&\sum_{k=0}^{N-1}\ket{k}\bra{\ominus k}
         =\sum_{k=0}^{N-1}\ket{e^i_k}\bra{e^i_{\ominus k}}\,,
\end{eqnarray}
where the value of the last summation does not depend on the basis label $i$. 
This is clearly the discrete analog of the continuous $\m$-dimensional parity
operator $P$ in (\ref{eq4:Wigner-9}),
\begin{equation}
  \label{eq4:Wigner-16}
   W_{0,0}=\sum_{k=0}^{N-1}\ket{\ominus k}\bra{k}
         =\sum_{k_0=0}^{p-1}\ket{-k_0}\bra{k_0}\otimes
          \sum_{k_1=0}^{p-1}\ket{-k_1}\bra{k_1}
          \otimes\cdots\,,
\end{equation}
the product of $\m$ factors of the analog of the one-dimensional parity
operator in (\ref{eq4:Wigner-10}).
And since the unitary shift operators factorize in accordance with 
(\ref{eq4:Weyl-product}), this factorization of the Wigner seed carries over
to all operators of the Wigner basis in virtue of property (W4). 
The limit $p\to\infty$, then, gives us the right-hand side of
(\ref{eq4:Wigner-9}) as desired.%
\footnote{This limit has its subtleties (see the references cited in
  Sec.~\ref{sec:WSlim}) and requires careful attention to the factor 
  of $2^\m$ in (\ref{eq4:Wigner-8}) which, roughly speaking, originates in
  \begin{displaymath}
    \tr{\ket{\ominus k}\bra{k}}=\delta_{2\odot k,0}=\delta_{k,0}
      \to\delta(x)=2^\m\delta(2 x)=\tr{\ket{-x}2^\m\bra{x}}\,.
  \end{displaymath}}

In summary, the basis composed of the operators $W_{m,n}$ as defined in
(\ref{eq4:Wigner-1}) obeys criteria (W1)--(W5) by construction, and also
criterion (W6) if the symmetry property (\ref{eq2:symmetry}) is imposed on the
phase factors $\alpha_l^i$ of (\ref{eq:defUil}).
We then have a genuine analog of the standard Wigner basis for continuous
degrees of freedom, and it is fair terminology to call the $W_{m,n}$s the
elements of the $N$-dimensional Wigner basis, as we have already been doing
above.  

It is worth remembering, however, that \emph{all} permissible choices for the
$\alpha_l^i$ give a good hermitian operator basis for which (W1)--(W5) are
true, and the limit $p\to\infty$ is of little concern for any particular value
of $N=p^\m$ at hand.
If one makes use of the option discussed in the paragraph after
(\ref{ortho-UU}) and multiplies the right-hand side of (\ref{conven}) by
$\gamma^{b_i\odot l}$ with $b_0=0$ and arbitrary field elements $b_i$ for
$i=1,2,\dots,N-1$, then 
\begin{equation}
  \label{eq4:Wigner-15'}
W^{(b)}_{0,0}=\frac{1}{N}\sum_{i,k=0}^{N-1}\gamma^{2\odot b_i\odot k}V^0_kV^i_0V^0_k
\end{equation}
replaces the $b_i\equiv0$ version of (\ref{eq4:Wigner-15}).
If one or more of the $b_i$s are nonzero, $W^{(b)}_{0,0}$ is different from all
$W_{m,n}$s and, therefore, the hermitian operator basis generated from the
seed $W^{(b)}_{0,0}$ is different from the Wigner basis --- the parity operator
(\ref{eq4:Wigner-16}) is not one of the basis operators.
There are in total $N^{N-1}$ different seeds $W^{(b)}_{0,0}$ and as many
hermitian operator bases and with suitable ${N\to\infty}$
limits for the $b_i$s the seeds will have well-defined limits themselves, 
but in our understanding only the $b\equiv0$ basis is a true
finite-dimensional analog of the Wigner basis.% 
\footnote{\label{fn:ringWig}%
  In arbitrary odd dimensions $N$, one can also introduce a
  Wigner-type operator basis by modifying the parity operator of 
  (\ref{eq4:Wigner-17}) through a replacement of the field arithmetic 
  by modulo-$N$ arithmetic ($\ominus\to\ominus_N$). 
  Consult Refs.~\refcite{Vourdas2,Durtmean,gross} for details.} 

We thus observe that the symmetric choice of (\ref{conven}) is the right
choice for obtaining a proper analog of the Wigner basis. 
It also endows the Wigner basis with certain elegant covariance
properties\cite{Durtmean} that will be discussed in
Sec.~\ref{sec:Wigner-covariance}.  

We further note that the property (W5) is sufficient to derive that
each Wigner operator is equal to the sum of projectors onto states from
different bases minus the identity operator as expressed by
(\ref{eq4:Wigner-1}); the explicit choice of MUB that we made in
Sec.~\ref{section2} is not crucial. 
Indeed, the sum of all the Wigner operators that belong to the $N+1$
(nonparallel) straight lines passing through a phase space point $(m,n)$ is
also equal to the sum of all Wigner operators plus $N$ times $W_{m,n}$; as a
consequence of (W5) it also equals $N$ times a sum of the projectors onto
states from different bases; now, the sum of all Wigner operators equals $N$
times the identity as noted in (\ref{eq4:Wigner-7}). 
It follows that each Wigner operator plus the identity operator
is equal to a sum of projectors onto states from different bases. 

This is how Wootters \textit{et al.} derived an
expression for (loosely analogous) Wigner operators similar to
(\ref{eq4:Wigner-1}),\cite{discretewigner} which may or may not possess
property (W6). 
Their approach is somewhat more general than ours in the sense that theirs
is valid whichever set of $N+1$ MUB is adopted, whereas
the expression (\ref{eq4:Wigner-1}) refers explicitly to the bases defined in
(\ref{xxx}) and specified unambiguously by the phase factors $\alpha^i_l$
that obey the constraints (\ref{eq:phase1}) and (\ref{eq2:phaseproduct}). 

In view of the properties (W1) to (W5) in (\ref{eq4:Wigner-4}), in particular
the marginals property (W5), it is natural to interpret the Wigner operators
as discrete phase-space localization operators.\cite{Wootters2,discretewigner}
Indeed, when the system is in a ``position'' eigenstate $\ket{e^N_k}$, 
the expectation value of $W_{m,n}$ equals $0$ for ${k\neq m}$, and $1/N$ for
${k=m}$, irrespective of the ``momentum label'' $n$.   
Similarly, when the system is prepared in a ``momentum'' eigenstate 
$\ket{e^0_l}$, the expectation value is $0$ for ${l\neq\ominus n}$, and $1/N$
for ${l=\ominus n}$, whatever the value of the ``position label'' $m$.
This situation is reminiscent of the uncertainty principle:\cite{uncertainty} 
When we have a state of sharp position, here: $\ket{e^N_k}$, then the value of
the position is definite while all values of the momentum label are equally
probable; and the analogous reverse case applies to states $\ket{e^0_l}$ of
sharp momentum.

As appealing as this picture is, it has a flaw: 
The expectation value of $W_{m,n}$ can be negative.
In fact, for odd $N$, we have
\begin{equation}
  \label{eq4:Wigner-17}
  W_{0,0}\bigl(\ket{k}\pm\ket{\ominus k}\bigr)
   =\pm\bigl(\ket{k}\pm\ket{\ominus k}\bigr)
\end{equation}
for $k=0,1,\dots,N-1$, so that $W_{0,0}$ has the $(N+1)/2$-fold eigenvalue
$+1$ and the $(N-1)/2$-fold eigenvalue $-1$.
In view of the unitary equivalence property (W4), explicitly stated in
(\ref{eq4:Wigner-5}), these are also the eigenvalues of all other $W_{m,n}$s.
It follows that the operators of the Wigner basis are not projectors, but each
of them is rather the difference between a projector onto a
$(N+1)/2$-dimensional subspace and a projector onto a $(N-1)/2$-dimensional
subspace. 

In (\ref{eq4:Wigner-1}) we have one projector for each of the $N+1$ MUB, and it
follows from (\ref{eq4:Wigner-8a}) that the expectation value of $W_{m,n}$ is
maximal for these states, 
\begin{equation}
  \label{eq4:Wigner-17a}
  \bra{e^N_m}W_{m,n}\ket{e^N_m}=1\quad\mbox{and}\quad
  \bra{e^i_{i\odot m\ominus n}}W_{m,n}\ket{e^i_{i\odot m\ominus n}}=1
\quad\mbox{for $i=0,1,\dots,N-1$}\,.
\end{equation}
They are, therefore, eigenstates to eigenvalue $+1$, and since they are $N+1$
states in a $(N+1)/2$-dimensional subspace, they are clearly linearly
dependent.
They are also assuredly complete because the projector on the $+1$ subspace of
$W_{m,n}$,
\begin{equation}
  \label{eq4:Wigner-17b}
  \frac{\mathbf{1}+W_{m,n}}{2}=\frac{1}{2}\biggl(\ket{e^{N}_m}\bra{e^N_m} 
          +\sum_{i=0}^{N-1}\ket{e_{i\odot m\ominus n}^{i}}
          \bra{e_{i\odot m\ominus n}^i}\biggr),
\end{equation}
is clearly spanned by those $N+1$ eigenstates, one from each basis.

A direct measurement of the expectation values of all Wigner basis operators
--- or, put differently, the experimental determination of the $N^2$ Wigner
coefficients $r_{m,n}$ of (\ref{eq4:Wigner-3}) --- would thus require the
realization of the $N^2$ binary observables (eigenvalues $\pm1$) that
distinguish the respective subspaces.
While possible in principle, such a procedure is not economical in practice,
because two different $W_{m,n}$s do not commute, and each $W_{m,n}$ 
must be measured separately.
 
Indeed, with one exception, all reports of experimentally determined Wigner
functions --- in the one-dimensional continuous case --- are actually Wigner
functions that are inferred from measured marginal distributions;
the said exception is the experiment of Refs.~\refcite{Wigner-exp1a} and 
\refcite{Wigner-exp1b}, which implemented the scheme introduced in
Ref.~\refcite{Wigner-exp2}. 
The measurements, reported in Ref.~\refcite{wignerdurtsing}, of the 
single--q-bit Wigner basis (\ref{eq4:Wigner-12a}) and a particular two--q-bit
Wigner basis of product form, exploited an optical implementation of a 
one--q-bit SIC~POVM that is optimal for single--q-bit
tomography.\cite{Jarda+2:04}   

The geometrical picture offered by the marginals and the corresponding sums
over affine straight lines, recall (\ref{eq4:Wigner-6}) and
(\ref{eq4:Wigner-8}), sheds some light on the solution of the mean king's
problem in Sec.~\ref{sec4.1}. 
As noted above, the correspondence (\ref{eq:correspond}) links
(\ref{eq4:Wigner-8}) to (\ref{eq4:MKmarginals}), and so we understand why the
preparation of the state $\ket{{e^i_k}^*,e^i_k}$ by the king's men 
is accompanied by the equiprobable firing of $N$ detectors that correspond to
the states $\ket{(i_1,i_2)}$ with $i_2=k$ when $i=N$ and $\ominus i_1\oplus
i\odot i_2=k$ otherwise.
The other detectors do not fire at all. 
If we re-express this property in terms of localization operators, in the
sense of the paragraph preceding (\ref{eq4:Wigner-17}), we find that the $N$
detectors that have a nonzero probability of firing correspond to
localization operators located on a straight line for which the marginal is
the projector $\ket{e^i_k}\bra{e^i_k}$.

\subsubsection{Covariance of the Wigner-type basis}
\label{sec:Wigner-covariance}
Upon projecting (\ref{eq4:MKstates}) onto the Bell basis we get 
\begin{eqnarray}
  \label{eq4:Wigner-18}
\ket{(i_1,i_2)}&=&\frac{1}{N}\sum_{m,n=0}^{N-1}\ket{B_{m,n}}
  \gamma ^{i_2\odot m \ominus i_1\odot n}   
   \Gamma_{m,n} 
 \nonumber\\ \mbox{with}\quad \Gamma_{m,n}&=&\left\{
   \begin{array}{c@{\mbox{\ for\ }}l}
     1 & m=0\,,\\[1ex]
     \alpha_m^{n\oslash m} & m>0\,,
   \end{array}\right.
\end{eqnarray}
where $\alpha^i_m$ is the phase factor of (\ref{eq:defUil}), explicitly stated
in (\ref{eq2:even-alpha}) for $N$ even and in (\ref{conven}) for 
$N$ odd, provided the symmetry property (\ref{eq2:symmetry}) is imposed, as we
assume throughout the present discussion.
Then ${\Gamma_m^n}^2=\gamma^{\ominus m\odot n}$, and we can regard the phase
factors $\Gamma_m^n$ as the appropriate square roots of 
$\gamma^{\ominus m\odot n}$. 

Making use of the transformation (\ref{defBmn}) that transforms Bell states
into displacement operators we get  an alternative expression for the Wigner
operator $W_{i_{1},i_{2}}$, 
\begin{equation} \label{eq4:Wigner-19}
W_{i_{1},i_{2}}=\frac{1}{N}\sum_{m,n=0}^{N-1}
   \gamma ^{ \ominus i_{1} \odot  n \oplus i_{2} \odot  m}
  \Gamma_{m,n}V^{n}_{m}\,. 
\end{equation}
In view of the symmetric choice (\ref{conven}), we can rewrite
(\ref{eq3:ith-Vmn}) for odd $N$ in the form
\begin{equation} \label{eq4:Wigner-20} 
\Gamma_{m,n}V_{m}^n=C_i^{\,}\Gamma_{m',n'}V_{m'}^{n'}C_i^{\dagger}
\quad\mbox{with}\ i\odot m\ominus n=m'\ \mbox{and}\ m=n'\,.
\end{equation}
This is the transformation law of the displacement operators under a change of
the underlying basis, the main ingredient on the right-hand side of
(\ref{eq4:Wigner-19}). 
It is sometimes referred to as the \emph{covariance} of the Heisenberg--Weyl
group. 
  
Similarly, the permutation invariance (\ref{eq3:differentBells}) of the Bell
basis under the action of  $C^*_i\otimes C^{\ }_i$ is sometimes referred to as
the covariance of the Bell basis. 
The other permutation invariance, noted in (\ref{Bell-2}), is of quite a
different kind.
But both reflect a general property: 
The Clifford group of unitary operators is the stabilizer of the
Heisenberg--Weyl group. 

In addition, the affine transformation (\ref{mappings}) that maps $(m,n)$ onto
$(m',n')$ is a symplectic transformation in the sense that it preserves the
symplectic form $m_{1}\odot n_{2}\ominus n_{1}\odot m_{2}$. 
Indeed, $m'_{1}\odot n'_{2}\ominus n'_{1}\odot m'_{2}=
m_{1}\odot n_{2}\ominus n_{1}\odot m_{2}$ so that 
\begin{equation}\label{eq4:Wigner-21}
C_i^{\,}W_{i_1,i_2}C_i^{\dagger}=W_{i'_1,i'_2}
 \quad \mbox{with}\ i\odot i_1\ominus i_2=i_1'\ \mbox{and}\ i_1=i_2'\,,
\end{equation}
which shows that the Clifford transformations $C^{\ }_i$ correspond to affine
reparameterizations of the phase-space labels of the operators in the Wigner
basis, the phase-space localization operators. 

The transformation laws (\ref{eq4:Wigner-20}) and
(\ref{eq4:Wigner-21}) hold for odd $N$ with the symmetric choice
(\ref{conven}). 
What about even prime power dimensions, ${N=2^\m}$? 
Here, the expression (\ref{eq2:even-alpha}) of the phase factors $\alpha^i_l$
is rather intricate and we do not know whether (\ref{eq4:Wigner-20}) and
(\ref{eq4:Wigner-21}) are valid.
It is an open question whether there is a set of field elements $b_i$ such
that, after supplementing the $\alpha^i_l$s of (\ref{eq2:even-alpha}) by factors
$(-1)^{b_i\odot l}$, they conspire to produce (\ref{eq4:Wigner-20}) and
(\ref{eq4:Wigner-21}). 

But one does know that other properties of Wigner operators, such as the
factorization (\ref{eq4:Wigner-16}) into a product of $\m$ Wigner operators of
dimension $p$, can only be had for odd $p$, 
not for ${p=2}$ and ${\m>2}$.\cite{laser,durtarxive} 
The two--q-bit case ${N=2^2}$ is an exception; 
there are q-quart Wigner operators that
factorize into products of two q-bit Wigner operators.
They have been realized experimentally for the purpose of biphoton 
polarimetry.\cite{wignerdurtsing} 

We emphasize that the requirements (W1) to (W5) in (\ref{eq4:Wigner-4}) are
obeyed by the $W_{m,n}$s for all prime power dimensions, even or odd,
irrespective of the convention chosen for the  $\alpha^i_l$s.
And (W6) is of no concern for even $N$.

Actually, it is easy to show that the different phase choices compatible with 
(\ref{ortho-UU}) preserve the MUB as a whole but shift the labels of their
basis states.\cite{Durtsept} 
The covariance of the Heisenberg--Weyl group (\ref{eq4:Wigner-20}) as well as
the elegant transformation law (\ref{eq4:Wigner-21}) are guaranteed, in odd
prime power dimensions, only for the symmetric phase-choice (\ref{conven}). 
This also concerns the phase point operators within the framework laid out by
Gibbons \textit{et al.},\cite{discretewigner} for which the bijection between
MUB and Wigner operators (\ref{eq4:Wigner-8}) also holds by construction,
independently of the choice of MUB and of the labeling of the MUB states.  
This result can be inferred in prime dimensions, for instance, 
from the study\cite{Appleby+2:08} of the properties of the Wigner operators
that correspond to different quantum nets in Wootters's terminology, or 
to different phase-choices compatible with (\ref{ortho-UU}) in ours.

Another elegant feature that singles out the symmetric phase-choice
(\ref{conven}) is that the corresponding Wigner function is well behaved with
regard to the composition law of Wigner operators, a property that was
remarked upon by Gibbons \textit{et al.} in Ref.~\refcite{discretewigner},
who noted that among all $N^{N-1}$ possible choices of quantum nets, there
exists a particular net that exhibits ``more than the required symmetry.'' 
This singled-out net corresponds to our symmetric phase choice in
(\ref{conven}).

\subsection{Mutually unbiased bases and finite affine planes}
\label{secaffin} 
The combinatorial structure that underlies the solution of the Mean King's
problem is known as a finite affine plane of order $N$.  
By definition an affine plane is an ordered 
pair of two sets, the first of which consists of elements $a_{\alpha}$, called 
points, and the second of which consists of subsets $L_{\omega}$ of the 
first, called lines. Two lines whose intersection is empty are called parallel. 
The following axioms hold:\cite{bennett95}
\begin{equation}  \label{eq4:APaxioms}
\begin{tabular}[b]{l@{:\ }p{0.7\textwidth}@{}}
A1&{If $a_{\alpha}$ and $a_{\beta}$ are distinct points, there is 
a unique line $L_{\omega}$ such that $a_{\alpha} \in L_{\omega}$ and 
$a_{\beta} \in L_{\omega}$.} \\
A2&{If $a_{\alpha}$ is a point not contained in the line $L_{\omega}$, 
there is a unique line $L_{\sigma}$ such that $a_{\alpha} \in L_{\sigma}$ 
and $L_{\sigma}\cap L_{\omega} = \emptyset $.}\\  
A3&{There are at least two points on each line, and there are 
    at least two lines.}
\end{tabular}
\end{equation}
To see how this works, think of an ordinary affine plane, and think of it 
as two sets, the set of points and the set of lines. 
Two points determine a unique line, while two lines either intersect in a
unique point, or else they are parallel and do not intersect at all. 
This is what the axioms (\ref{eq4:APaxioms}) say. 

If the number of points is finite the affine plane is also said to be finite,
and it is assigned a finite number $N$, called its order. 
A finite affine plane of order $N$ has exactly $N^2$ points and $N^2+N$ lines. 
Each line contains $N$ points, and $N+1$ lines intersect in each point. 
There are altogether $N+1$ \emph{pencils} of parallel lines containing $N$
lines each. 
If we label the lines of every pencil with a set of $N$ letters, we can use
two of the pencils to provide a ``coordinate system'' for the affine plane.  
Each remaining pencil then defines what is known as a Latin square --- a 
square array of $N^2$ symbols, such that there are $N$ different kinds of
symbols, and such that the same symbol never occurs twice in a row or in a
column of the array.\footnote{Sudokus are ${9\times9}$ Latin squares of a
  restricted kind.}\   
Examples for such arrays are the two addition tables 
in Table~\ref{tbl:4-field},
but by no means all Latin squares arise in such an orderly manner.

To see how this works, consider ${N = 3}$. 
Pick two pencils of parallel lines, and label their lines with $0,1,2$ and
$0', 1', 2'$. 
The nine points of the affine plane can then be arranged in an array with
points on the lines of the first pencil making up the columns, and those of
the second pencil making up the rows. 
The lines of the remaining two pencils of parallel lines are 
labelled by $A,B,C$ and $\alpha, \beta, \gamma$. 
Marking all points in the array that occur on line $A$ with this letter, and
so on for the other lines, will give rise to two Latin squares: 
\begin{equation} \begin{array}{c|ccc|} \ & 0 & 1 & 2 \\ \hline 
0' & A & B & C \\ 1' & B & C & A \\ 2' & C & A & B \\ \hline 
\end{array}  \ \hspace{15mm} 
\begin{array}{c|ccc|} \ & 0 & 1 & 2 \\ \hline 
0' & \alpha & \gamma & \beta \\ 1' & \beta & \alpha & \gamma \\ 
2' & \gamma & \beta & \alpha \\ \hline \end{array}  
\end{equation}
The squares must be Latin because the line labelled $A$, say, 
intersects each of the lines in the two pencils we started out with 
exactly once, and similarly for all other lettered lines.
Now recall that the line labeled $A$ intersects the line labelled $\alpha$ 
in a unique point. 
This explains why the two Latin squares we obtain must have the interesting
property of being orthogonal Latin squares;
another name for such a pair is a Graeco-Latin pair.\cite{bennett95}  
By definition this means that picking a pair of symbols, one Latin and one
Greek --- one from each of the two Latin squares --- determines a unique point
in the original array. 
To check that we did things right we simply superpose the two 
squares, and check that the pair of symbols $A\alpha$ occurs once and once 
only, and similarly for all other pairs. 
Incidentally, we see another interesting thing, namely that we could just as 
well have used the Latin letters to label the columns and the Greek letters 
to label the rows. 
The symbols we used in the first place will then 
distribute themselves into another Graeco-Latin pair: 
\begin{equation} \begin{array}{c|ccc|} \ & 0 & 1 & 2 \\ \hline 
0' & A\alpha & B\gamma & C\beta \\ 1' & B\beta & C\alpha & A\gamma \\
 2' & C\gamma & A\beta & B\alpha \\ \hline \end{array} \hspace{8mm} 
 \leftrightarrow 
\hspace{8mm} \begin{array}{c|ccc|} \ & A & B & C \\ \hline 
\alpha & 00' & 22' & 11' \\ \beta & 12' & 01' & 20' \\
 \gamma & 21' & 10' & 02' \\ \hline \end{array} 
\end{equation}
Given the facts about finite affine planes that were recited 
above, it is clear that all of this works for every finite affine plane, 
and regardless of what pencils of parallel lines we pick. 
Setting two of the pencils aside to define the array, 
the remaining $N-1$ pencils always define $N-1$ 
mutually orthogonal Latin squares. 
This much is guaranteed by the intersection properties of the affine plane. 
Conversely, $N-1$ 
mutually orthogonal Latin squares will define an affine plane of order $N$. 

But finite affine planes come with an existence problem of their own; 
indeed already Euler raised the question whether it is at all possibe 
to find a pair of orthogonal Latin squares when ${N = 6}$. 
He phrased it as a problem concerning 36 officers. 
More than a hundred years later it was proved that the answer is ``no.'' 
This important result was reported in 1900 by the mathematician 
Tarry,\cite{Tarry:00} who proved by means of an exhaustive calculation 
that Euler's problem does not possess a solution, in agreement with 
Euler's conjecture.  
It follows that finite affine planes of order 6 do not exist. 
Progress since then has been slow.
Finite affine planes do exist if $N = p^\m$, where $p$ is a prime number. 
They do not exist if $N = 4k +1$ or $N = 4k + 2$ and $N$ is not the sum of two
squares, or if $N = 10$. 
All other cases are open. 
If $N = p^\m$, a finite affine plane can be constructed using the 
methods of analytical geometry, with the finite field of order $p^\m$ as the
field of scalars, but examples not of this form are known as well. 

A finite affine plane can be turned into a finite projective plane through the
addition of an extra line ``at infinity.'' 
It should be emphasized that finite planes, whether affine or projective, are
much more than just interesting toys --- in classical computer 
science they play prominent roles, for instance in the theory of error
correcting codes, and we have already seen that they have quantum mechanical
applications. 
 
The relation between MUB and finite affine planes can be seen already at the
level of the MUB polytope discussed in Sec.~\ref{section0}. 
The idea is to represent the lines by the $N^2+N$ vertices of the polytope, and
the points by a subset of its $N^{N+1}$ facets. 
Two points are to lie on a line if the corresponding vertices are 
vertices of the same facets, and two lines intersect in a point if the
corresponding facets share a common vertex. 
It turns out\cite{BE05} that if an affine plane exists such a 
correspondence can always be set up, and the $N^2$ selected facets will then
be placed in such a way that their centers form a regular simplex
in $\mathbf{R}^{N^2-1}$. 
This construction needs neither finite fields nor the special feature that the
vertices of the polytope correspond to one-dimensional projectors on Hilbert
space. 
But when they do, it is possible to choose 
--- following Wootters\cite{Wootters87,Wootters2,discretewigner} --- 
the special set of Wigner operators that we have discussed in 
Sec.~\ref{sec4.2.3}, and to relate the construction to the partition of the
Heisenberg--Weyl group that is associated with the MUB:\cite{india} 
Then each basis is associated with a straight line that passes through the
origin in the plane.  

Whether there is a deeper relation between the existence problem for MUB and
the existence problem for finite affine planes is not known today. 
It has been conjectured that such a relation exists,\cite{saniga04,Paterek+2:09}
but a recent attempt to use a pair of Graeco-Latin squares that does exist
when $N=10$ to construct a set of four MUB in this dimension
failed.\cite{Paterek+3:09} 
It is interesting to notice that if $N$ mutually orthogonal Latin squares
exist, then there always exist $N+1$ of them.
Similarly, if $N$ MUB exist, then there always exist ${N+1}$ of
them.\cite{Weiner09} 

In the 19th century, the combinatorial structures now known as finite
geometries were studied more concretely by geometers, who realized them as
configurations of lines and points, or more generally as configurations of
subspaces of a complex projective space.\cite{klein26} 
In 1844 Hesse, following earlier work by Pl\"ucker, studied a configuration of
9 lines and 12 points in the projective plane, such that each line contains 4
points and each point lies on 3 lines.\cite{hesse44}. 
Translated into the language of quantum theory, where the projective plane is
the set of rays in a three-dimensional Hilbert space ($N=p=3$), Hesse's twelve
points are indeed the twelve kets that compose the four MUB of three kets
each.   
His construction was generalized to the case of arbitrary prime $N$ by
Segre,\cite{segre86} who therefore in a sense discovered the maximal sets of
MUB in prime dimensions --- although some necessary ingredients, including the
quantum mechanical significance of the construction, were very naturally
missing.   

Segre's starting point was an elliptic curve in complex projective
space,\cite{hulek86} whose symmetry group consists of the Heisenberg--Weyl
group together with an extra reflection, an element of order~2. 
When $N$ is an odd prime, there are $N^2$ such reflections, 
since the Heisenberg--Weyl group acts on them in accordance with
(\ref{eq4:Wigner-5}), which corresponds to the condition (W4) in
(\ref{eq4:Wigner-4}). 
In our terminology this means that he introduced a discrete parity operator 
with the matrix representation%
\footnote{Since $N$ is an odd prime, the field addition $\oplus$ is modulo-$N$
  addition.}
\begin{equation} 
[W_{0,0}]_{a,b} = \delta_{0,a\oplus b} \,. 
\end{equation}
This operator is both hermitian and unitary, with eigenvalues $\pm 1$, and in
fact it splits the Hilbert space into two subspaces, of
dimension $n$ and $n-1$ respectively, where $N = 2n-1$ is an odd prime. 
There are altogether $N^2$ such subspaces of dimension $n$, 
and Segre observed that there exists $N^2 + N$ vectors such that each subspace
contains $N+1$ of the vectors, and each vector lies in exactly $N$ of the
subspaces. 
In the notation used to describe such things, we have a configuration of type 
\begin{equation} 
\bigl(N^2_{N+1}, N(N+1)_N\bigr) \,. 
\end{equation}
These incidence relations are exactly those of a finite affine plane. They are 
clearly quite remarkable: In ${N = 2n-1}$ dimensions two $n$-dimensional
subspaces intersect in (at least) a single vector, but the remarkable thing is
that only ${N^2 + N}$ distinct vectors are needed for the entire configuration. 
And, of course, once we have chosen the standard representation of the
Heisenberg--Weyl group, these $N^2 +N$ vectors are precisely the kets that
make up the MUB.

To see why this is so, let us go back to the definition of the face point
operators in (\ref{embed2}). 
The first face point operator is defined by picking one projector from each
MUB. 
Any choice will do. 
Then the combinatorics of the affine plane --- or 
alternatively the action of the Heisenberg--Weyl group --- 
will define a definite \mbox{$N^2$-plet} of face point operators. 
Now consider the kets corresponding to the ${N+1}$ projectors we picked. 
Typically, ${N+1}$ kets will span the $N$-dimensional Hilbert space. 
But let us pick ``the first vector in each basis'' (referring to the standard
set of MUB of \ref{sec:app1}), that is: the kets represented by the columns
\begin{equation} 
\psi^{(0)} = \frac{1}{\sqrt{N}}\left(\begin{array}{c} 
                  1 \\ 0 \\ 0 \\ \vdots \\ 0 \\ 0 
                  \end{array}\right),\qquad 
\psi^{(r)} = \frac{1}{\sqrt{N}}\left(\begin{array}{c} 
                   1 \\ \gamma_N^{r1^2} \\ \gamma_N^{r  2^2} \\ 
                        \vdots \\ \gamma_N^{r(N-2)^2} \\ \gamma_N^{r(N-1)^2} 
                   \end{array} \right), 
\quad 1 \leq r \leq N \,. 
\end{equation}
By inspection we see that they span an $n$-dimensional subspace only, and
indeed that they are all eigenvectors of $W_{0,0}$ with eigenvalue $+1$. 
Since the face point operators, and the choices of MU vectors made for them,
are related by the Heisenberg--Weyl group, there will be altogether $N^2$
subspaces of this kind, and they will necessarily have the intersection
properties discovered by Segre. 
But to him this was a statement about the geometry of an elliptic curve in
projective space, not about quantum mechanics --- the latter was still several
decades into his future.  

Segre's observation holds true in all odd prime power dimensions. 
In particular, as observed above in the context of
(\ref{eq4:Wigner-17})--(\ref{eq4:Wigner-17b}), 
all  Wigner basis operators in odd prime power dimensions possess 
a $n=\frac{1}{2}(N+1)$-dimensional subspace to eigenvalue $+1$ and
a $n-1=\frac{1}{2}(N-1)$-dimensional subspace to eigenvalue $-1$.

In marked contrast, no similar construction is known for even $N$. 
In this case there is no parity operator available, a fact that also
causes well studied complications when one tries to define analogs of the
Wigner function.\cite{chatu05}

         %% Section 4
%%%% file name: MUB-5.tex
%%%% input file for MUB.tex 
%%%%
%%%% last changes on 20 April 2010 by Berge
%%%% minor correction on 27 April 2010
%%%%
%%%%%%%%%%%%%%%%%%%%%%%%%%%%%%%%%%%%%%%%%%%%%%

\section{Mutually unbiased Hadamard matrices}
\label{section5}

\subsection{Pairs of mutually unbiased bases and Hadamard matrices}
\label{sec5.1}
Let us look at the problem of finding MUB from a different perspective. 
As in Sec.~\ref{section0} we represent kets as column vectors.
The kets $\ket{u_0},\ket{u_1},\dots,\ket{u_{N-1}}$ of an orthonormal basis
then correspond to the $N$ columns of a unitary matrix $U$. 
By convention, the computational basis is represented by the unit
matrix~$\mathbbm{1}$.
Then,
\begin{equation}
  \label{eq5:base2U}
  U=\left(\begin{array}{c}
          \bra{0}\\ \bra{1} \\ \vdots \\ \bra{N-1}
         \end{array}\right)
    \bigl(\ket{u^{\ }_0},\ket{u^{\ }_1},\dots,\ket{u^{\ }_{N-1}}\bigr)
\end{equation}
turns the basis kets into the unitary matrix, and
\begin{equation}
  \label{eq5:U2base}
  \bigl(\ket{u^{\ }_0},\ket{u^{\ }_1},\dots,\ket{u^{\ }_{N-1}}\bigr)
= \bigl(\ket{0},\ket{1},\dots,\ket{N-1}\bigr)U
\end{equation}
recovers the basis from $U$.

If the columns of a unitary matrix are permuted, or multiplied with phase 
factors, the corresponding basis as a whole is unaffected. 
Therefore, we say that two unitary matrices are 
equivalent if and only if they can be related in this way,  
\begin{equation} \label{ekvivalens1}  
U_1 \sim U_2 \hspace{5mm} \Leftrightarrow \hspace{5mm} U_2 = 
U_1PE \,. 
\end{equation}
Here $P$ is a permutation matrix and $E$ is a diagonal unitary matrix. 

There is a second, stronger notion of equivalence in which matrices 
that are related by permutations and rephasings of rows are also regarded 
as equivalent,  
\begin{equation} \label{ekvivalens2} 
U_1 \approx U_2 \qquad \Leftrightarrow \qquad 
U_2 = E_2P_2U_1P_1E_1 \,. 
\end{equation}
In particular this means that we can present every unitary matrix in 
\emph{dephased form}: with all entries in the first row and the 
first column chosen to be real and nonnegative. 
In this respect, the second equivalence relation reminds us of how particle
physicists treat their Kobayashi--Maskawa mixing matrix. 
If the matrix is not dephased it is said to be \emph{enphased}. 
The \emph{core} of a dephased matrix is its lower right square submatrix of
size $N-1$.    

Any basis that is unbiased with respect to the computational basis is now 
represented by a complex \emph{Hadamard matrix} $H$. 
This is a rescaled unitary matrix all of whose matrix elements have unit
modulus,   
\begin{equation} 
  \label{Hadam1}
  |H_{i,j}|^2 = 1\,,\quad i,j=0,\dots,N-1  
  \quad \mbox{and} \quad 
  HH^{\dagger} = N\mathbbm{1} \,.
\end{equation}
An example which works for any $N$ is the \emph{Fourier matrix} 
whose matrix elements are 
\begin{equation}  \label{Fourier} 
[F_N]_{j,k} = \gamma_N^{jk}\,,\qquad j,k = 0, 1, \dots , N-1 \,,
\end{equation} 
with $\gamma_N = \Exp{2\pi\I/N}$ as in (\ref{eq1:AB-eigen}). 
This matrix is used to define the discrete Fourier transform. 
We recall from Sec.~\ref{sec:WSexist} that its 
existence means that pairs of MUB exist in all dimensions.  
Another example, for ${N=p^\m}$, is the Galois--Fourier matrix 
${[G_N]_{j,k}=\gamma^{j\odot k}}$ with $\gamma=\Exp{\I 2\pi/p}$ that plays a
central role in the construction of the dual basis in Sec.~\ref{sec:dual}.

Further examples include the Hadamard matrices $H_i^{(p)}$ for the
prime-dimensional bases associated with the unitary operators $XZ^i$ of
(\ref{eq1:prime1}) with ${i=0,1,\dots,p-1}$.
In accordance with (\ref{eq1:prime3}), their matrix elements are
\begin{equation}\label{eq5:primeHi1}
  \bigl[H_i^{(p)}\bigr]_{j,k}=\gamma^{-jk}\gamma^{\frac{1}{2}ij(j-1)}
\end{equation}
and their unique dephased forms
\begin{equation}
  \label{eq5:primeHi2}
   \bigl[H_i^{(p)}\bigr]_{j,k}\Bigm/\bigl[H_i^{(p)}\bigr]_{j,0}=\gamma^{-jk}
\end{equation}
are all equal to the inverse Fourier matrix.
As a set, the matrices in (\ref{eq5:primeHi1}) are equivalent to the
\emph{standard set} of \ref{sec:app1} in the stronger sense 
of (\ref{ekvivalens2}).

Our terminology is a bit unusual: 
In most of the literature a Hadamard matrix is required to have real entries
only.  
Such \emph{real Hadamard matrices} have many applications in computer science,
and in quantum information too.  
Sylvester\cite{Sy67} constructed examples for all $N = 2^{\m}$, and
Hadamard\cite{Ha93} proved that real Hadamard matrices do not exist unless 
${N= 2}$ or ${N = 4k}$. 
It was conjectured by Paley\cite{Paley} that 
they do exist in all cases not excluded by Hadamard. 
This conjecture has been verified for all ${N \leq 664}$.\cite{KT04} 
By the way, the non-existence of real 
Hadamard matrices in dimensions not divisible by 4 means that pairs of real 
MUB do not exist in real Hilbert spaces unless their dimension
equals $2$ or $4k$.\cite{BSTW05}  
Another special class of Hadamard matrices are those of 
\emph{Butson type},\cite{Bu63} 
which by definition have all matrix elements equal to rational roots of
unity. 
The Fourier matrix, the Galois--Fourier matrix, and the
matrices $H^{(p)}_i$ of (\ref{eq5:primeHi1}) are obvious examples.
For an overview of the theory of Hadamard matrices and their many
applications, consult Horadam's book.\cite{Horadam:07}

For our purposes a pair of MUB that can be transformed into each other by an 
overall unitary matrix will be regarded as equivalent.
The problem of classifying all such unbiased bases was first raised by
Kraus.\cite{Kraus:87}  
It will be convenient to distinguish ordered and unordered pairs.
Let $(M_0, M_1)$ denote an \emph{ordered pair} of MUB, with each basis 
represented as the columns of a unitary matrix. 
We identify pairs that can be transformed into each other by means of a single
unitary matrix. 
Therefore, two ordered pairs of bases will be considered equivalent, written 
\begin{equation} 
(M_0^\prime, M_1^\prime) \sim (M_0, M_1) \,,
\end{equation} 
if and only if there exist permutations $P_0, P_1$, diagonal unitary matrices 
$E_0, E_1$, and a unitary matrix $U$ such that
\begin{equation} \label{equiv1}
(UM^\prime_0P_0E_0, UM_1^\prime P_1E_1) = (M_0,M_1) \,.
\end{equation}
By using the freedom to perform overall unitary transformations, we can 
bring any pair of MUB into the \emph{standard form} $(\mathbbm{1}, H)$, 
where $H$ stands for a complex Hadamard matrix. 
But this still leaves some freedom 
to perform permutations and rephasings from the left, because 
\begin{equation}  \label{equival} 
({\mathbbm 1},H_1) \sim (EP{\mathbbm 1}P^{-1}E^{-1}, EPH_1P_1E_1) = 
({\mathbbm 1}, EPH_1P_1E_1) \,.
\end{equation}
The conclusion is that two pairs of ordered MUB, written in standard 
form, are equivalent if and only if the two Hadamard matrices are equivalent 
in the sense of (\ref{ekvivalens2}),  
\begin{equation} 
 ( \mathbbm{1}, H_1) \sim (\mathbbm{1}, H_2)  
\qquad\Leftrightarrow \qquad H_1 \approx H_2 \,.  
\end{equation}
Haagerup\cite{Ha96} devised a useful way of testing for this kind 
of equivalence. 
The matrices $H^{(p)}_i$ of (\ref{eq5:primeHi1}) are equivalent to each other.

Now consider \emph{unordered pairs} of MUB, denoted by
$\{ M_0, M_1 \}$. 
The freedom to perform overall unitary transformations implies that  
$(\mathbbm{1}, H) \sim (H^\dagger , \mathbbm{1})$. 
It follows that 
\begin{equation} 
\{ \mathbbm{1}, H\} \sim \{ \mathbbm{1},H^\dagger \} \,. 
\end{equation}
Therefore unordered pairs of MUB may be equivalent even when the 
ordered pairs are not. 
Indeed 
\begin{equation} \label{eqmub}
\{ \mathbbm{1}, H_1 \} \sim \{ \mathbbm{1}, H_2 \} \qquad
\Leftrightarrow \qquad
\left\{\begin{array}{l} 
\mbox{either $H_1 \approx H_2\,,$}\\[1ex] 
\mbox{or $H_1 \approx H_2^\dagger\,.$} \end{array} \right.
\end{equation}

\subsection{Triplets of mutually unbiased bases and circulant matrices}
\label{sec5.2}
The question when two MUB triplets, say, are equivalent is a 
little bit involved. 
In an \emph{ordered triplet} the first two bases 
are kept fixed, one of them being the standard basis and 
the other some fixed Hadamard matrix $H_1$. Then 
the freedom to perform further permutations and rephasings from the left 
is severely restricted, and we can only say that 
\begin{equation} 
 (\mathbbm{1}, H_1, H_2) \sim ( \mathbbm{1}, H_1, H_3)
\qquad \Rightarrow \qquad H_2 \approx H_3 \,.    
\end{equation}
The converse is false. 
Equivalence of \emph{unordered sets} of $k+1$ MUB can be discussed similarly,
but becomes harder and harder to check in practice because there are $k+1$
different choices of the basis to be represented by the unit matrix. 
Keeping this limitation in mind, a collection of  $k+1$ ordered MUB
$(\mathbbm{1}, H_1, \dots, H_k )$ is called \emph{homogeneous} if
all the Hadamard matrices $H_i$, $i=1,\dots, k$,  
are equivalent, and \emph{heterogeneous} if there is a pair of
inequivalent matrices among the Hadamard matrices.\cite{BBELTZ07}

Two Hadamard matrices $H_1$ and $H_2$ are said to be MUHM if 
\begin{equation}  \label{MUH} 
\frac{1}{\sqrt{N}}H_1^\dagger H_2 = H_3 \,,
\end{equation}
where $H_3$ is a Hadamard matrix too. 
This is interesting because it implies that the triplet 
$(\mathbbm{1}, H_1, H_2)$ represents three MUB. 
More generally a set of $N$ MUHM is equivalent to a collection of $N+1$ 
MUB. 

Triplets of MUB that include the Fourier matrix have an 
interesting interpretation in terms of the discrete Fourier transform. 
Given a sequence of complex numbers $z_i$, $0 \leq i \leq N-1$, its 
Fourier transform is  
\begin{equation} 
\tilde {z} = Fz \qquad \Leftrightarrow \qquad 
z = F^{\dagger}\tilde{z} \,.  
\end{equation} 
The column vector whose components are $\tilde{z}_i/\sqrt{N}$ is unbiased with
respect to the Fourier basis if and only if the sequence $z_i$ is unimodular,
$|z_i|^2 = 1$, and it is unbiased with respect to the standard basis if and
only if $\tilde{z}_i$ is unimodular. 
Hence vectors that are unbiased with respect to both the standard basis and
the Fourier basis are in one-to-one correspondence to sequences obeying 
\begin{equation} 
|z_i|^2 = |\tilde{z}_i|^2 = 1 
\end{equation}
for all values of $i$. Such sequences are called 
\emph{biunimodular}.\cite{Bjorck,Ha96} 
The first examples were in effect produced by Gauss.
When $N$ is odd they are 
\begin{equation} \label{Gaussbi} 
z_j^{(n,m)} = 
\Exp{\frac{2\pi \I}{N}(mj^2+nj) }
\end{equation}
where $m,n$ are integers modulo $N$ and the greatest common divisor of $m$ and
$N$ equals~$1$.
To prove that these sequences are biunimodular we must perform a Gauss sum, as
discussed in \ref{sec:app2}. 

Biunimodular sequences have an interesting property that emerges when 
one studies the autocorrelation function 
\begin{equation} 
\Gamma_a = \frac{1}{N}\sum_{i=0}^{N-1}\tilde{z}^*_i\tilde{z}_{a+i} \,.  
\end{equation}
An easy calculation shows that 
\begin{equation} 
\Gamma_a = \frac{1}{N}\sum_{i=0}^{N-1}|z_i|^2\gamma_N^{ai} \,.  
\end{equation}
Hence, if the sequence is biunimodular it obeys  
\begin{equation} 
\Gamma_a = \delta_{a,0} \,.  
\end{equation} 
Therefore $\tilde{z}_i$ and $\tilde{z}_{a+i}$, with $a$ fixed and 
nonzero, make up orthogonal columns. 

Any column vector can be used to define a \emph{circulant matrix}, where each 
column is obtained from the preceding one by shifting all its elements 
cyclically in such a way that all the diagonal elements are the
same.\cite{mehta77}
For an explicit example see (\ref{cirkulanter}) below. 
The matrix elements are 
\begin{equation} 
C_{ij} = \tilde{z}_{i-j\,(\mathrm{mod}\,N)} \,.  
\end{equation} 
With this definition a circulant matrix is a Hadamard matrix if and only if 
the sequence $z_i$ is biunimodular. 
It follows that all vectors unbiased with respect 
to both the standard basis and the Fourier basis can be collected into 
a set of circulant Hadamard matrices whose columns form bases that are 
unbiased with respect to the standard and Fourier bases. 
There can be no ``stray'' unbiased vectors not belonging to an unbiased
basis. 
We observe that any circulant matrix is diagonalized by the Fourier matrix. 
More precisely, if the first column of the circulant matrix $C$ is defined 
by the sequence $\tilde{z}_i$, then  
\begin{equation} 
F^\dagger C F = \mathrm{diag}(z_0, z_1, \dots , z_{N-1}) \,.  
\end{equation}
It follows that all circulant matrices commute. 
Moreover, via (\ref{MUH}) this confirms that $F$ and $C$ represent a pair of
unbiased bases.   

An example of a MUB triplet of this type is the triplet consisting of 
the eigenvectors of the three cyclic subgroups of the Heisenberg--Weyl 
group which exist in all dimensions: 
the three abelian subgroups composed of the powers of $X$, $Z$, and $XZ$
of Sec.~\ref{sec:WSprime}, for instance.
When $N = p$ is prime, one known solution for a complete set of MUHM
consists of $\mathbbm{1}$, $F$, and $N-1$ circulant matrices constructed from
the biunimodular sequences (\ref{Gaussbi}) given by Gauss; see \ref{sec:app1}. 

It is natural to ask if there are other solutions. 
In fact this is a discrete version of the Pauli problem:\cite{PauliProblem} 
Given the modulus of a function and that of its Fourier transform, 
is the function uniquely determined? 
Bj\"orck and coworkers looked into this question,\cite{Bjorck}
and they found all biunimodular sequences for ${N \leq 8}$. 
Equivalently, they found all vectors unbiased to the Fourier matrix in these
dimensions. 
For ${N = 5}$ there are $20$ vectors, all of them given by Gauss's formula, 
for ${N = 6}$ there are $48$ vectors, including $12$ given by Gauss, 
for ${N = 7}$ there are $532$ vectors, including $42$ given by Gauss, 
and for $N = 8$ there is an infinite number of solutions. 
This is true whenever $N$ contains a square factor,\cite{Faugere:01} while the
number of solutions is always finite for prime $N$.\cite{Haagerup:08}.

There are also MUHM triplets that do not include the Fourier or the
Galois--Fourier matrix.  
We will see examples later.

\subsection{Classification of Hadamard matrices of size $N \leq 5$}
\label{sec5.3}
For $N \leq 5$ the classification of all Hadamard matrices under 
the equivalence relation (\ref{ekvivalens2}) is complete. 
All complex $2 \times 2$ Hadamard matrices are equivalent to the 
Fourier matrix $F_2$, here without the $1/\sqrt{2}$ factor of
(\ref{eq1:qbitHada}), 
\begin{equation}
  F_2 =   
  \left(
    \begin{array}{cc} 
      1 & {\ }{\ }1   \\ 
      1 & -1  \\ 
    \end{array} 
  \right) \,. 
\label{f2}
\end{equation}
This is a real Hadamard matrix. 
When $N=3$, the set of all inequivalent Hadamard matrices contains the only
element
\begin{equation}
  F_3 =
  \left(
    \begin{array}{ccc} 
      1 & 1  & 1 \\ 
      1 & \gamma & \gamma^2 \\
      1 & \gamma^2 & \gamma 
    \end{array} 
  \right) \,,
\label{f3}
\end{equation}
where $\gamma =\Exp{2\pi\I/3}$ as usual.\cite{Ha96}   
When $N=5$, all complex Hadamard matrices are again equivalent
to the Fourier matrix $F_5$.\cite{Ha96}  
The known maximal sets of MUB in these dimensions, and 
indeed in all prime dimensions, consist of the standard basis together with 
equivalent Hadamard matrices of the form $H = E F$, for $p$ 
different choices of a diagonal unitary matrix $E$. 

This remark about prime dimensions is illustrated by the matrices in
(\ref{eq5:primeHi1}) except that the inverse Fourier matrix appears there, but
that is only one permutation away from the Fourier matrix itself.
Indeed, we could have the Fourier matrix just as well, simply by interchanging
the roles of $X$ and $Z$ in (\ref{eq1:prime1}) and using the eigenstates of
$X$ as the computational basis.
Since $X$ and $Z$ are unitarily equivalent, the two sets of MUB are as well.

For $N=4$ the situation is different: 
There exists a one-parameter family of equivalence classes, 
\begin{equation}
  F_4(a) =  
  \left(
    \begin{array}{rrrr}
      1 & 1 & {\ }{\ }1 & {\ }{\ }1 \\
      1 &  \Exp{\I a} & \ \ -1 &  -  \Exp{\I a} \\
      1 & -1 & {\ }{\ }1 & -1 \\
     1 &   - \Exp{\I a} & -1 &   \Exp{\I a} \\
    \end{array}
  \right) . 
\label{f4a}
\end{equation}
Hadamard\cite{Ha93} himself proved that all $N = 4$ Hadamard matrices are 
equivalent to a member of this family, for some value $0 \leq a < \pi$ of 
the phase $a$. 
If $a = \frac{\pi}{2}$, this is the standard Fourier matrix $F_4$. 
Choosing $a= 0$ produces the Galois--Fourier 
matrix $F_4(0) \approx F_2 \otimes F_2$, which is a real Hadamard matrix.

\subsection{Affine families and tensor products}
\label{sec5.4}
Why does the continuous family appear when $N = 4$? 
To analyze this question we keep $N$ arbitrary, 
multiply the matrix elements of the core of the dephased form of a given
Hadamard matrix by arbitrary phase factors, and expand to first order in the
angles: 
\begin{equation} 
H_{ij}\rightarrow H_{ij}\Exp{\I\phi_{ij}} \simeq H_{ij}
(1 + \I\phi_{ij}) \,,\qquad 1 \leq i,j \leq N-1 \,.  
\end{equation}
Then we solve the unitarity equations to first order in the angles $\phi_{ij}$. 
This is a linear system, but the number of equations exceeds the 
number of unknowns.
 
The number of free parameters in the solution of this linearized problem is
called  the \emph{defect} of the matrix $H$.
It can be explicitly determined by computing the rank of a certain 
matrix.\cite{TZ08}
The defect gives an upper bound on the dimension of any continuous set of
inequivalent Hadamard matrices containing $H$. 
If the defect is nonzero it can happen that the solution to the linearized
unitarity equations holds to all orders, in which case we speak of an
\emph{affine family} of Hadamard matrices.\cite{TZ06a}
It can also happen that the full unitarity equations are obeyed if the angles
become nonlinear functions of each other, and then we have a \emph{nonaffine
family}. 
If the defect is zero the matrix is said to be \emph{isolated}.   

It is known that the defect of the Fourier matrix is zero whenever $N$ is a 
prime number, hence there are no continuous families containing the Fourier 
matrix in these dimensions.\cite{TZ08} 
On the other hand, whenever $N=N_1N_2$ is a composite 
number one can produce continuous affine families from any
choice of Hadamard matrices in dimensions $N_1$ and $N_2$.\cite{Ha96,Di04} 
If both $N_1$ and $N_2$ are prime, $N = p_1p_2$, the construction gives a
$(p_1-1)(p_2-1)$-dimensional orbit of inequivalent Hadamard matrices including
the Fourier matrix, which explains what happens for $N = 4$.  

A more basic, and quite important, fact about tensor product Hilbert spaces is
the following:  
Let $\{H^A_1, \dots, H^A_k\}$ be a set of $k$ MUHM of size $N_A$, 
while $\{H^B_1, \dots, H^B_k\}$ denotes a set of $k$ MUHM of size $N_B$. 
Then the tensor products $\{H^A_1 \otimes H^B_1, \dots, H^A_k \otimes H^B_k\}$
form a set of  $k$ unbiased Hadamard matrices in  $\mathbf{C}^{N_A N_B}$. 
To prove this it is enough to check that condition (\ref{MUH}) is obeyed. 
When $k = 2$, we have
\begin{equation}
 \frac{1}{\sqrt{N}} 
(H^A_1 \otimes H^B_1)^{\dagger} (H^A_2 \otimes H^B_2)=
\frac{1}{\sqrt{N_A}} {H^A_1}^{\dagger} H^A_2 \otimes
\frac{1}{\sqrt{N_B}} {H^B_1}^{\dagger} H^B_2 \,.
\label{mulbas1}
\end{equation}
The matrix on the right-hand side is a Hadamard matrix by assumption, 
and we are done. Note that the pair with cross terms
$\{H^A_1 \otimes H^B_2,  H^A_2 \otimes H^B_1\}$
is also unbiased, but these Hadamard matrices are not unbiased
with respect to the pair used in (\ref{mulbas1}).
Hence by tensoring two sets of $k$ MUHM 
of dimension $N_A$ and $N_B$ we will obtain exactly $k$ MUHM of the product
structure in the extended space of size $N=N_A N_B$, but not more of them.
This is the construction mentioned at the end of Sec.~\ref{sec:WSprime} for
$N_A=2$, $N_B=3$, and $k=3$.

We say that the Hadamard matrix $H$ is \emph{separable}
if it is equivalent to any matrix of the product form
\begin{equation}
  H \approx H_{N_1} \otimes H_{N_2} \,,
 \label{sephad}
\end{equation}
where $H_{N_1}$ and $H_{N_2}$ are $N_1\times N_1$ and $N_2\times N_2$ Hadamard
matrices, respectively. 
If this is not the case, the Hadamard matrix $H$
of size $N_1N_2$ will be called \emph{entangled}. 
This concept requires that a concrete tensor product decomposition is given
beforehand. 
One may find a Hadamard matrix of size $N=12$
which is separable with respect to the $2 \times 6$ factorization,
but entangled with respect to the $3 \times 4$ splitting. 
An example is the matrix $F_2 \otimes S_6$, where $S_6$ is the Tao matrix that
will be discussed in the next section.

\subsection{Hadamard matrices of size $N = 6$}
\label{sec5.5}

${N=6}$ is the smallest composite number for which the two factors are
different, the smallest integer that is not a power of a prime.
It is the smallest dimension for which the MUB existence problem is open, and
it is also the smallest dimension for which the classification of all Hadamard
matrices is an unsolved question. 
But the hunt for ${N = 6}$ Hadamard matrices is ongoing, and was brought
to a sunny plateau recently by Karlsson.\cite{Karlsson:10a,Karlsson:10b} 

We begin by defining an $H_2$-\emph{reducible} Hadamard matrix as a 
Hadamard matrix for which all its $2\times2$ submatrices are themselves 
Hadamard matrices. 
Karlsson proved the theorem that a $6\times6$ Hadamard matrix is
$H_2$-reducible if and only if it contains a single $2\times2$ Hadamard
submatrix. 
As a simple corollary, $H_2$-reducible Hadamard matrices are very easy to
recognize:  
A $6\times6$ Hadamard matrix is $H_2$-reducible if and only if its dephased
form contains a matrix element equal to $-1$.\cite{Karlsson:10a}  
With the sole exception of the Tao matrix,\cite{Tao} 
all analytically known examples take this form. 
Moreover, the set of such Hadamard matrices belong to a three-parameter family
that was explicitly constructed by Karlsson.\cite{Karlsson:10b}

Karlsson starts with the ansatz 
\begin{equation} 
H = \left( \begin{array}{cc|cc|cc} 1 & 1 & 1 & 1 & 1 & 1 \\ 
1 & - 1 & z_1 & -z_1 & z_2 & - z_2 \\ \hline 
1 & z_3 & \bullet & \bullet & \bullet & \bullet \\ 
1 & - z_3 & \bullet & \bullet & \bullet & \bullet \\ 
\hline 1 & z_4 & \bullet & \bullet & \bullet & \bullet \\ 
1 & - z_4 & \bullet & \bullet & \bullet & \bullet 
\end{array} \right), 
\end{equation}
where the $z_i$ are phase factors and the $2\times2$ blocks that have not been 
written out are guaranteed to be Hadamard matrices. 
We used the fact that four phase factors that add to zero form a rhombus in
the complex plane, which is why  they pair up in the way indicated. 
This ansatz is rewritten as 
\begin{equation} \label{BKAnsatz} 
H = \left( \begin{array}{c@{\quad}c@{\quad}c} F_2 & Z_1 & Z_2 \\[1ex]
 Z_3 & \frac{1}{2}Z_3AZ_1 & \frac{1}{2}Z_3BZ_2 \\[1ex] 
Z_4 & \frac{1}{2}Z_4BZ_1 & \frac{1}{2}Z_4AZ_2 \end{array} \right). 
\end{equation}
This matrix will be unitary if and only if 
\begin{equation} 
A + B = F_2 \,,\qquad A - B = \sqrt{3}F_2\I\Lambda \,,\qquad 
\Lambda^\dagger \Lambda = \mathbbm{1} \,,\qquad 
\Lambda^\dagger = \Lambda \,. 
\end{equation}
The unitary $2\times2$ matrix $\Lambda$, 
and \emph{a fortiori} the matrices $A$ and $B$, will 
therefore depend on two free parameters that parameterize a sphere 
--- which can be thought of as the equator of the group $SU(2)$. 
We find 
\begin{equation} 
A = \left( \begin{array}{cc} A^{\ }_{11} & A^{\ }_{12} \\ 
{A}^*_{12} & -{A}^*_{11} \end{array} \right)
\end{equation} 
with
\begin{equation}
  A_{11} = -\frac{1}{2} + \I\frac{\sqrt{3}}{2}(x_1+ \I x_2 + x_3) \,,\qquad 
  A_{12} = -\frac{1}{2} + \I\frac{\sqrt{3}}{2}(x_1 -\I x_2 - x_3) 
\end{equation}
and
\begin{equation} 
   B(x_1,x_2,x_3) = A(-x_1,-x_2,-x_3) \,, 
\end{equation}
where the three real parameters $(x_1, x_2, x_3)$ are constrained by 
\begin{equation}
   x_1^2 + x_2^2 + x_3^2 = 1 \,. 
\end{equation}
Actually, another solution is ${\Lambda = \pm \mathbbm{1}}$, but 
in the end this gives only Hadamard matrices that are equivalent to
one of the above.

It remains to ensure that all matrix elements are unimodular. 
The conditions for this can be written in an elegant form using M\"obius 
transformations that take the unit circle to the unit circle. 
Indeed 
\begin{equation} \label{Mobius1} 
z_3^2 = \mathcal{M}_A(z_1^2)= \mathcal{M}_B(z_2^2) \,,\qquad
z_4^2 = \mathcal{M}_A(z_2^2)= \mathcal{M}_B(z_1^2)\,, 
\end{equation}
where 
\begin{equation} 
\mathcal{M}(z) = \frac{\alpha z - \beta}{\beta^*z - \alpha^*}
\end{equation}
with the respective parameter values
\begin{equation}\label{Mobius3} 
\alpha_A = A_{12}^2\,,\quad\beta_A = A^2_{11}\,,\qquad 
\alpha_B = B_{12}^2\,,\quad\beta_B = B^2_{11} 
\end{equation}
for $\mathcal{M}_A$ and $\mathcal{M}_B$. 
Provided that at least one of these M\"obius transformations is 
non-degenerate these equations can be solved (up to a sign) for 
$z_2,z_3,z_4$ in terms of $z_1$, say, so together they contribute only 
one real parameter to the family of $H_2$-reducible Hadamard matrices. 
There are four points where both transformations are degenerate, namely 
\begin{equation} 
(x_1,x_2,x_3) = (0,0, \pm 1) \quad\mbox{and}\quad 
(x_1,x_2,x_3) = (\pm 1, 0 , 0) \,. 
\end{equation}
Hence the parameter space has three dimensions, and can be roughly 
described as a circle bundle over a two-dimensional sphere, but with four 
special points where the circle has been blown up to a torus.

It would be desirable to work out exactly what choices of the three parameters 
lead to equivalent Hadamard matrices. 
This problem has been solved only partially. 
Changing the sign of any $z_i$ leads to equivalent Hadamard matrices.
It is also known that the transformations 
\begin{equation} 
   (x_1,x_2,x_3) \rightarrow (-x_1, -x_2,x_3) \rightarrow 
   (x_1,-x_2,-x_3) \rightarrow (-x_1,-x_2,-x_3) 
\end{equation}
lead to equivalent Hadamard matrices if supplemented by appropriate 
transformations of the phase factors $z_i$. 
Hence at most one octant of the sphere is needed in the parameterization. 

It remains to describe some examples of special interest. 
The first family to be discovered was the affine 
\emph{Fourier family}\cite{Ha96}
\begin{equation} {F}(a,b) = 
\left( \begin{array}{rrrrrr} 1 & 1 & 1 & 1 & 1 & 1 \\ 
1 & \gamma z_1 & \ \gamma^2z_2 & \hphantom{\ z_2}\gamma^3 
& \ \gamma^4z_1 & \ \gamma^5z_2 \\
1 & \gamma^2 & \gamma^4 & 1 & \gamma^2 & \gamma^4 \\
1 & \ \gamma^3z_1 & z_2 & \gamma^3 & z_1 & \gamma^3z_2 \\
1 & \gamma^4 & \gamma^2 & 1 & \gamma^4 & \gamma^2 \\
1 & \gamma^5z_1 & \gamma^4z_2 & \gamma^3 & \gamma^2z_1 & \gamma z_2 
\end{array}\right) 
\label{Fourier6} 
\end{equation}
with $\gamma=\gamma_6=\Exp{\I2\pi/6}$ here
while $z_1 = \Exp{\I2\pi a}$ and $z_2 = \Exp{\I2\pi b}$.
In the construction the two free parameters $a,b$ arise because a 
six-dimensional space can be written as a tensor product. For this subfamily 
the equivalence problem has been fully understood.\cite{BBELTZ07} Thus there 
is a discrete group acting on the square, or torus, parameterized by $a,b$.
It is a semi-direct group of a dihedral group with a discrete
translation group. 
This dihedral group is the symmetry group of a regular hexagon.  
The result is that the original square is divided into
$144$ equivalent triangles of equal area. 
One of them has corners at $(0,0)$, $(\frac{1}{6},0)$ and 
$(\frac{1}{6},\frac{1}{12})$, and every affine $F(a,b)$ is equivalent to one
for which $(a,b)$ lies within this triangle.

The twin family of transposed matrices ${F}^{\mathrm{T}}(a,b)$
can be parameterized in an analogous way. 
These two families intersect at the Fourier matrix itself,  
\begin{equation} 
F_6 = F(0,0) = F^{\mathrm{T}}(0,0) \approx F_2 
\otimes F_3 \approx  F_3 \otimes F_2 \,.  
\end{equation}
The equivalence happens because the factors of $6 = 2\cdot 3$ are 
relatively prime; see Ref.~\refcite{Ta07} for a general discussion of
equivalences between tensor products of Fourier matrices. 

In the family of $H_2$-reducible Hadamard matrices one finds the Fourier 
family at the special point $(x_1, x_2,x_3) = (0,0,1)$, while the transposed 
Fourier family sits at $(x_1,x_2,x_3) = (1,0,0)$; recall that both 
M\"obius transformations of (\ref{Mobius1})--(\ref{Mobius3}) 
become degenerate at these points. 
Curiously the one parameter family $F^{\mathrm{T}}(0,b)$ also sits at $(0,0,1)$, 
and similarly $F(a,0)$ also sits at $(1,0,0)$. 

One more affine family is known, namely the Di\c{t}\u{a} family,\cite{Di04} 
which in dephased form is given by   
\begin{equation} 
D(a) = 
\left( \begin{array}{rrrrrr} 
1 & 1 & 1 & 1 & 1 & 1 \\
1 & - 1 & \I  & - \I  & - \I  & \I  \\
1 & \hphantom{-z^*}\I  & - 1 & \I z & - \I z & - \I  \\
1 & - \I  & \I z^* & - 1 & \hphantom{-z^*}\I  & - \I z^* \\ 
1 & - \I  & - \I z^* & \hphantom{-z^*}\I  & - 1 & \I z^* \\
1 & \I  & - \I  & - \I z & \I z & - 1 \end{array} 
\right) \qquad\mbox{with $z=\Exp{\I2\pi a}$}\,.  
\label{Dita2} 
\end{equation}
We obtain all inequivalent examples if we impose the restriction 
$-\frac{1}{8}< a \leq \frac{1}{8}$. 
It includes the Butson-type matrix $D_6(0)$, known as the Di\c{t}\u{a} matrix, 
and composed of fourth roots of unity. 
This can be found in several different places within the three-parameter
family, reflecting the fact that the equivalence problem for the latter is
unsolved. 
One possibility is to set $x_1 = x_2 = x_3$, in which case the Di{t}\u{a} 
family is parameterized by the phase factor $z_1$. 

Another Hadamard matrix of special interest is the circulant 
matrix\cite{Bjorck} 
\begin{equation}
  C_6 = \left( \begin{array}{rrrrrr} 
1 & \I d & - d & - \I & - d^* & \I d^* \\
\I d^* & 1 & \I d & - d & - \I & - d^* \\
- d^* & \I d^* & 1 & \I d & - d & -\I \\
- \I & - d^* & \I d^* & 1 & \I d & - d \\
- d & - \I & - d^* & \I d^* & 1 & \I d \\
\I d & - d & - \I & - d^* & \I d^* & 1 \end{array} \right),
\label{Bjorck} 
\end{equation}
where
\begin{equation} 
d = \frac{1-\sqrt{3}}{2} + \I\sqrt{\frac{\sqrt{3}}{2}}\,,\quad d^*d=1\,. 
\label{d} 
\end{equation}
The unimodular number $d$ solves the equation $d^2 - (1-\sqrt{3})d + 1 = 0$. 
It is known that every circulant Hadamard matrix is equivalent to either $F_6$
or $C_6$.

\begin{figure}[t]
\centerline{\includegraphics{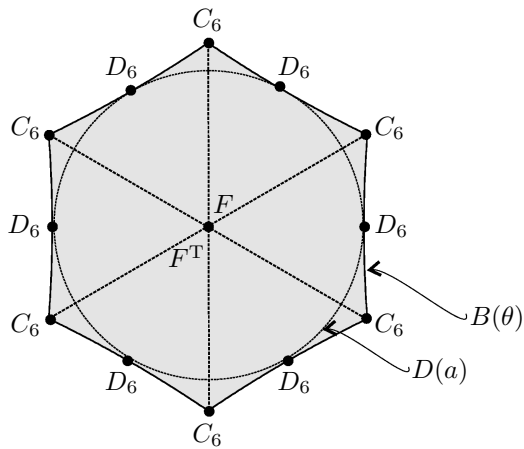}}
\caption{Sz{\"o}ll{\H{o}}si's two-dimensional family of $N=6$ complex Hadamard
  matrices interpolates between the generalized Fourier matrix 
  $F= F\bigl(\frac{1}{6},0\bigr)$
  and the  hermitian  family $B(\theta)$, which includes $C_6$ and $D_6$. 
It is parametrized by the common interior of two deltoids. 
There are actually several ``leaves'' over the interior, and it is divided
into six equivalent sectors. 
Di\c{t}\u{a}'s affine family $D(a)$, see (\ref{Dita2}), is represented by a
circle inscribed into the figure. }
\label{fighad7}
\end{figure}

Before Karlsson's work several non-linear subfamilies of Hadamard matrices 
were known. 
The first to be found (by Beauchamp and Nicoara\cite{BN08}) was 
the one-parameter family $B(\theta )$ containing all Hadamard matrices
equivalent to a hermitian matrix. 
It interpolates between $C_6$ and $D_6(0)$ in a complicated way. 
It is included as the boundary of a two-parameter family of \emph{bicirculant}
Hadamard matrices found by Sz{\"o}ll{\H{o}}si.\cite{Sz09}  
By definition, a bicirculant matrix is divided into four blocks of equal size,
each block being a circulant matrix in itself. 
Sz{\"o}ll{\H{o}}si's family contains all bicirculant 
matrices with two independent blocks only, according to the pattern  
\begin{equation} \label{eq5:bicirc}
X_6 = \left( \begin{array}{rr} A & B \\ B^\dagger & - A^\dagger \end{array} 
\right) \,, 
\end{equation}
where $H$ is bicirculant because $A$ and $B$ are circulant,  
\begin{equation} 
A = \left( \begin{array}{ccc} a & b & c \\ c & a & b \\ b & c & a 
\end{array} \right) \,, \qquad 
B = \left( \begin{array}{ccc} d & e & f \\ 
f & d & e \\ e & f & d \end{array} \right) \,.  
\label{cirkulanter} 
\end{equation}
The individual entries are unimodular phase factors. 
Since any two circulant matrices commute the unitarity conditions are quite 
simple to state. 
Sz{\"o}ll{\H{o}}si ended up with an appealing picture of the resulting
two-parameter family.  
In the complex plane the parameter space is bounded by two \emph{deltoids}
related by a reflection. 
By definition a deltoid is a 3-hypocycloid, that is the curve traced out if
you place the tip of your pen at the rim of a wheel, and then let this wheel
roll inside a larger wheel whose inner rim has three times the radius of the
rolling wheel; see Fig.~\ref{fighad7}. 
The picture is that of an umbrella, and in 
fact of two superposed umbrellas because above each point there are two 
inequivalent matrices that can be represented as the transposes of each other. 
Thus we have two two-parameter families $X_6(\alpha )$ 
and $X_6^{\rm T}(\alpha )$ coming together at their common boundary. 
One can easily check that they are subfamilies of Karlsson's family.

Another one-parameter family of symmetric Hadamard matrices\cite{MS08} was 
extended to a two-parameter family by Karlsson.\cite{Karlsson:09} 
This family can be obtained by setting $z_1 = z_2$ and $z_3 = z_4$ in the
ansatz (\ref{BKAnsatz}). 
Interestingly it is then possible to solve explicitly 
for the matrices $A$ and $B$ in terms of the phases $z_1$ and $z_3$. 

The elegance of the available constructions is very encouraging, but they are 
not the end of the story. It has been conjectured\cite{BBELTZ07} that a 
four-parameter family exists. One reason for this is 
that the defect of all included matrices has been found to be four, whenever 
it has been checked,\cite{BBELTZ07} and moreover there is by now strong 
numerical evidence for the conjecture.\cite{SNS08,PC:Karlsson2010} 
Yet, the set of inequivalent ${N = 6}$ Hadamard matrices is disconnected, 
because there is also an isolated matrix that does not belong to any
continuous family.  
This is a symmetric Butson-type Hadamard matrix composed of third roots of
unity only, known as \emph{Tao's matrix}.\cite{Moorhouse,Tao}  
It is isolated because its defect vanishes. 
One does not know if other isolated matrices exist.

\subsection{Hadamard matrices for $N \geq 7$}
\label{sec5.6}
Some general facts are known also in higher dimensions, in 
particular affine families stemming from known Hadamard matrices have been
much studied.  
As we have already mentioned, the Fourier matrix is an isolated matrix if 
and only if $N$ is a prime number.\cite{TZ08}
When $N$ is a power of a prime, $N = p^\m$, 
all affine orbits stemming from the Fourier matrix are explicitly known. The 
dimension of these orbits reads $d=p^{\m-1}[(p-1)\m-p] + 1$
and is equal to the defect of $F_N$.\cite{TZ08} 
It is also known that every real Hadamard matrix admits an affine
orbit if $N \geq 12$.\cite{Sz08} 
In prime dimensions, affine orbits cannot pass through the 
Fourier matrix, but Petrescu found an example for $N = 7$ which contains a 
Butson-type matrix built from sixth roots of unity.\cite{Pe97} 
 
All circulant Hadamard matrices up to $N \leq 9$ have been found.\cite{Bjorck} 
When $N$ contains a square factor this includes a continuous
family,\cite{Faugere:01} 
whereas the number is finite for all prime $N$.\cite{Haagerup:08} 
Many block circulant examples are also known.\cite{Craigen}
Special methods for constructing Hadamard matrices include one based on tiling
abelian  groups,\cite{MRS07} 
one based on $N$ equiangular vectors in $N/2$ dimensions,\cite{HP04} 
as well as a method for constructing Hadamard matrices of size $N$ from
matrices of size $N/2$.
This gives a rich supply of examples with ${N=8}$.\cite{Di08,Dita:10}
And, of course, there are many ad hoc constructions. 
A catalog of known Hadamard matrices for $N\le 16$ is available,\cite{TZ06a} 
also as an updated Internet version.\cite{TZ06b}

\subsection{All mutually unbiased bases for $N \leq 5$}
\label{sec:allMUB-Nle5}
Since we know that the Hadamard matrix in dimensions $2$, $3$, and $5$ 
is unique up to equivalences it seems reasonable to expect that the maximal 
set of MUB is also unique up to an overall unitary transformation. 
When $N = 2$ a maximal set of MUB can be thought of --- as we did in
Sec.~\ref{section0} --- as a regular octahedron inscribed in the Bloch sphere,
and the uniqueness follows from the fact that all such octahedra are related
by a rotation, corresponding to a unitary transformation in the $N = 2$
Hilbert space.  
Equivalently, there is the observation of Sec.~\ref{sec:WSprime} that q-bit
operators are associated with directions in $\mathbf{R}^3$ and complementary
observables  must refer to orthogonal directions.

Uniqueness continues to hold for $N=3$ and $N=5$, although a complicated 
calculation is needed to see this.\cite{Kostrikin} 
The explicit form of unbiased Hadamard matrices forming one maximal 
set of MUHM for any prime ${N=p}$ is provided in \ref{sec:app1}.
Another, equivalent, maximal set is composed of the matrices $H^{(p)}_i$ in
(\ref{eq5:primeHi1}). 

The case $N = 4$ is more interesting because of its one-parameter 
family of inequivalent Hadamard matrices. 
It is also simple enough that 
the calculations can be done by hand.\cite{BeBrWe} 
We begin by looking for ordered MUB triplets of the form 
$(\mathbbm{1}, F_4(a), H)$, where $F_4(a)$ is written in the standard form
(\ref{f4a}) and $H$ is some Hadamard matrix obtained by enphasing $F_4(a)$,
possibly with its rows permuted. 
After going through all the possibilities, one finds that there are exactly
three families of ordered triplets of MUB, with $2$ or $1+2$ free parameters
each:  
\begin{equation} 
\bigl(\mathbbm{1}, F_4(a), H^{(1)}(\phi_1; \alpha_1) \bigr)\,,\quad
\bigl(\mathbbm{1}, F_4(0), H^{(2)}(\phi_2; \alpha_2) \bigr)\,,\quad 
\bigl(\mathbbm{1}, F_4(0), H^{(3)}(\phi_3; \alpha_3) \bigr) 
\,.  
\end{equation}
The third members of these triplets are given by 
\begin{eqnarray} 
H^{(1)}(\phi_1; \alpha_1) &=& \left( \begin{array}{cccc} 
1 & 1 & 1 & 1 \\ 
   \Exp{\I\alpha_1} & \Exp{\I(\alpha_1 + \phi_1 )} 
& -\Exp{\I\alpha_1} & -\Exp{\I(\alpha_1 + \phi_1 )} \\ 
-1 & 1 & -1 & 1 \\ 
   \Exp{\I\alpha_1} & -\Exp{\I(\alpha_1 + \phi_1 )} 
& -\Exp{\I\alpha_1} & \Exp{\I(\alpha_1 + \phi_1 )} 
\end{array} \right) ,\nonumber\\
H^{(2)}(\phi_2; \alpha_2) &=& \left( \begin{array}{cccc} 
1 & 1 & 1 & 1 \\ 
   \Exp{\I\alpha_2} & \Exp{\I(\alpha_2 + \phi_2 )} 
& -\Exp{\I\alpha_2} & -\Exp{\I(\alpha_2 + \phi_2 )} \\ 
 -\Exp{\I\alpha_2} & \Exp{\I(\alpha_2 + \phi_2 )} 
& \Exp{\I\alpha_2} & -\Exp{\I(\alpha_2 + \phi_2 )} \\ 
1 & -1 & 1 & - 1 
\end{array} \right),\nonumber\\
H^{(3)}(\phi_3; \alpha_3) &=& \left( \begin{array}{cccc} 
1 & 1 & 1 & 1 \\ 
1 & - 1 & 1 & - 1 \\ 
 -\Exp{\I\alpha_3} & -\Exp{\I(\alpha_3 + \phi_3 )} 
& \Exp{\I\alpha_3} & \Exp{\I(\alpha_3 + \phi_3 )} \\ 
   \Exp{\I\alpha_3} & -\Exp{\I(\alpha_3 + \phi_3 )} 
& -\Exp{\I\alpha_3} & \Exp{\I(\alpha_3 + \phi_3 )} 
\end{array} \right),  
\end{eqnarray}
respectively.
Regarded as unordered triplets, the last two are actually special cases of the
first, so there is a single $1+2$ parameter family of unordered triplets. 

It is straightforward to check that none of these families contains a 
quartet of MUB.
The only way to obtain a quartet is to pick the third member of two 
different ordered triplets. 
Moreover, there is only one way in which this can be done, namely to set  
\begin{equation} 
\alpha_1 = \alpha_2 = \alpha_3 = \frac{\pi}{2} \,, 
\qquad a = \phi_1 = \phi_2 = \phi_3 = 0 \,.  
\end{equation} 
This leads to the standard solution for a maximal set of MUB, which 
is thereby shown to be unique up to an overall unitary transformation. 
For $N=5$ there are two inequivalent triplets.\cite{BeBrWe}

\begin{table}[tb]
\tbl{%
One choice for the five MUB of a two--q-bit system
($N=2^2$) can be characterized as the bases of common eigenstates to five sets
of three commuting period-2 observables each, or as the eigenstate bases of
five period-4 observables. 
Bases 0--2 consist of product states; bases 3 and 4 consist of maximally
entangled states. 
Together with the identity ${\mathbf{1}\otimes\mathbf{1}}$ and phase factors
$\pm1$, $\pm\I$, the 15 observables in the middle column constitute the
two--q-bit Heisenberg--Weyl group; their 15 expectation values determine the
state of the two--q-bit system uniquely. 
The five unitary observables in the right column are pairwise complementary; see
Sec.~\ref{sec2.4}.  
The period-5 unitary transformation of (\ref{eq5:period-5}) permutes the five
period-4 observables cyclically: ${0\to1\to2\to3\to4\to0}$.
\label{tbl:qbitpair}}
{\rule{4em}{0pt}\begin{tabular}{cl@{\ \ }r@{\ \ }cl}\toprule
  & \multicolumn{3}{c}{Set of three commuting} &
 \multicolumn{1}{c}{Complementary}\\
Basis & \multicolumn{3}{c}{period-2 observables} &
 \multicolumn{1}{c}{period-4 observables} 
\\ \colrule 
0 & $\sigma_z\otimes\mathbf{1}$ & $\mathbf{1}\otimes\sigma_z$ 
  & $\sigma_z\otimes\sigma_z$ 
  & $\ds\frac{1+\I}{2}(\sigma_z\otimes\mathbf{1}
                       -\I\,\mathbf{1}\otimes\sigma_z)$ \\[2ex] 
1 & $\sigma_x\otimes\mathbf{1}$ & $\mathbf{1}\otimes\sigma_x$ 
  & $\sigma_x\otimes\sigma_x$   
  & $\ds\frac{1+\I}{2}(\mathbf{1}\otimes\sigma_x
                       -\I\,\sigma_x\otimes\sigma_x)$  \\[2ex]
2 & $\sigma_y\otimes\mathbf{1}$ & $\mathbf{1}\otimes\sigma_y$
  & $\sigma_y\otimes\sigma_y$ 
  & $\ds\frac{1+\I}{2}(\sigma_y\otimes\mathbf{1}
                       -\I\,\mathbf{1}\otimes\sigma_y)$ \\[2ex]   
3 & $\sigma_x\otimes\sigma_y$ & $\sigma_y\otimes\sigma_z$ 
  & $\sigma_z\otimes\sigma_x$ 
  & $\ds\frac{1+\I}{2}(\sigma_y\otimes\sigma_z
                       -\I\,\sigma_z\otimes\sigma_x)$ \\[2ex]
4 & $\sigma_y\otimes\sigma_x$ & $\sigma_z\otimes\sigma_y$ 
  & $\sigma_x\otimes\sigma_z$ 
  & $\ds\frac{1+\I}{2}(\sigma_z\otimes\sigma_y
                       -\I\,\sigma_y\otimes\sigma_x)$ \\
\botrule
\end{tabular}\rule{4em}{0pt}}
\end{table}

Since $N = 4$ gives the Hilbert space for two q-bits it is interesting to ask 
how the MUB behave with respect to entanglement. 
In fact three of them can be chosen to consist of separable states only, 
while the remaining two are constructed out of maximally entangled 
Bell states.\cite{Zeil,Qinfo}
One can understand these five MUB as bases composed of the common
eigenstates to three two--q-bit observables with period~2 or, equivalently, as
the eigenstate bases of pairwise complementary period-4 operators; see
Table~\ref{tbl:qbitpair}.\cite{EngMet} 
Alternatively we can use the magic basis for the two--q-bit Hilbert space, 
so that real vectors are maximally entangled.\cite{magi} 
It is easy to see that there is a MUB triplet consisting of three 
real bases, although this is a triplet that cannot be extended to a maximal 
set. 
Incidentally the three real MUB form a maximal set for a real
four-dimensional Hilbert space, and this observation is closely related to the
existence of a platonic body in $\mathbf{R}^4$, called the 24-cell.  
The Segre configuration (mentioned in Sec.~\ref{secaffin}) has an analog
known as Reye's configuration: 
If we pick a pair of vectors from two distinct bases, there is a unique vector
in the third basis which is linearly dependent on the first
two.\cite{Aravind:00} 

We note that the unitary transformation that is defined by the mapping
\begin{equation}
  \label{eq5:period-5}
  \bigl(\sigma_x\otimes\mathbf{1},\sigma_z\otimes\mathbf{1},
        \mathbf{1}\otimes\sigma_x,\mathbf{1}\otimes\sigma_z\bigr)
  \longrightarrow
  \bigl(\sigma_y\otimes\sigma_y,\textbf{1}\otimes\sigma_x,
        \sigma_y\otimes\textbf{1},\sigma_x\otimes\sigma_x\bigr)
\end{equation}
is of period~5 and permutes the period-4 observables in the last column of
Table~\ref{tbl:qbitpair} cyclically, which is why the five bases are listed in
this particular order.
We have here an illustration of the
observation\cite{Go07,Wootters+1:07,Kern+2:09} that, in the case
of $\m$--q-bit systems ($N=2^\m$), a maximal set of ${N+1}$~MUB can be generated
from the computational basis by repeated application of a suitable unitary
operator with period~${N+1}$.
When $N=p^{\m}$ with ${p=3\ (\mbox{mod}\ 4)}$ this can be done with an
anti-unitary operator.\cite{Appleby:09b}.

\subsection{Triplets of mutually unbiased bases in dimension $6$} 
\label{sec5.8}
Since a complete list of all possible sets of five MUB in $N = 4$ 
can be constructed by hand, one might guess that the case of $N = 6$ could
easily be settled with a computer. 
Numerical searches have been performed by many, but it seems that the first
published account is the one by Zauner,\cite{Zauner} 
who was led to conjecture that at most three MUB can be found. 
By now the evidence for his conjecture is overwhelming, but not quite
conclusive, which tells us something about how fast the complexity of a
Hilbert space grows with dimension.  

The problem of classifying all pairs of MUB is equivalent to the problem of 
classifying Hadamard matrices. 
With partial results on this problem available,
one can go on to ask what pairs can be extended to triplets of MUB, and in how 
many ways this can be done. 
For the Fourier family of Hadamard matrices (and its transpose), 
a clear picture has emerged.\cite{BW09,JMM09,Brthesis} 
There is very strong evidence that the number of kets unbiased to the bases
represented by the pair $\bigl(\mathbbm{1},F(a,b)\bigr)$ equals $48$,
regardless of the values taken by the parameters $a,b$, with $F(a,b)$ as
introduced in (\ref{Fourier6}).
For generic values of the parameters these vectors can be collected into 
eight different unbiased bases which, however, are not MU. 
Some values of the parameters are special in this regard: 
The Fourier matrix $F(0,0)$ admits $16$ unbiased bases,\cite{Grassl}
and $F(\frac{1}{6}, 0)$ admits up to 70. Note that these 
values of the parameters are special also because they correspond to singular 
points in the moduli space of all Hadamard matrices of this type, and that 
$F(\frac{1}{6},0)$ is very special because it is also included in the
bicirculant family  $X_6(\alpha )$. 

The evidence consists in computer calculations for a large number of members 
of the family\cite{BW09}, and also a proof that there exists a vicinity
of $(a,b) =(0,0)$ where the number of unbiased vectors is constant\cite{JMM09}
and equal to $48$.  
In one version, the procedure begins with the observation that the condition
for a ket to be unbiased with respect to the bases pair corresponding to 
$({\mathbbm 1}, H)$, for some Hadamard matrix $H$, is a set of multivariate
polynomial equations that can in principle be brought to ``diagonal'' form (in
the way one would do Gauss elimination for linear equations) by means of
Gr\"obner bases for the polynomials. 
In the end polynomial equations in single variables are solved to high
enough accuracy.  
The procedure works nicely for all of the affine families, while results for
the nonaffine families are somewhat uncertain because of more stringent
demands on computer memory.

\begin{table}[tb]
\tbl{%
Number $N_\mathrm{v}$ of kets unbiased with respect to a given complex
Hadamard matrix and the number $N_\mathrm{t}$ of bases (not mutually unbiased)
which can be formed out of them, obtained for generic values of the parameters
$a$ and $b$ as well as for ${|a_1|<a_*<|a_2|\leq\frac{1}{8}}$.
\label{tbl:triplets}}  
{\rule{4em}{0pt}\begin{tabular}{lccccccc} \toprule
 Matrix & $F(a,b)$ & $F(0,0)$ & $F(\frac{1}{6},0)$ & $D(0)$ 
        & $D(a_1)$ & $D(a_2)$ & $S_6$ 
\\ \colrule
    $N_{\rm v}$ & $48$ & $48$ & $48$ & $120$ & $120$ & $48$ & $90$ \\
    $N_{\rm t}$ & $8$ & $16$ & $70$ & $10$ & $4$ & $4$ & $0$\\
\botrule 
\end{tabular}\rule{4em}{0pt}}
\end{table}

In Table~\ref{tbl:triplets}, we show the number $N_\mathrm{v}$ of 
kets unbiased to the computational basis and one additional listed basis, as
well as the number $N_\mathrm{t}$ of bases (or triplets of MUB) that can be
formed from these vectors.\cite{BW09}  
The results for the twin families $F(a,b)$ and $F^\mathrm{T}(a,b)$ are the
same, and hence results for the latter are not given explicitly. 
For the Di\c{t}\u{a} family $D(a)$ of (\ref{Dita2}) one finds that the result
depends on the parameter value; 
if ${|a|<a_*\simeq 0.0177}$ there are $120$ unbiased 
vectors, and if ${a_*< |a| \leq \frac{1}{8}}$ there are $48$ of them. 
This takes care of all inequivalent values of $a$. 
Note that the Butson-type matrix $D(0)$ is quite exceptional; 
moreover, in this case the phases that define the unbiased kets 
are known exactly. 
The isolated Butson-type matrix $S_6$ does not admit even a single  
triplet of MUB.

Exactly what makes the unbiased vectors collect into bases in some, but not 
all cases, is imperfectly understood. 
For triplets of MUB involving $F(0,0)$, we have given the explanation in terms
of the discrete Fourier transform,\cite{BBELTZ07} and for the affine family
$F(a,b)$ some partial understanding exists.\cite{JMM09}

Some continuous families of triplets of MUB are known. 
In particular, Zauner showed that any bicirculant Hadamard matrix gives rise
to a triplet because (\ref{MUH}) can be solved for $H_1$ and $H_2$ if
$H_3$ is a specified bicirculant Hadamard matrix.\cite{Zauner}   
In fact, the entire set of triplets in $N = 4$ dimensions can 
be shown to arise in this way. 
For $N = 6$, this means that Sz{\"o}ll{\H{o}}si's bicirculant family
$X_6(\alpha)$ gives rise to a two-parameter set of triplets. 
Another continuous family of the form $\bigl(\mathbbm{1}, F(0,b(t)),
H(t)\bigr)$ has been constructed by Jaming \textit{et  al.};\cite{JMM09} 
the third member of their triplet family belongs to the Fourier family.

\subsection{A maximal set of mutually unbiased bases when $N = 6$?}
\label{sec5.9}
We now ask whether any of the explicitly known triplets of MUB can be 
extended to a quartet. 
The answer is that none of them can,\cite{Grassl} and the failure can be 
expressed quantitatively.\cite{BBELTZ07}  
If a quartet involving the Fourier matrix did exist, one would be able to
find a pair of bases among the 16 bases unbiased with respect to 
$(\mathbbm{1},F)$ such that the Grassmannian distance between them is equal to
unity.
However, the best one can do is $D^2_c = 0.93$.
Remembering that a random pair of bases are 
situated at a distance given by $D^2_c = 0.86$, this is not impressive. 
Other pairs of MUB have not been treated in quite that much detail, but
Jaming \textit{et al.} recently proved that no quartets of MUB including any
member of the Fourier family $F(a,b)$ can exist.\cite{JMM09} 
The proof involves approximations of the elements of the columns that
represent the kets by rational roots of unity, exhaustive computer
searches, and careful estimates of the errors involved.

Direct numerical searches for maximal sets have been carried
out,\cite{Zauner,Br08} but relatively 
few such investigations have been published. 
Butterley and Hall\cite{BH07} have conducted a 
search based on the minimization of a suitable function. 
The minimization proceeds by picking a point at random in some parameter
space, and changing it until a minimum is reached. 
The problem is that this minimum may not be the global minimum, 
so the procedure could miss its target even if the target --- in this case a
quartet of MUB --- is there. 
Indeed, the success rate was 60.4\% when $N = 5$, 
but only 0.9\% when $N =7$. 
No quartets were found for $N = 6$. 
This result is suggestive but not definitive. 

Brierley and Weigert\cite{BW08} concentrated on finding 
\emph{MU constellations}, defined as up to $N+1$ sets of orthogonal kets that
are MU with respect to each other. 
It is not required that the sets have $N$ members. 
In fact, for $N = 6$ they were able to find seven sets with two members each. 
This constellation is denoted by $\{ 2^7\}_6$, while a quartet of MUB is the
constellation $\{ 5^4\}_6$, in a notation that should now be obvious (given
the fact that five orthogonal vectors automatically define a sixth, unbiased to
all vectors that are unbiased with respect to the original five). 
They then proceeded to search for constellations that necessarily exist if the
quartet exists, such as $\{ 6, 3, 3, 3\}_6$, $\{ 6, 4, 3, 2\}_6$, and so on.  
Altogether they found $17$ examples of such constellations for which 
their success rate in dimension $6$ was zero. 
The advantage of the procedure is that the parameter spaces in which the
search is conducted are comparatively small --- in the two quoted examples
there are $40$ parameters, as opposed to $70$ parameters for a quartet of MUB. 
The success rates for similar calculations in $N = 7$ were high.    

Hence we feel that the answer to the question in the title of this subsection
must be ``no.'' 
It is fair to say, however, that a structural understanding of this negative
result is missing. 
A precise translation into Euler's problem of the $36$ officers (see
Sec.~\ref{secaffin}) could provide this --- if there is one, and if the
translation provides a structural understanding of the latter problem.

\subsection{Heisenberg--Weyl group approach for $N=6$}
\label{sec5.10}
We have seen how the abelian subgroups of the Heisenberg--Weyl group identify
the maximal set of MUB if $N$ is a power of a prime, whereby the construction
of the MUB relies heavily on the properties of the Galois field with $N$
elements. 
As noted earlier, this construction is not applicable for other values of $N$,
simply because there is no corresponding Galois field.
The failure of this approach, therefore, says nothing about the existence of
maximal sets of MUB in non--prime-power dimensions. 
As noted repeatedly, this existence problem is open, even in the most
intensely studied case of ${N=6}$.\cite{Archer,Grassl,JMM09,BBELTZ07,BH07,BW08}

Since the Galois--Fourier construction of the Heisenberg--Weyl group, which
works so well for prime power dimensions, cannot be
applied for ${N=6,10,12,14,\dots}$, one could try to repeat the procedure with
operations that do not form a field; for instance, we could try to use
distributive rings with $N$ elements, possibly the modulo-$N$ ring that
suffices for statements like (\ref{eq1:AB-trace}).%
\footnote{Recall footnote `\ref{fn:field}': In marked contrast to a field, a
  ring may have zero products of nonzero elements, such as $2\odot_63=0$.}\   
For ${N=6}$ the only ring is the modulo-6 ring, and we have the usual $N^2=36$
Heisenberg--Weyl unitary operators of Sec.~\ref{sec:WSgroups}.

Let us see.
The powers of the ${N+1=7}$ operators of (\ref{eq1:prime1}) do form seven
abelian subgroups, but they do not exhaust all $36$ products $X^jZ^k$ because
quite a few of these products belong to more than one subgroup. 
For example, we have $\gamma_6X^2Z^2=(XZ)^2=-(XZ^4)^2$ and, therefore, 
the operators $XZ$ and $XZ^4$ are not complementary.

\begin{table}[tb]
\tbl{%
The twelve abelian subgroups of order six of the modulo-$6$ Heisenberg--Weyl
group of unitary operators. 
The six elements of each subgroup are given by the powers
of the period-$6$ unitary operator that generates the subgroup.
These generators $X^mZ^n$ are listed in the second column without, however,
displaying the phase factors $\Exp{\I(\pi/6)mn}$ that are needed when the
product $mn$ is odd to compensate for the $(-1)^{mn}$ 
factor in (\ref{eq1:HWgroup6}).
The last column shows which six other generators are complementary partners. 
\label{tbl:6-ring}}
{\rule{9em}{0pt}\begin{tabular}{ccl} \toprule
& Period-$6$ & Complementary \\
Subgroup & observable & partners   \\ \colrule
 0\qquad &              $X$                         & 1,\,5,\,6,\,7,\,9,\,10\\
 1 &   \makebox[2em][r]{$X$}\makebox[2em][l]{$Z$}   & 0,\,2,\,6,\,7,\,8,\,11\\
 2 &   \makebox[2em][r]{$X$}\makebox[2em][l]{$Z^2$} & 1,\,3,\,6,\,8,\,9,\,10\\
 3 &   \makebox[2em][r]{$X$}\makebox[2em][l]{$Z^3$} & 2,\,4,\,6,\,7,\,9,\,11\\
 4 &   \makebox[2em][r]{$X$}\makebox[2em][l]{$Z^4$} & 3,\,5,\,6,\,7,\,8,\,10\\
 5 &   \makebox[2em][r]{$X$}\makebox[2em][l]{$Z^5$} & 0,\,4,\,6,\,8,\,9,\,11\\
 6 &                                         $Z$    & 0,\,1,\,2,\,3,\,4,\,5\\
 7 & \makebox[2em][r]{$X^2$}\makebox[2em][l]{$Z$}   & 0,\,1,\,3,\,4,\,10,\,11\\
 8 & \makebox[2em][r]{$X^2$}\makebox[2em][l]{$Z^3$} & 1,\,2,\,4,\,5,\,10,\,11\\
 9 & \makebox[2em][r]{$X^2$}\makebox[2em][l]{$Z^5$} & 0,\,2,\,3,\,5,\,10,\,11\\
10 & \makebox[2em][r]{$X^3$}\makebox[2em][l]{$Z$}   & 0,\,2,\,4,\,7,\,8,\,9\\
11 & \makebox[2em][r]{$X^3$}\makebox[2em][l]{$Z^2$} & 1,\,3,\,5,\,7,\,8,\,9\\
\botrule
\end{tabular}\rule{9em}{0pt}}
\end{table}

In total, there are twelve abelian subgroups of six elements each, the
identity plus five more interesting ones, obtained as powers of period-6 
unitary operators. 
In Table~\ref{tbl:6-ring} we see that each of the these twelve ``generators''
has six complementary partners, so that the corresponding bases are MU.
But there are not more than three bases that are pairwise MU.
For instance, the bases `0' and `1' are MU and are both MU with bases `6' and
`7', but these are not MU themselves, so that `0,1,6' and `0,1,7' are MUB
triplets whereas `0,1,6,7' is \emph{not} a MUB quartet.  

Similarly, the modulo-$4$ ring construction fails for ${N=4}$.\cite{Durtmutu}
The modification that replaces the Galois field shifts by modulo-$N$ shifts
simply does not work, except when $N$ is prime (Sec.~\ref{sec:WSprime}) and
the two ways of shifting coincide.

         %% Section 5
%%%% file name: MUB-6.tex
%%%% input file for MUB.tex 
%%%%
%%%% last changes on 20 April 2010 by Berge
%%%%
%%%%%%%%%%%%%%%%%%%%%%%%%%%%%%%%%%%%%%%%%%%%%%

\section{Brief summary and concluding remarks}  
\label{section6}
We used the Galois-shift based Heisenberg--Weyl group to construct
first maximal sets of MUB in prime power dimensions and then the generalized
Bell states associated with them.
Several applications to quantum information processing were discussed, some in
considerable detail: dense coding and teleportation, quantum cryptography and
cloning machines, the Mean King's problem and state tomography.
Owing to the somewhat unconventional parameterization in terms of numbers that
are both field elements and ordinary integers, the approach we presented is
relatively new, and some results are rather recent.\cite{Durtmutu,Durtsept}
There are yet other applications of these techniques, including the discrete
phase operators\cite{Planat2} (that would correspond to the dual group in our
terminology), and there are interesting connections between MUB and
SIC~POVMs\cite{Zauner,Renes+al,Grassl,Colin+3:05,Appleby+2:07,Appleby:09} 
that present appealing applications in the framework of tomography and deserve
further study.   
 
Some  of these applications do not require the basic operations (addition
and multiplication) of a field, a ring structure suffices, as is the case
for instance for the SIC~POVMs, teleportation, dense coding, or the discrete
Weyl-type phase space function. 
All of them can be realized by use of the usual modulo-$N$ operations for
Hilbert spaces of arbitrary dimension.
For the construction of maximal sets of MUB, the modulo-$N$ rings are good
enough in prime dimensions only, that is: when they are fields.
This fact enables us to design the
prime-distinguishing function described in \ref{sec:app2}. 

For what concerns the construction of MUB, the dimensionality seems to play a
crucial role.  
The reasons why prime power dimensions are so special are not clearly
understood as yet, and it is certainly worth investigating this
problem in the future. 
We offer a speculation below that is suggested by the significance of the
Hilbert space dimension in quantum physics.   
  
It is worth emphasizing that the search for maximal sets of inequivalent MUB
in each dimension is related to several different mathematical problems.
The literature contains, beside the aforementioned constructions  
that use orthogonal unitary matrices\cite{india,Vlasov:04,Pittenger+1:04} and  
discrete phase space,\cite{discretewigner,BRKS07,Paz+2:05}
an abundance of valuable papers on the MUB problem related to group  
theory,\cite{india,discretewigner,Ki06a,ST07,AK07} 
angular momentum,\cite{Ki06b,Kibler09} 
finite fields and affine planes,\cite{Planat2,BE05}
and mutually orthogonal Latin squares.\cite{Beth,Paterek+3:09}
Nearly all of these constructions rely on properties of primes and  
prime powers and on an underlying finite 
field.\cite{Wootters,Kibler09,Daoud+1:10,Chau:02,Chaturvedi:02,%
Klappenecker+1:04,Vourdas:07}
After some translation the problem is equivalent to that of
finding mutually orthogonal Cartan subalgebras in the Lie algebra of
$\mathrm{SL}(N)$.\cite{Kostrikin,Boykin+3:07}.
The problem also occurs in radar science,\cite{Alltop80}
operator algebra,\cite{Popa83} and coding theory.\cite{errorcorr}
Geometric approaches to the problem are developed in
Refs.~\refcite{BE05,KRBS07}, and \refcite{KRBS08}. 

Let us try to collect here the information concerning the number of known
MUB, which depends on the number-theoretic properties of the dimension $N$. 
The following list, which by its nature is unavoidably incomplete, contains
statements about MUB and MUHM, which are easily translated into the
respectively other terminology with the aid of (\ref{eq5:base2U}) and
(\ref{eq5:U2base}). 
\begin{enumerate}\renewcommand{\theenumi}{\alph{enumi}}
\item\label{list6:1}
Maximal sets of MUB exist for all prime power dimensions, ${N=p^\m}$.
\item\label{list6:2}
If the dimension is not a power of a prime, ${N\neq p^\m}$, maximal sets of
MUB are not known.
It is highly unlikely that there are sets of MUB for ${N=6}$ with more than
three bases.
\item\label{list6:3}
For any $N\ge 2$ there exists at least one triplet of MUB. 
This is equivalent to the statement that there exists at least a pair of MUHM.
\item\label{list6:4}
For $N=2$, $3$, $4$, and $5$, all maximal sets of $N+1$ MUB are equivalent. 
\item\label{list6:6}
For $N>3$, there are certain sets of MUHM that cannot be extended to a
maximal set. 
\item\label{list6:7}
In prime dimension, $N=p$, all known sets of MUHM are equivalent to the standard
set of \ref{sec:app1}. 
It seems unlikely that there are other nonequivalent sets of MUHM, but we are
not aware of a formal proof that they do not exist. 
\item\label{list6:8}
The case of a prime power dimension, $N=p^\m$, 
can be naturally interpreted as a system of $\m$ quantum degrees of freedom,
each described in its own $p$-dimensional Hilbert space.
In this case several sets of MUB may exist with different 
entanglement properties of the basis states.
A complete set of MUB represented by block-circulant Hadamard matrices was
constructed by Combescure.\cite{Combescure:09}
For two, three, or four q-bits ($N=4$, $8$, or $16$)
the number of sets of MUB obtained from
the Heisenberg--Weyl group but differing in their entanglement properties is  
one, four, and seventeen, respectively.\cite{Zeil,RBKS05,BRKS07}
\item\label{list6:10}
In certain square dimensions, $N=d^2$, the known sets of MUB 
are larger than would follow from the factorization of $d$.  
Wocjan and Beth\cite{Beth} show that $k$ mutually orthogonal Latin squares of
order $N$ enable one to construct ${k+2}$ MUB in dimension $N^2$.
For $N=26^2$ this yields six MUB.
On the other hand, the product-ket construction described at the end of
Sec.~\ref{sec:WScomp} yields at most ${p_1^{a_1}+1}$ MUB in dimension
${N=p_1^{a_1} \cdots p_n^{a_n}}$ with $p_1<p_2<\cdots<p_n$, whenever the
dimension is not a prime power. 
It has been established\cite{Aschbacher+1:07} that one cannot do better by
using any other group to replace the Galois-shift Heisenberg--Weyl group. 
\end{enumerate}
Finally, a list entry about continuous degrees of freedom:
\begin{enumerate}\renewcommand{\theenumi}{\alph{enumi}}\setcounter{enumi}{8}
\item\label{list6:11}
Continuous degrees of freedom have, as a rule, a continuum of MUB. 
An exception is the periodic degree of freedom (``motion along a circle'') 
for which only one pair of MUB is known.  
\end{enumerate}

Items (\ref{list6:1}), (\ref{list6:2}), and (\ref{list6:8}) invite a speculation
about the difference between Hilbert space dimensions $N$ that are a power
of a prime and those that are other composite numbers. 
In the spirit of Sec.~\ref{sec:WScomp}, we follow
Schwinger's\cite{Schwinger,KinDyn,SchwingerQMbook} guidance and 
associate one quantum degree of freedom with each prime factor of $N$.
If different primes occur, we surely have a physical system composed of
different components.
But if there is only one prime, we could have indistinguishable
components, in which case the physical system behaves as one whole and the
separation into the $\m$ subsystems of (\ref{list6:8}) is artificial because
the labels $m=0,1,\dots,\m-1$ are physically meaningless.
From this physical point of view, then, it is quite satisfactory that prime
power dimensions are not so different from prime dimensions (maximal sets of
MUB for both) while other dimensions are not on the same footing (relatively few
bases that are MU).
A clear-cut demonstration that, indeed, there are no maximal sets of MUB for
${N\neq p^\m}$ is surely desirable.

         %% Section 6
%%%% file name: MUB-End.tex
%%%% input file for MUB.tex 
%%%%
%%%% last changes on 20 April 2010 by Berge
%%%%
%%%%%%%%%%%%%%%%%%%%%%%%%%%%%%%%%%%%%%%%%%%%%%

\section*{Acknowledgments}
\addcontentsline{toc}{section}{Acknowledgments}

It is a pleasure to thank  W.~Bruzda, {\AA}.~Ericsson, J.-\AA.~Larsson, and
W.~Tadej for a long-term collaboration on research projects related to
mutually unbiased bases and for allowing us to mention some of their
unpublished results. 
We are also grateful to V.~Cappellini, M.~Grassl, Z.~Jelonek, M.~Matolcsi, 
A.~Scott, A.~Uhlmann, and S.~Weigert for inspiring discussions
and to  C.W.~Chin, P.~Di{\c t}{\v a}, R.~Nicoara, A.~Schinzel, A.J.~Skinner, 
and F.~Sz{\"o}ll{\H{o}}si for helpful correspondence.
We also thank S.~Chaturvedi for explaining much of Segre's construction before
we knew it was already known.
Sincere thanks to P.~Cara for patiently answering our questions about finite
fields, and to A. Eusebi  for attracting our attention to a sign error in
Ref.~\refcite{Durtsept} and correspondence on this subject.   

The authors gratefully acknowledge support from 
the ICT Impulse Program of the Brussels Capital Region (project Cryptasc), 
the Interuniversity Attraction Poles program of the Belgian Science Policy
Office, under grant IAP~P6-10 (photonics@be),
and the Solvay Institutes for Physics and Chemistry (TD); 
the A$^*$Star Grant 012-104-0040 (BGE);
VR, the Swedish Research Council (IB); 
the  grant DFG-SFB/38/2007 of Polish Ministry of Science and Higher Education,
Foundation for Polish Science and European Regional Development Fund, under
agreement no MPD/2009/6 (K{\.Z}). 
Centre for Quantum Technologies is a Research Centre of Excellence funded by
Ministry of Education and National Research Foundation of Singapore.

\appendix
\addtocontents{toc}{\protect\setlength{\seclabwidth}{6em}}

\section{Generalized position and momentum operators for spherical
  coordinates}
\label{sec:app0}
We denote the cartesian coordinate operators by $A_1$, $A_2$, and $A_3$, and
their complementary partners are the (linear) momentum operators $B_1$, $B_2$,
and $B_3$. 
The Heisenberg commutation relations
\begin{equation}
  \label{app0:1}
  \bigl[A_j,B_k\bigr]=\I\delta_{j,k}\mathbf{1}
\end{equation}
state that we have three independent copies of the $A,B$ pair of
Secs.~\ref{sec:WSlim} and \ref{sec:WScont1}.
The operators $R,\Theta,E$ for the spherical coordinates, introduced in
Secs.~\ref{sec:WScont2}--\ref{sec:WScont4} are related to the cartesian $A_j$s
in the familiar way,
\begin{eqnarray}
  \label{app0:2}
  A_1&=&\displaystyle R\,\sin\Theta\,\frac{E+E^\dagger}{2}\,,\nonumber\\
  A_2&=&\displaystyle R\,\sin\Theta\,\frac{E-E^\dagger}{2\I}\,,\nonumber\\
  A_3&=&R\,\cos\Theta\,,
\end{eqnarray}
so that the relations
\begin{eqnarray}
  \label{app0:3}
  R&=&\displaystyle\sqrt{A_1^2+A_2^2+A_3^2}\,,\nonumber\\
 \tan\frac{\Theta}{2}&=&\displaystyle\sqrt{\frac{R-A_3}{R+A_3}}\,,
\nonumber\\
  E&=&\displaystyle\frac{A_1+\I A_2}{\displaystyle\sqrt{A_1^2+A_2^2}}
\end{eqnarray}
express the spherical coordinate operators in terms of the cartesian
coordinate operators.

Their complementary partners are linear functions of the cartesian momenta, 
\begin{eqnarray}
  \label{app0:4}
  S&=&\sum_{j=1}^3\frac{1}{2}(A_jB_j+B_jA_j)\,,\nonumber\\
  \Omega&=&\displaystyle\frac{1}{R}\Bigl(\bigl(A_1B_1+B_2A_2\bigr)A_3
                       -\bigl(A_1^2+A_2^2\bigr)B_3\Bigr)\,,\nonumber\\
  L&=&A_1B_2-A_2B_1\,.
\end{eqnarray}
Of these, the generator $S$ of scaling transformations and the generator $L$
of rotations around the $A_3$ axis are familiar operators, whereas $\Omega$ is
not standard textbook fare.

Of the fifteen commutators that involve two different ones of the operators in
(\ref{app0:3}) and (\ref{app0:4}) all vanish except for 
\begin{equation}
  \label{app0:5}
\bigl[R,S]=\I R\,,\qquad
\Bigl[\tan\frac{\Theta}{2},\Omega\Bigr]=\I \tan\frac{\Theta}{2}\,,
\qquad  \bigl[E,L\bigr]=-E\,,
\end{equation}
as one can verify with the aid of
\begin{equation}
  \label{app0:6}
  \bigl[f(A_1,A_2,A_3),B_k\bigr]
  =\I\frac{\partial}{\partial A_k}f(A_1,A_2,A_3)\,.
\end{equation}
The numerical spherical coordinates $(x,y,z)=
(r\sin\vartheta\,\cos\varphi,r\sin\vartheta\,\sin\varphi,r\cos\vartheta)$
are singular for $z=\pm r$ and, in particular, for $r=0$ and these
singularities are inherited by the corresponding operators.
The factors $R$ in (\ref{eq1:contc10}) and $\sin\Theta$ in (\ref{eq1:contd7})
bear witness thereof.

\section{Standard sets of mutually unbiased Hadamard matrices  
for prime dimension}
\label{sec:app1}
For completeness we provide here an explicit form of a maximal set of $N$ MUHM 
in the case of an arbitrary odd prime dimension, $N=p\ge 3$.
It is different from, and supplements, the example of (\ref{eq5:primeHi1}).

As a first element in the set of MUHM
let us choose the Fourier matrix (\ref{Fourier}), $H^{(0)}=F_N$.
Then introduce the diagonal unitary $N\times N$ matrix $E_N$ with matrix
elements 
\begin{equation}
  [E_N]_{jk} =\delta_{j,k}\,
  \Exp{\I \frac{2 \pi}{N} j^2}
  \quad \mbox{where} \quad j,k=0,1,2,\dots,N-1\,.
\label{D_N}
\end{equation}
It allows us to define a sequence of $N$ matrices $\left( H^{(0)},\ H^{(1)},\
  \ldots,\ H^{(N-1)} \right)$, where
\begin{equation}
  \label{Hr}
  H^{(r)} =E_N^r H^{(0)} 
  \quad \mbox{for} \quad   r=0,1,2,\dots,N-1\, .
\end{equation}
By construction all these matrices are complex Hadamard matrices. 
Furthermore, the products   
\begin{equation}
  \label{HsHr}
  X_{r-s}=
  \frac{1}{\sqrt{N}} {H^{(s)}}^{\dagger} H^{(r)} = 
  \frac{1}{\sqrt{N}} F_N^\dagger E_N^{r-s} F_N^{\ }
\end{equation}
are Hadamard matrices for all $r\ne s$ from the set $\{0,1,\dots , N-1 \}$
if and only if the dimension $N$ is an odd prime.

Hence the set $\{ H^{(0)}, H^{(1)},\dots, H^{(N-1)}  \}$
is a set of $N$ MUHM, 
referred to as the \emph{standard set of MUHM},
which generates the \emph{standard set of $N+1$ MUB}, 
according to (\ref {MUH}). 
We observe that, just like the set (\ref{eq5:primeHi1}), 
this set of Hadamard matrices is homogeneous,
since all its members arise by enphasing the same Fourier matrix $F_N$,
hence they are equivalent and  share the same core.
The equivalence of the standard set of MUHM and the set of (\ref{eq5:primeHi1})
is shown with the aid of the identity
\begin{equation}
  \label{app1:identity}
  \frac{1}{2}j(j-1)=\frac{1}{2}q(q-1)+q(j-q)^2\quad(\mathrm{mod}\ N)
\qquad\mbox{with $\ds q=\frac{1}{2}(N+1)$}\,,
\end{equation}
which is valid for all odd $N$ values.

\section{A prime-distinguishing function}
\label{sec:app2}
We return to Sec.~\ref{sec:WSprime}, but now consider the ${N+1}$ operators of
(\ref{eq1:prime1}) for arbitrary values of $N\geq2$.
In accordance with
\begin{equation}
  \label{app2:1}
  \bigl(XZ^n\bigr)^N=\left\{
    \begin{array}{cl}
      \mathbf{1} &\mbox{\ if $N$ is odd}\\[1ex]
     (-1)^n\mathbf{1} & \mbox{\ if $N$ is even}
    \end{array}\right\}=(-\mathbf{1})^{(N-1)n}
\end{equation}
for $n=0,1,2,\dots,N-1$, the eigenkets $\ket{n,k}$ of $XZ^n$ obey the
eigenvalue equation 
\begin{equation}
  \label{app2:2}
  XZ^n\ket{n,k}=\ket{n,k}\beta_N^{\ }(n)\gamma_N^k\qquad
  \mbox{with $\beta_N(n)^N=(-1)^{(N-1)n}$}\,.
\end{equation}
For $n=0$, we have the eigenkets of $X$ and choose $\beta_N(0)=1$ to
enforce consistency with (\ref{eq1:q-Fourier})--(\ref{eq1:-defXZ}), that is:
$\ket{0,j}=\ket{\widehat{j}}$;
for $n=1,2,\dots,N-1$ we choose a convenient convention for $\beta_N^{\ }(n)$ in
(\ref{app2:10}) below.
As always, we have ${\gamma^{\ }_N=\Exp{\I 2\pi/N}}$ here, and we recall the
Weyl commutation rule ${\gamma^{\ }_NXZ=ZX}$, the central algebraic property
of the period-$N$ unitary operators $X$ and $Z$ introduced in
Sec.~\ref{sec:WSexist}. 

The projector on the $k$th eigenstate of $XZ^n$ is given by
the appropriate analog of (\ref{eq1:projectors}),
\begin{equation}
  \label{app2:3}
  \ket{n,k}\bra{n,k}=\frac{1}{N}\sum_{l=0}^{N-1}
      \biggl(\frac{XZ^n}{\beta^{\ }_N(n)\gamma_N^k}\biggr)^l\,.
\end{equation}
We use this to evaluate the transition probability between 
$\ket{n,k}$ and $\ket{0,j}$ in terms of a trace,
\begin{eqnarray}
  \label{app2:4}
  \bigl|\braket{0,j}{n,k}\bigr|^2
  &=&\mathrm{tr}{\left\{\frac{1}{N^2}\sum_{l,l'=0}^{N-1}
        \biggl(\frac{XZ^n}{\beta^{\ }_N(n)\gamma_N^k}\biggr)^l
        \biggl(\frac{X}{\gamma_N^j}\biggr)^{l'}\right\}}
\nonumber\\
  &=&\frac{1}{N^2}\sum_{l=0}^{N-1}\beta_N(n)^{-l}\gamma_N^{(j-k)l}
                  \tr{\bigl(XZ^n\bigr)^lX^{-l}}\,,
\end{eqnarray}
where we have recognized that only terms with ${l+l'=0\ (\mbox{mod $N$})}$
contribute to the double sum.

As an immediate consequence of 
${\bigl(XZ^n\bigr)^l=\gamma_N^{\frac{1}{2}nl(l-1)}X^lZ^{nl}}$, we get
\begin{equation}
  \label{app2:5}
 \tr{\bigl(XZ^n\bigr)^lX^{-l}}=N\gamma_N^{\frac{1}{2}nl(l-1)}\delta^{(N)}_{nl,0}\,,
\end{equation}
where we meet the modulo-$N$ Kronecker symbol that is defined by
\begin{equation}
  \label{app2:6}
  \delta^{(N)}_{j,k}=\left\{
    \begin{array}{cl}
      1 & \mbox{\ if $j=k$ (mod $N$)}\,,\\[1ex]
      0 & \mbox{\ if $j\neq k$ (mod $N$)}\,.
    \end{array}\right.
\end{equation}
To proceed further, we write 
\begin{eqnarray}
  \label{app2:6a}
N&=&N_1N_2\geq2\,,\nonumber\\ n&=&mN_1\geq1\,,  
\end{eqnarray}
where $N_1$ is the greatest common divisor of $n$ and $N$, 
${N_1=\gcd(n,N)\geq1}$, 
which implies that $m$ and $N_2$ are co-prime, ${\gcd(m,N_2)=1}$.
For ${l=l_1N_2+l_2}$ with ${l_1=0,1,\dots,N_1-1}$ and ${l_2=0,1,\dots,N_2-1}$,
we then have
\begin{equation}
  \label{app2:7}
  \delta^{(N)}_{nl,0}=\delta^{(N)}_{ml_2N_1,0}=\delta^{(N_2)}_{ml_2,0}
                    =\delta^{(N_2)}_{l_2,0}\,,
\end{equation}
so that 
\begin{eqnarray}
  \label{app2:8}
 \tr{\bigl(XZ^n\bigr)^lX^{-l}}
  &=&N\gamma_N^{\frac{1}{2}ml_1N(l_1N_2-1)}\delta^{(N_2)}_{l_2,0}\nonumber\\
  &=&N(-1)^{ml_1(l_1N_2-1)}\delta^{(N_2)}_{l_2,0}\nonumber\\
  &=&N(-1)^{(N_2-1)l_1}\delta^{(N_2)}_{l_2,0}\,,
\end{eqnarray}
where we encounter a distinction between even and odd $N_2$ values that is
quite similar to the even-odd distinction in (\ref{app2:1}).
The last equality in (\ref{app2:8}) recognizes that ${l_1(l_1N_2-1)}$ is even
when $N_2$ is odd and that $m$ is odd when $N_2$ is even.

After combining the various ingredients, (\ref{app2:4}) turns into
\begin{equation}
  \label{app2:9}
    \bigl|\braket{0,j}{n,k}\bigr|^2
   =\frac{1}{N}\sum_{l_1=0}^{N_1-1}(-1)^{(N_2-1)l_1}\beta_N(n)^{-l_1N_2}
                \gamma_{N_1}^{(j-k)l_1}\,,
\end{equation}
and upon imposing
\begin{equation}
  \label{app2:10}
  \beta_N(n)^{N_2}=(-1)^{N_2-1}
\end{equation}
we arrive at
\begin{equation}
  \label{app2:11}
     \bigl|\braket{0,j}{n,k}\bigr|^2=\frac{1}{N_2}\delta^{(N_1)}_{j,k}\,,
\end{equation}
with the slightly frivolous convention of $\delta_{j,k}^{(1)}=1$ for all $j,k$.
Inasmuch as 
\begin{equation}
  \label{app2:12}
  \beta_N^{\ }(n)=\left\{
    \begin{array}{ll}
      \displaystyle\gamma^{\ }_{2N_2}=\gamma_{2N}^{N_1}=\Exp{\I\pi/N_2}
      &\mbox{\ if $N_2$ is even}\\[1ex]
      1 &\mbox{\ if $N_2$ is odd}
    \end{array}\right\}\qquad\mbox{for}\ n=1,2,\dots,N-1
\end{equation}
obeys the requirement in (\ref{app2:2}) and also meets the constraint
(\ref{app2:10}), it is indeed permissible to impose the latter. 
Other choices for $\beta_N^{\ }(n)$, as permitted by (\ref{app2:2}), differ
from this $\beta_N^{\ }(n)$ by a power of $\gamma_N^{\ }$, equivalent to a
cyclic relabeling of the states in the $n$th basis. 

In summary, we have
\begin{equation}
  \label{app2:13}
  \sqrt{N}\, \bigl|\braket{0,j}{n,k}\bigr|=\sqrt{N_1}\delta^{(N_1)}_{j,k}
\qquad\mbox{with}\ N_1=\gcd(n,N)
\end{equation}
for $n=1,2,\dots,N-1$.
It follows that the $0$th basis and the $n$th basis are MU only if 
$\gcd(n,N)=1$, which can be true for all $n$ only if $N$ is prime:
The $N+1$ bases of eigenstates of the operators in (\ref{eq1:prime1}) do not
constitute a maximal set of MUB if $N$ is composite.  

The general-$N$ version of (\ref{eq1:prime3}) is
\begin{equation}
  \label{app2:14}
  \braket{l}{n,k}=\frac{1}{\sqrt{N}}\beta_N(n)^{-l}
                   \gamma_N^{-kl}\gamma_N^{\frac{1}{2}nl(l-1)}\,,
\end{equation}
which follows from (\ref{app2:2}) upon recalling that
${\bra{l}Z=\gamma_N^l\bra{l}}$ and ${\bra{l+1}X=\bra{l}}$. 
This agrees with (\ref{eq1:prime3}) for odd $N$ values, 
for which ${\beta_N^{\ }(n)=1}$.  
For ${k=j+a}$ in (\ref{app2:13}) we then have
\begin{equation}
  \label{app2:15}
  \sqrt{N}\,\bigl|\braket{0,j}{n,j+a}\bigr|
=\frac{1}{\sqrt{N}}\Biggl|\sum_{l=0}^{N-1}
  \Bigl(\beta_N^{\ }(n)\gamma_N^a\Bigr)^{-l}
        \gamma_N^{\frac{1}{2}nl(l-1)}\Biggr|=\sqrt{N_1}\delta^{(N_1)}_{a,0}\,,
\end{equation}
where we choose
\begin{equation}
  \label{app2:16}
  a=\left\{
    \begin{array}{c@{\quad}l}
    \displaystyle \frac{1}{2}(N-1)n & \mbox{if $N_2$ is odd,}\\[2ex]  
    \displaystyle \frac{1}{2}(N-1)n-\frac{1}{2}N_1 
     & \mbox{if $N_2$ is even,}\\[1ex]  
    \end{array}\right.
\end{equation}
so that
\begin{equation}
  \label{app2:17}
  \beta_N^{\ }(n)\gamma_N^a=\gamma_{2N}^{(N-1)n}
\end{equation}
and, therefore,
\begin{equation}
  \label{app2:18}
  \frac{1}{\sqrt{N}}\Biggl|\sum_{l=0}^{N-1}
        \gamma_{2N}^{-(N-1)nl}
        \gamma_N^{\frac{1}{2}nl(l-1)}\Biggr|
 =\frac{1}{\sqrt{N}}\Biggl|\sum_{l=0}^{N-1}
        \gamma_{2N}^{(l-N)ln}\Biggr|
=\sqrt{N_1}\delta^{(N_1)}_{a,0}
\end{equation}
for all $N=2,3,4,\dots$ and $n=1,2,\dots,N-1$.
After taking into account that
\begin{eqnarray}
  \label{app2:19}
  &a\neq0\ (\mbox{mod}\ N_1)
&\quad\mbox{if $N$ is even and $N_2$ is odd and $m$ is odd}\nonumber\\
\mbox{whereas}\quad &a=0\ (\mbox{mod}\ N_1)
&\quad\mbox{otherwise,}
\end{eqnarray}
this states that
\begin{equation}
  \label{app2:20}
  \frac{1}{\sqrt{N}}\Biggl|\sum_{l=0}^{N-1}
        \gamma_{2N}^{(N-l)ln}\Biggr|=\left\{
\begin{array}{l}
0\quad\mbox{if $N$ even with both 
{\small$\displaystyle\frac{N}{\gcd(n,N)}$} % 
and {\small$\displaystyle\frac{n}{\gcd(n,N)}$} odd,}\\[2ex]
\sqrt{\gcd(n,N)}\quad\mbox{else,}
\end{array}\right.
\end{equation}
which can also be verified by expressing the sum over $l$ in terms of
standard Gauss sums;\cite{PC:Chin} see, for example, pages 85--90 in
Ref.~\refcite{Lang-book}. 

It follows from (\ref{app2:20}) that the function $N\mapsto g(N)$ that is
defined by 
\begin{equation}
  \label{app2:21}
  g(N)=\sum_{n=1}^{N-1}
     \Biggl(\frac{1}{\sqrt{N}}\Biggl|\sum_{l=0}^{N-1}\gamma_{2N}^{(N-l)ln}\Biggr|
            -1\Biggr)
\end{equation}
for $N>1$ can be evaluated as
\begin{equation}
  \label{app2:22}
  g(N)=\mathop{\sum{}'}\limits_{n=1}^{N-1}\sqrt{\gcd(n,N)}-(N-1)\,,
\end{equation}
where the primed summation omits all even-$N$ terms for which both
${N/\gcd(n,N)}$ and ${n/\gcd(n,N)}$ are odd.
There are no omissions if $N$ is odd or a power of $2$.

The $g(\ )$ of (\ref{app2:21})
is a \emph{prime-distinguishing function} in the sense of
\begin{eqnarray}
  \label{app2:23}
   g(N)&=&0\quad\mbox{if $N$ is prime,}\nonumber\\
   g(N)&\neq&0\quad\mbox{if $N$ is composite,}
\end{eqnarray}
because ${\gcd(n,N)=1}$ for all $n$ when $N$ is prime, whereas ${\gcd(n,N)>1}$
for some $n$ when $N$ is composite.
In the latter situation, the right-hand side of
(\ref{app2:22}) contains a sum of the irrational square roots of the prime
factors of $N$, and possibly products of these square roots, with positive 
integer weights, and no such sum can be rational, so that $g(N)$ is
irrational and ${g(N)=0}$ is impossible.%
\footnote{We owe this argument to M.~Grassl.} 

For odd $N$, equivalent forms of $g(N)$ are
\begin{eqnarray}
  \label{app2:24}
    g(N)&=&\sum_{n=1}^{N-1}
  \Biggl(\frac{1}{\sqrt{N}}
         \Biggl|\sum_{l=0}^{N-1}\gamma_N^{\frac{1}{2}(l-1)ln}\Biggr|-1\Biggr)
\nonumber\\
        &=&\sum_{r=1}^{N-1}
           \Biggl(\frac{1}{\sqrt{N}}
           \Biggl|\sum_{l=0}^{N-1}\gamma_N^{rl^2}\Biggr|-1\Biggr)\,,
\end{eqnarray}
of which the first is obtained from (\ref{app2:21}) by the shift 
${l\to l+\frac{1}{2}(N-1)}$ in the sum over $l$, and the second identity
follows from (\ref{app1:identity}) with ${r=nq}$.
Here we make contact with \ref{sec:app1}, inasmuch as
\begin{equation}
  \label{app2:25}
     g(N)=\sum_{r=1}^{N-1}\biggl(\Bigl|[X_r]_{ii}^{\ }\Bigr|-1\biggr)
         =\sum_{r=1}^{N-1}\Biggl(\frac{1}{\sqrt{N}} \biggl|
           [F_N^\dagger E_N^{r-s} F_N^{\ }]_{ii}^{\ }\biggr|-1\Biggr)\,,
\end{equation}
in accordance with (\ref{HsHr}); the index $i$ is arbitrary here because the
$X_r$s are circulant matrices. 
For $N=p$, an odd prime, we encounter in (\ref{app2:24}) the well known Gauss
sum\cite{Lang-book} 
\begin{equation}
  \label{app2:26}
      \Biggl|  \sum_{j=0}^{p-1} \gamma^{rj^2} \Biggr|=\sqrt{p}\,.
\end{equation}
As just demonstrated, it is needed to check explicitly that the Hadamard
matrices given in \ref{sec:app1} really are MU when the dimension is prime
(see, for instance, Refs.~\refcite{Ivanovic} and \refcite{Durtsept}),
and (\ref{app2:26}) is also a key ingredient for conceiving a maximally
entangling quantum gate that generalizes the two--q-bit \textsc{cnot} gate in
arbitrary dimension.\cite{DKK08}  

Concerning the composite-$N$ case of (\ref{app2:23}), we can be more specific
about ${g(N)\neq0}$.
In fact,
\begin{eqnarray}
  \label{app2:27}
  g(N)>0 && \mbox{if $N$ is odd and composite}\nonumber\\
         && \mbox{or $N$ is a multiple of $4$,}
\end{eqnarray}
and $g(N)<0$ can only occur when $N$ is an odd multiple of $2$.
The case of composite odd $N$ is immediate because there are no terms omitted
in (\ref{app2:22}).
For even $N$, we exploit the identity
\begin{eqnarray}
  \label{app2:28}
  g\bigl(2^\m\nu\bigr)&=&g\bigl(2^\m\bigr)
                         +\frac{2^\m-2^{(\m-1)/2}}{2-\sqrt{2}}g(\nu)
              +\frac{2^\m-2^{\m/2}}{2-\sqrt{2}}\bigl(\sqrt{\nu}-1\bigr)
\nonumber\\&&
              +2^{(\m-1)/2}\bigl(2^{\m/2}-1-2^{-1/2}\bigr)(\nu-1)\,,
\end{eqnarray}
which is valid for ${\m=1,2,\dots}$ and odd ${\nu\geq3}$; it holds also for
${\nu=1}$ if we adopt the convention that ${g(1)=0}$.
The first three summands on the right-hand side of (\ref{app2:28}) cannot be
negative, whereas the fourth is positive for ${\m>1}$ and negative for
${\m=1}$. 
Indeed, we have
\begin{equation}
  \label{app2:29}
  g(2p)=\frac{\sqrt{p}-1}{2+\sqrt{2}}\bigl(\sqrt{2}+1-\sqrt{p}\bigr)
\end{equation}
when $p$ is an odd prime.

\begin{figure}[t]
\centerline{\includegraphics{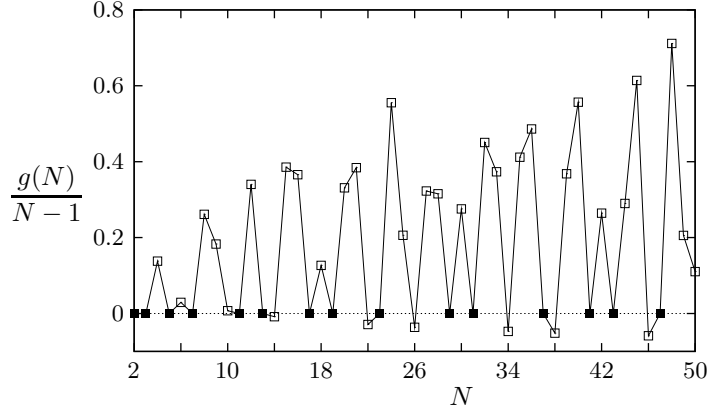}}
\caption{The prime-distinguishing function $g(N)$ of (\ref{app2:20}) for
  ${2\leq N\leq50}$; 
  for normalization, the function values are divided by ${N-1}$. 
  Straight lines connect successive values of ${g(N)/(N-1)}$ to guide the eye.
  Filled squares show where ${g(N)=0}$, which happens when $N$ is prime. 
  Empty squares indicate ${g(N)\neq0}$ and so identify composite $N$ values.
  Consistent with (\ref{app2:27}), we have ${g(N)<0}$ for $N=14$, $22$, $26$,
  $34$, $38$, and $46$, with the respective $g(N)$ values given by 
  (\ref{app2:29}).
}
\label{app2:gNfig}
\end{figure}

The value of ${g\bigl(2^\m\bigr)}$, needed in (\ref{app2:28}), is available as
the ${p=2}$ version of the general prime-power value of $g(N)$ that is given by
\begin{equation}
  \label{app2:30}
  g(p^\m)=\bigl(p^{\m/2}-1\bigr)\bigl(p^{(\m-1)/2}-1\bigr)
\quad\mbox{for $p$ prime}.
\end{equation}
We have, in particular, ${g(2p)<0}$ for ${p\geq7}$,  
${g(2p^2)<0}$ for ${p\geq29}$, and ${g(2p^\m)<0}$ for ${p\ge37}$ when ${\m>2}$.
A survey for $N$ up to $2\times10^6$ established that there are $92$, $676$,
$6\,949$, $77\,310$, and $155\,150$ $N$ values not exceeding $10^3$, $10^4$,
$10^5$, $10^6$, and $2\times10^6$, respectively, for which ${g(N)<0}$. 
These matters are illustrated in Fig.~\ref{app2:gNfig}.

Equations (\ref{app2:28}) and (\ref{app2:29}) are particular cases of the
general factorization formula
\begin{equation}
  \label{app2:31}
  h(N_1N_2)=h(N_1)h(N_2)\quad \mbox{if}\ \gcd(N_1,N_2)=1\,,
\end{equation}
where the auxiliary function ${N\mapsto h(N)}$ is defined by
\begin{eqnarray}
  \label{app2:32}
  h(N)&=&g(N)+\left\{
    \begin{array}{l@{\quad}l}\displaystyle
      N+\sqrt{N}-1 & \mbox{if $N$ is odd}\\[1.5ex]\displaystyle
      N+\frac{1}{2}\sqrt{N}-1 & \mbox{if $N$ is even}\\
    \end{array}\right\}\nonumber\\
   &=&\frac{1}{4}\bigl(3-(-1)^N\bigr)\sqrt{N}
      +\sum_{n=1}^{N-1}
     \frac{1}{\sqrt{N}}\Biggl|\sum_{l=0}^{N-1}\gamma_{2N}^{(N-l)ln}\Biggr|      
\end{eqnarray}
for ${N=1,2,3,\dots}$; consistent with the convention ${g(1)=0}$, we have
${h(1)=1}$. 
One establishes (\ref{app2:31}) by an exercise in counting that exploits the
explicit form of ${g(N)}$ in (\ref{app2:22}).

We observe, as an immediate consequence of (\ref{app2:31}), that
\begin{equation}
  \label{app2:33}
h(N)=h\bigl(p_1^{\m_1}\bigr)h\bigl(p_2^{\m_2}\bigr)h\bigl(p_3^{\m_3}\bigr)\cdots
\end{equation}
if ${N=p_1^{\m_1}p_2^{\m_2}p_3^{\m_3}\cdots}$ is the prime-factor
decomposition of $N$.
In conjunction with (\ref{app2:30}) this facilitates the computation of
${g(N)}$ without an actual evaluation of the summations in (\ref{app2:21}) or
(\ref{app2:22}). 

As a final remark we note that the derivation of (\ref{app2:20}) with
quantum-mechanical reasoning in the context of searching for MUB in dimension
$N$ seems to indicate that the existence
problem of maximal sets of MUB and MUHM is related to number-theoretical
properties of the dimension. 
We leave the matter at that.

\section{Mutually unbiased bases for $N=4$}\label{sec:app3}
In accordance with (\ref{xxx}), the set of MUHM for the maximal set of MUB for
$N=p^\m$ of Sec.~\ref{section2} is given by
\begin{equation}
  \label{eq:app3-1}
  \bigl[H^{(N)}_j\bigr]_{k,l}=\sqrt{N}\braket{e^N_k}{e^j_l}
  ={\alpha^j_{\ominus k}}^*\gamma^{\ominus k \odot l}
\end{equation}
for $j,k,l=0,1,\dots,N-1$, so that $H^{(N)}_j=A^{(N)}_jG_N^{-1}$ 
is the product of the inverse Galois--Fourier matrix with matrix elements
\begin{equation}
  \label{eq:app3-2}
  \bigl[G_N^{-1}\bigr]_{k,l}=\gamma^{\ominus k \odot l}
\end{equation}
and the diagonal matrix of phase factors
\begin{equation}
  \label{eq:app3-3}
  \bigl[A^{(N)}_j\bigr]_{k,l}=\delta_{k,l}{\alpha^j_{\ominus k}}^*
\end{equation}
with $A^{(N)}_0=\mathbbm{1}_N^{\ }$ and $H^{(N)}_0=G^{-1}_N$ in particular for
the $0$th basis, the dual basis.
The conventional choices for $\alpha^j_l$ are found in (\ref{conven}) for odd
$N$ and in (\ref{eq2:even-alpha}) for even $N$. 
For even $N=2^\m$, we note that $\ominus l=l$ for all field elements
and $G_N^{-1}=G^{\ }_N$ since $\gamma=-1$ . 

As an example, we consider $N=4$ with the field addition and multiplication
tables of Table~\ref{tbl:4-field}(a). 
The Fourier--Galois matrix $G_4$ is the tensor product of $G_2$ with itself,
\begin{equation}
  \label{eq:app3-4}
 H^{(4)}_0= G_4^{-1}=G_4^{\ }=\left(
    \begin{array}{rrrr}
    1 & 1 & 1 & 1 \\ 1 & -1 & 1 & -1 \\ 1 & 1 & -1 & -1 \\ 1 & -1 & -1 & 1  
    \end{array}\right)=
    \left(\begin{array}{rr}G_2 & G_2 \\ G_2 & -G_2\end{array}\right)
    =G_2\otimes G_2\,,
\end{equation}
where $G_2$ is the $2\times2$ Hadamard matrix of (\ref{eq1:qbitHada}).
We are reminded here of the sign sequences in (\ref{eq4:Wigner-12a}).
The binary components $l=(l_0,l_1)$ of the four field elements $0=(0,0)$,
$1=(1,0)$, $2=(0,1)$, and $3=(1,1)$ are needed for the calculation of the phase 
factors
\begin{equation}
  \label{eq:app3-5}
  N=4:\qquad {\alpha^j_{\ominus l}}^*={\alpha^j_l}^*
        =\prod_{m,n=0}^1(-\I)^{j\odot(l_m2^m)\odot(l_n2^n)}
\end{equation}
along with ${2^0\odot2^0=1}$, ${2^0\odot2^1=2^1\odot2^0=2}$,  ${2^1\odot2^1=3}$.
This gives
\begin{equation}
  \label{eq:app3-6}
  {\alpha^j_0}^*=1\,,\quad  {\alpha^j_1}^*=(-\I)^{j\odot1}=(-\I)^j\,,\quad
  {\alpha^j_2}^*=(-\I)^{j\odot 3}\,,
\end{equation}
and
\begin{equation}
  \label{eq:app3-7}
   {\alpha^j_3}^*=(-\I)^{j\odot1}\bigl[(-\I)^{j\odot2}\bigr]^2(-\I)^{j\odot3}
                =(-\I)^{j+j\odot3}(-1)^{j\odot2}\,.
\end{equation}
The resulting phase matrices are $A^{(4)}_0=\mathbbm{1}_4$ and
\begin{equation}
  \label{eq:app3-8}
  A^{(4)}_1=\left(\begin{array}{rrrr}
1 & \phantom{-}0 & \phantom{-}0 & \phantom{-}0\\ 
0 & -\I & 0 & 0 \\ 0 & 0 & \I  & 0\\ 0 & 0 & 0 & 1  
    \end{array}\right),\quad
  A^{(4)}_2=\left(\begin{array}{rrrr}
1 & \phantom{-}0 & \phantom{-}0 & \phantom{-}0\\ 
0 & -1 & 0 & 0 \\ 0 & 0 & -\I & 0\\ 0 & 0 & 0 & -\I  
    \end{array}\right),\quad
  A^{(4)}_3=\left(\begin{array}{rrrr}
1 & \phantom{-}0 & \phantom{-}0 & \phantom{-}0\\  
0 & \I & 0 & 0 \\ 0 & 0 & -1 & 0\\ 
0 & 0 & 0 &  \I  
    \end{array}\right),
\end{equation}
and the Hadamard matrices are $H^{(4)}_0=G_4$ as well as
\begin{equation}
  \label{eq:app3-9}
  H^{(4)}_1=\left(\!\begin{array}{rrrr}
1 & \phantom{-}1 & \phantom{-}1 & \phantom{-}1\\ 
-\I & \I & -\I & \I \\ \I & \I & -\I & -\I \\ 1 & -1 & -1 & 1  
    \end{array}\right),\ 
  H^{(4)}_2=\left(\!\begin{array}{rrrr}
1 & \phantom{-}1 & \phantom{-}1 & \phantom{-}1\\ 
-1 & 1 & -1 & 1 \\ -\I & -\I & \I & \I \\ -\I & \I & \I & -\I  
    \end{array}\right),\ 
  H^{(4)}_3=\left(\!\begin{array}{rrrr}
1 & \phantom{-}1 & \phantom{-}1 & \phantom{-}1\\ 
\I & -\I & \I & -\I \\ -1 & -1 & 1 & 1\\ \I & -\I & -\I & \I  
    \end{array}\right).
\end{equation}
When multiplied by $\frac{1}{\sqrt{N}}=\frac{1}{2}$, the columns of $H^{(4)}_j$
represent the kets of the $j$th bases with reference to the computational
basis, the $4$th basis.%
\footnote{$H_4^{(4)}=\mathbbm{1}_4^{\ }$, so to say, but no factor of
  $\frac{1}{2}$ for the $4$th basis.}\ 
Up to relabeling, they coincide with those derived by Bandyopadhyay
\textit{et al.\/} in a similar fashion.\cite{india}

       %% Acknowledgments, references, appendixes

\end{document}